\documentclass{aa}    

\usepackage{times}                          
\usepackage{txfonts}
\usepackage{booktabs}
\usepackage{caption}
\usepackage{subcaption}
\newcommand{\Si}[5]{\mbox{$#1\,^#2{\rm #3}^{{\rm #4}}_{\rm #5}$}}

\newcommand{\Teff}{T_{\rm eff}}
\newcommand{\kms}{km s$^{-1}$}
\newcommand{\ms}{m s$^{-1}$}

\newcommand{\gfe}{\log gf\varepsilon_\odot}

\newcommand{\lgesi}{\log \varepsilon_{\text{Si}}}

\newcommand{\cobold}{\texttt{CO$^5$BOLD}}
\newcommand{\linfor}{\texttt{LINFOR3D}}
\newcommand{\stagger}{\texttt{STAGGER}}
\newcommand{\balder}{\texttt{BALDER}}
\newcommand{\chisq}{$\chi^2$}

\newcommand{\abufinal}{$7.57\pm0.04$}

\newcommand{\lineAng}[1]{$#1~\mathrm{\AA}$}

\newcommand{\wk}{$w_\mathrm{K}$}

\newcommand{\ddx}{\ensuremath{\alpha\,|x|}}
\newcommand{\ddxsq}{\ensuremath{\alpha^2\,|x|^2}}
\newcommand{\beq}{\begin{equation}}
\newcommand{\eeq}{\end{equation}}
\newcommand{\x}{\ensuremath{x}}
\newcommand{\xx}{\ensuremath{x^\prime}}
\newcommand{\eref}[1]{\mbox{(\ref{#1})}}

\newcommand{\Ac}[1]{\ensuremath{\mu_\mathrm{#1}}}
\newcommand{\Aa}[1]{\ensuremath{\varepsilon_\mathrm{#1}}}
\newcommand{\Acm}[1]{\ensuremath{\mu^\diamond_\mathrm{#1}}}
\newcommand{\Aam}[1]{\ensuremath{\varepsilon^\diamond_\mathrm{#1}}}

\newcommand{\Aas}[1]{\ensuremath{\varepsilon^\odot_\mathrm{#1}}}
\newcommand{\nd}[1]{\ensuremath{n_\mathrm{#1}}}
\newcommand{\nds}[1]{\ensuremath{n^\odot_\mathrm{#1}}}
\newcommand{\ndm}[1]{\ensuremath{n^\diamond_\mathrm{#1}}}
\newcommand{\mfX}{\ensuremath{\mathrm{X}}}
\newcommand{\mfY}{\ensuremath{\mathrm{Y}}}
\newcommand{\mfZ}{\ensuremath{\mathrm{Z}}}

\usepackage{graphics,graphicx,epsfig}	
\graphicspath{{figs/}}

\usepackage{amsmath} 						

\usepackage{etoolbox}                       
\makeatletter
\patchcmd{\frontmatter@RRAP@format}{(}{}{}{}
\patchcmd{\frontmatter@RRAP@format}{)}{}{}{}
\makeatother

\usepackage{fancyhdr}
\usepackage{gensymb}

\usepackage{wrapfig}

\usepackage[compact]{titlesec}         
\titlespacing{\section}{0pt}{2pt}{2pt} 

\begin{document}

\title{The solar photospheric silicon abundance according to
    CO$^5$BOLD}
\subtitle{Investigating line broadening, magnetic fields, and model
    effects}
\date{Submitted: \today{}}
\author{S.A.\,Deshmukh \inst{1}
    \and H.-G.\,Ludwig\inst{1,5}
    \and A.\,Ku\v{c}inskas\inst{2}
    \and M.\,Steffen\inst{3}
    \and P.S.\,Barklem\inst{4}
    \and E.\,Caffau\inst{5}
    \and V.\,Dobrovolskas\inst{2}
    \and P.\,Bonifacio\inst{5}
}

\institute{
Zentrum f\"ur Astronomie der Universit\"at Heidelberg,
Landessternwarte, K\"onigstuhl 12,
69117 Heidelberg, Germany\\
\email{sdeshmukh@lsw.uni-heidelberg.de}
\and
Astronomical Observatory of Vilnius University,
Saul\.{e}tekio al. 3, Vilnius LT-10257, Lithuania
\and
Leibniz-Institut f\"ur Astrophysik Potsdam, An der Sternwarte
16, D-14482 Potsdam, Germany
\and
Theoretical Astrophysics,
Department of Physics and Astronomy, Uppsala
University, Box 516, SE-751 20 Uppsala, Sweden
\and
GEPI, Observatoire de Paris, PSL Research University, CNRS,
Place Jules Janssen, 92190 Meudon, France
}

\date{Received 23 August 2021; accepted 1 August 2022}

\abstract
{In this work, we present a photospheric solar silicon
    abundance derived using \cobold\ model atmospheres and the \linfor\
    spectral synthesis code. Previous works have differed in their
    choice of a spectral line sample and model atmosphere as well as their
    treatment of observational material, and the solar silicon
    abundance has undergone a downward revision in recent years. We
    additionally show the effects of the chosen line sample, broadening
    due to velocity fields, collisional broadening, model spatial
    resolution, and magnetic fields.}
{Our main aim is to derive the photospheric solar silicon
    abundance using updated oscillator strengths and to mitigate model
    shortcomings such as over-broadening of synthetic spectra. We also
    aim to investigate the effects of different line samples,
    fitting configurations, and magnetic fields on the fitted
    abundance and broadening values.}
{\cobold\ model atmospheres for the Sun were used in conjunction with
    the \linfor\ spectral synthesis code
    to generate model spectra, which were then fit to observations in the
    Hamburg solar atlas. We took
    pixel-to-pixel signal correlations into account by means of a
    correlated noise model. The choice of line
    sample is crucial to determining abundances, and we present a sample
    of 11 carefully selected lines (from an initial choice of 39 lines) in both
    the optical and
    infrared, which has been made possible with newly determined oscillator
    strengths for the majority of these lines. Our final sample
    includes seven optical \ion{Si}{I} lines, three infrared \ion{Si}{I} lines,
    and one optical \ion{Si}{II} line.}
{We derived a photospheric solar silicon abundance of
    $\lgesi$ = \abufinal , including a $-0.01$ dex correction from Non-Local
    Thermodynamic Equilibrium (NLTE) effects. Combining
    this with meteoritic abundances and previously determined photospheric
    abundances results in a metal mass fraction $\mfZ/\mfX=0.0220~\pm~0.0020$.
    We found a tendency of obtaining overly broad synthetic lines. We mitigated
    the impact of this by devising a de-broadening procedure. The
    over-broadening of synthetic lines does not substantially affect the
    abundance determined in the end. It is primarily the line selection that
    affects the final fitted abundance.}
{}


\maketitle
%

\section{Introduction}
\label{sec:introduction}
\noindent
The chemical composition of the solar photosphere serves as a widely applied
yardstick in astronomy, since it is considered largely representative of the
chemical make-up of the present-day Universe. The determination of solar
abundances has a long history, dating back at least  to the 1920s
\citep{payneStellarAtmospheres1925,unsoeldUeberStruktur1928,
    russellCompositionSun1929}. Among the elements which are
spectroscopically accessible in the solar photosphere, silicon plays a
somewhat special role. Besides being relatively abundant, it is used as a
reference to relate the solar photospheric composition to the composition of
type-I carbonaceous chondrites which are believed to constitute fairly
pristine samples of material from the early Solar System
\citep[e.g.][]{andersAbundancesElements1989,loddersSolarSystem2003,
    loddersAbundancesElements2009,palmeSolarSystem2014}.
This important aspect of the silicon abundance led to many investigations trying
to establish an ever more precise and accurate solar reference value.
Over the years, atomic and observational data have improved, and increasingly
sophisticated modelling techniques
have been applied, including time-dependent three-dimensional (3D) model
atmospheres \citep[e.g.][]{caffauSolarChemical2011, pereiraHowRealistic2013,
    asplundLineFormation2000} and treatment of departures
from thermodynamic equilibrium (NLTE)
\citep[e.g.][]{bergemannObservationalConstraints2019, amarsi3DNonLTE2019,
    steffenPhotosphericSolar2015}.

From the above, it is tempting to conclude that the determination of
the abundance of a particular element in the solar photosphere is an entirely
objective process. Unfortunately, this is not
the case, due to several judicious decisions a researcher must make along the
way, namely: \textbf{i)} which observational material to use, \textbf{ii)}
where to place the continuum when normalising
the spectrum, \textbf{iii)} which lines have accurate atomic data (meaning
oscillator strengths and broadening
constants), and \textbf{iv)} which lines are largely unaffected by blends.
These aspects influence the final outcome of an analysis, and we
list some previous works' results below to illustrate the evolution of the
fitted solar silicon
abundance with time. Indeed, as we shall later show, statistical uncertainties
are of secondary importance here, and it is primarily the line selection that
dominates the final outcome (including
uncertainties). Moreover, in the present analysis process, we found that more
sophisticated model atmospheres do not necessarily give a more coherent picture,
since they can bring to light modelling shortcomings which were not recognised
before.

\citet{holwegerSolarAbundance1973} determined an LTE solar silicon abundance of
$\lgesi = 7.65 \pm 0.07$ based on 19 \ion{Si}{I} lines whose oscillator
strengths were measured by \citet{garzAbsoluteOscillator1973}.
\citet{wedemeyerStatisticalEquilibrium2001}
derived a 1D NLTE correction of -0.010\,dex for silicon. Together with a
correction of the scale of Garz by \citet{beckerSolarMeteoritic1980}, they
arrived at a silicon abundance of $\lgesi = 7.550 \pm 0.056$. Some of the first
multi-dimensional (multi-D) studies were carried out by
\citet{asplundLineFormation2000} and
\citet{holwegerPhotosphericAbundances2001a}, who arrived at
$\lgesi = 7.51\pm0.04$ and $\lgesi=7.536\pm0.049$, respectively.
The statistical equilibrium of silicon and
collisional processes were investigated by
\citet{shiStatisticalEquilibrium2008}, and
an abundance of $\lgesi = 7.52 \pm 0.06$ was found, taking an extended line
sample into account. They also specified that the NLTE effects on
optical silicon lines are weak, but it was later found that near-infrared
lines have sizeable NLTE effects \citep{shiSiliconAbundances2012,
    bergemannRedSupergiant2013}. \citet{shchukinaNonlteDetermination2012}
conducted an NLTE analysis of 65 \ion{Si}{I} lines, and found
$\lgesi = 7.549 \pm 0.016$. \citet{shaltoutAbundanceSilicon2013} obtained a
3D~LTE solar silicon abundance of $\lgesi = 7.53 \pm 0.07$ and a 1D~NLTE
abundance of $\lgesi = 7.52 \pm 0.08$, using the aforementioned -0.010 dex
correction.

More recently, \citet{scottElementalComposition2015} conducted a 3D~LTE study
and found an abundance as low as $\lgesi = 7.52 \pm 0.03$. The result was later
corroborated by \citet{amarsiSolarSilicon2017} who derived a 3D~NLTE
correction of $-0.01$\,dex to this abundance. This
analysis used nine \ion{Si}{I} lines and one \ion{Si}{II} line in the optical
wavelength range. The final abundance was calculated by means of a weighted
average. All in all, over the last twenty years, a slight downward trend of the
derived solar abundance of silicon has become apparent, but, on a level which is
on the edge of being statistically significant.

In the present work, we apply \cobold\ model atmospheres
\citep{freytagSimulationsStellar2012} to derive the photospheric abundance of
silicon in the Sun. Primarily, the motivation to do so was the availability of
new data for the oscillator strengths of silicon lines
\citep{pehlivanrhodinExperimentalComputational2018}.
This enlarged the set of silicon lines that could be potentially useful in an
abundance analysis, adding oscillator strengths for near-infrared lines.
Additionally,  we intended to relate solar abundances so far derived with
\cobold\ models \citep[for a summary, see][]{caffauSolarChemical2011} to the
meteoritic abundance scale.

The necessary spectral synthesis calculations were performed with the
LINFOR3D code\footnote{https://www.aip.de/Members/msteffen/linfor3d} in
LTE approximation, with only a few exceptions. To compare this with
observations, we
developed a custom spectral fitting routine that accounts for correlated
photometric noise in the observations. Our analysis stands out by using
a sizeable number of lines with carefully
calculated line broadening constants, new oscillator strengths, and
investigating systematic
shortcomings of our 3D model atmosphere. The last point became important
since we found that our model was
predicting systematically overly broadened lines, seen also in
\citet{caffauPhotosphericSolar2015} for oxygen lines.

The rest of the paper is structured as follows:
Section~\ref{sec:method} discusses the details of the model atmospheres used in
this study, the methodology of line selection, and spectral synthesis.
It also touches on the role of magnetic field effects pertaining to these
topics. Section~\ref{sec:fitting} describes the fitting routine and the
implemented
correlated noise model. Our results, and the differences due to model choice,
are shown in Section~\ref{sec:results},
discussed in Section~\ref{sec:discussion}, and summarised in
Section~\ref{sec:conclusion}. Finally, the appendices explain model
atmosphere differences in abundance and broadening
(Appendix~\ref{appendix:anish}), differences in regard to NLTE
(Appendix~\ref{appendix:matthias}), the choice of the broadening kernel
(Appendix~\ref{appendix:ddsmooth}), the centre-to-limb variation of the
continuum (Appendix~\ref{appendix:clv}), and the partition functions used in
\linfor\ (Appendix~\ref{appendix:partition}) in further detail.
\section{Stellar atmospheres and spectral synthesis}
\label{sec:method}

\subsection{Model atmospheres}
\label{subsec:model_atmospheres}
Systematic errors from spectral synthesis and fitting come from
the use of 1D hydrostatic model atmospheres, and the assumption of LTE when
it is not valid to do so. 1D hydrostatic model atmospheres
rely on mixing-length theory
\citep{bohm-vitenseUberWasserstoffkonvektionszone1958,
    henyeyStudiesStellar1965}, and introduce additional free parameters such
as micro- and macro-turbulence \citep{grayObservationAnalysis2008}.
Spectral lines generated from these model atmospheres are too narrow to fit
observations, if they do not take macroscopic broadening into account,
necessitating the use of free parameters to fit observations. 3D model
atmospheres, on the other hand, should in principle
be able to reproduce line shapes, shifts and asymmetries, and spectral
lines synthesised with these can be directly fit to observations.

The prominent state-of-the-art radiative-convective
equilibrium 1D models of solar and stellar atmospheres such as
ATLAS \citep{kuruczATLAS12SYNTHE2005}, MARCS
\citep{gustafssonGridModel1975, gustafssonGridMARCS2008} and PHOENIX
\citep{allardModelAtmospheres1995, allardLimitingEffects2001} use classical
mixing-length theory, and the efficiency of convective energy transport here
is controlled by a free parameter $\alpha_{\text{MLT}}$. This, along with
the requirement of fitting the micro- and macro-turbulence free parameters
during line synthesis, is a major drawback of 1D models.

3D atmospheres account for the time-dependence and
multi-dimensionality of the flow, and a
spectral synthesis using these atmospheres reproduces line shapes and
asymmetries. Prominent examples include \stagger\
\citep{magicStaggergridGrid2013}, MuRAM
\citep{voglerSimulationsMagnetoconvection2005},
Bifrost \citep{gudiksenStellarAtmosphere2011}, and
Antares \citep{leitnerStructureSolar2017}. In this work, we use \cobold\ ,
a conservative hydrodynamics solver able to model surface convection,
waves, shocks and other phenomena in stellar objects
\citep{freytagSimulationsStellar2012}.
As 3D atmospheres are understandably expensive to run,
their output can be saved as a sequence where flow properties are recorded,
commonly called a sequence of `snapshots'.
These snapshots are then used in spectral synthesis codes, such as
\linfor\ (used in this study) or \textrm{MULTI3D}
\citep{leenaartsMULTI3DDomainDecomposed2009}. The parameters of the
\cobold\ model atmospheres used in this work are summarised in
Table~\ref{table:model_atm}. We use 20 model snapshots to compute line
syntheses with \linfor. Throughout this work, the `Model ID'
(see Table~\ref{table:model_atm}) for each model will be used to refer to
it.


The msc600, m595, b000 and b200 models all use a short characteristics
scheme with double Gauss-Radau quadrature and 3 $\mu$-angles. The n59 model
uses a long characteristics Feautrier scheme with Lobatto quadrature and 4+1
$\mu$-angles. In the current version of \cobold\, the former scheme is given the
name ``MSCrad'' and the latter is given the name ``LCFrad''
\citep{freytagSimulationsStellar2012,steffenRadiationTransport2017}.

\begin{table*}
    \caption{
        3D model properties
    }
    \centering
    \begin{tabular}{lllllll}
        \noalign{\smallskip}\hline\hline\noalign{\smallskip}
        Model ID    & Box Size                  & Resolution                     & Grid Points                                          & $T_{\text{eff}}$ & <$B_\textrm{z}$> & Rad. Trans. \\
        (d3gt57g44) & [Mm$^3$]                  & [km$^3$]                       & $N_\mathrm{x}\times N_\mathrm{y}\times N_\mathrm{z}$ & [K]              & [G]              & Module      \\
        \noalign{\smallskip}\hline\noalign{\smallskip}
        msc600      & $8.0\times 8.0\times 2.3$ & $32\times 32\times 10\ldots15$ & $250\times 250\times 207$                            & $5773 \pm 9.4$   & --               & MSCrad      \\
        n59         & $5.6\times 5.6\times 2.3$ & $40\times 40\times 15$         & $140\times 140\times 150$                            & $5774 \pm 16$    & --               & LCFrad      \\
        m595        & $5.6\times 5.6\times 2.3$ & $40\times 40\times 15$         & $140\times 140\times 150$                            & $5775 \pm 15$    & --               & MSCrad      \\
        \midrule
        b000        & $5.6\times 5.6\times 2.3$ & $40\times 40\times 15$         & $140\times 140\times 150$                            & $5750 \pm 19$    & 0                & MSCrad      \\
        b200        & $5.6\times 5.6\times 2.3$ & $40\times 40\times 15$         & $140\times 140\times 150$                            & $5793 \pm 17$    & 200              & MSCrad      \\
        \noalign{\smallskip}\hline\noalign{\smallskip}
    \end{tabular}
    \tablefoot{
        ``Model ID'' lists the name of a 3D model
        used in this paper (note all models share the prefix ``d3gt57g44''),
        ``Box Size'' the geometrical size of its
        computational domain, ``Resolution'' the applied grid spacing,
        ``Grid Points'' the number of mesh points per dimension,
        ``$T_{\text{eff}}$'' the effective temperature
        of the model, ``<$B_\mathrm{z}$>'' the mean magnetic
        field component in vertical direction, and ``Rad. Trans. Module'' the
        name of the radiation transport module used (see Sec.
        \ref{subsec:model_atmospheres} for
        an explanation). Different
        from the other models, the msc600 model uses a variable grid spacing in the
        vertical direction; for that reason the covered range is provided. We note
        that `$\pm$'
        in ``$T_{\text{eff}}$'' should not be interpreted as an uncertainty: it is the natural
        dispersion of effective temperature in the time series illustrating fluctuations between
        snapshots. Each model has $\log(\mathrm{g})=4.44$ and solar metallicity.
    }
    \label{table:model_atm}
\end{table*}

\subsection{Line sample}
Silicon is an interesting element as it is an important electron donor in
late-type stars.  Though it seems that \ion{Si}{I}'s ionisation
potential of 8.15~eV would mean a significant amount of silicon
to be present in the form of \ion{Si}{II}, the higher
ionisation potential of \ion{Si}{II} is unfavourable for the appearance of
strong lines in the solar spectrum at wavelengths longer than 3000~\AA\
\citep{mooreSiliconSun1970, russellCompositionSun1929}.
We therefore primarily consider \ion{Si}{I} lines and two carefully chosen
\ion{Si}{II} lines.

The \ion{Si}{I} and \ion{}{II} line list below
(Table~\ref{table:line_list}) was compiled from the line lists used in the
solar abundance determination by
\citet{holwegerPhotosphericAbundances2001a},
\citet{wedemeyerStatisticalEquilibrium2001},
\citet{shiStatisticalEquilibrium2008}, and \citet{amarsiSolarSilicon2017}.
The initial line selection was performed by
synthesising line profiles and comparing to observations, thus checking
these for blends.

We solely used the observational data by \citet{neckelSolarRadiation1984}
(hereafter the 'Hamburg spectrum') and have worked with the disc-centre and
disc-integrated spectrum. \citet{doerrHowDifferent2016} show the resolution
of the Hamburg spectrum to be up to $520,000$ compared to the often-used
Liege spectrum \citep{delbouilleAtlasPhotometrique1973},
which practically was shown to have a resolution of $\sim 216,000$.
The higher spectral resolution allows for a more continuous representation
of the pixel-to-pixel variations in the spectrum, and we utilise a noise
model that represents the covariance between pixels.
Additionally, the Liege spectrum covers a range of $3000 - 10000$\,\AA,
while the Hamburg spectrum covers a range of $3290 - 12510$\,\AA,
affording us access to very clean near-infrared lines. Furthermore, it remains
unclear whether the Liege spectrum is a true disk-centre spectrum or an integral
over a narrow range of angles around disk-centre, and the available
documentation does not precisely establish this.

From the original 39 lines, we chose a subset of 11 based on comparisons
between disk-centre and disk-integrated
spectra, the line shape and the precision of oscillator strength. We
focus on lines that do not have strong line
blends that can interfere with abundance determination. Each line was
weighted on a scale of $1 - 3$ based on these
criteria, and the resulting weights were used to compute the mean
abundance. This idea of a weighted mean was
inspired by the work of \citet{amarsiSolarSilicon2017}.

There are various sources of error that can enter during the spectroscopic
fitting, but the choice of oscillator strength for each spectral line is often
the largest one. We use semi-empirical $gf$-values from
\citet{pehlivanrhodinExperimentalComputational2018} (hereafter also PR18)
for every line in our chosen sub-sample, which have
been updated with respect to previous experimental values measured by
\citet{garzAbsoluteOscillator1973} with accurate lifetime renormalisations by
\citet{obrianRadiativeLifetimes1991,obrianVacuumUltraviolet1991} (hereafter also
G73+OBL91). The $gf$-values from PR18 were obtained by combininb calculated
level lifetimes with experimental branching ratios. Moreover, the calculations
were validated against existing measurements of level lifetimes and oscillator
strengths. An advantage of the PR18 data is the homogeneity of the extensive
set of line transitions, ranging from UV to infrared wavelengths.
On average across all lines in Table~\ref{table:line_list}, the new values give
a $\Delta\log{gf}=0.024$ decrease with respect to the previous measurements, and
lead to a respective increase of the mean abundance by the same value when
using all lines. Using the new values of the oscillator strengths lowers the
overall scatter in fitted abundance values by $0.07$ dex and the formal
uncertainty by $0.01$ dex.

\begin{figure*}
    \centering
    \begin{subfigure}{0.49\textwidth}
        \includegraphics[width=\textwidth]{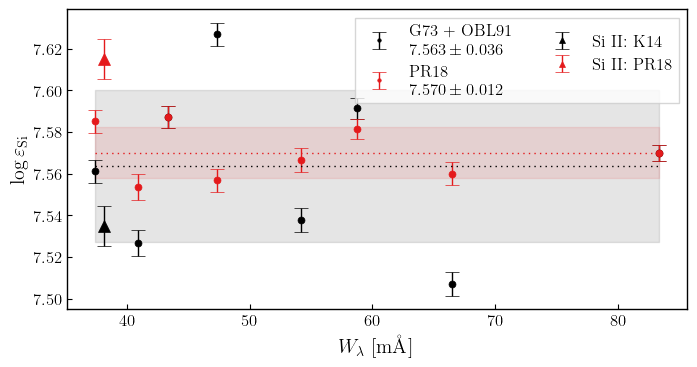}
    \end{subfigure}
    \hfill
    \begin{subfigure}{0.49\textwidth}
        \includegraphics[width=\textwidth]{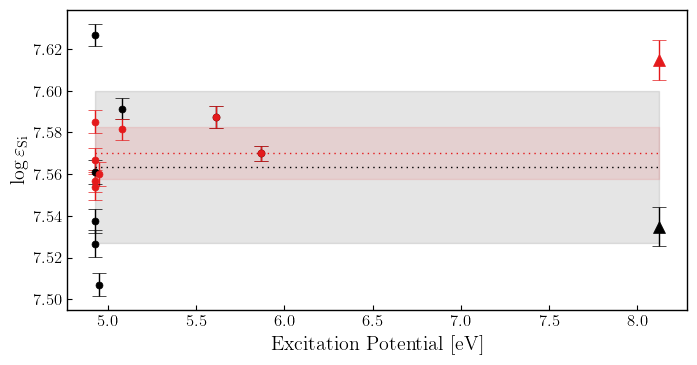}
    \end{subfigure}
    \caption{Individual fitted Si abundances for a set of eight \ion{Si}{I} lines
        and one \ion{Si}{II} line as a function of equivalent width
        (\textbf{left}) and excitation potential (\textbf{right}) for eight
        \ion{Si}{I} lines and one \ion{Si}{II} line. Only the \ion{Si}{I} lines
        are used to determine the mean and RMSE (see definition in
        Eq.~\ref{eq:rmse}), given in the legend in the left panel. G73+OBL91
        indicates oscillator strengths
        from \citet{garzAbsoluteOscillator1973} normalised according to the
        results of \citet{obrianRadiativeLifetimes1991}; the oscillator strength
        for the \ion{Si}{II} line for comparison was taken from \citet{K14};
        PR18 indicates oscillator strengths from
        \citet{pehlivanrhodinExperimentalComputational2018}.}
    \label{fig:abu_diff}
\end{figure*}

We investigated the overall performance of the PR18 $gf$-values on the set of
lines used in \citet{scottElementalComposition2015} and
\citet{amarsiSolarSilicon2017}.
Fig. \ref{fig:abu_diff} depicts the resulting Si abundances. The use of the PR18
data reduces the overall scatter in the Si abundances from $0.036$ dex
(applying the data of G73+OBL91) to $0.018$ dex. In all cases, there is no
discernible dependence on
line strength or excitation potential. The only \ion{Si}{II} line at
\lineAng{6371} that we consider sufficiently reliable for abundance analysis
may indicate a slight ionisation imbalance. However, this imbalance is of
similar magnitude for both sets of oscillator strengths. Moreover, the line
has a high excitation potential and is partially blended (see
Fig.~\ref{fig:line_syn}) so that the apparent imbalance could be the result of
systematics in the analysis.

In previous studies, strong infrared \ion{Si}{I} lines above
$10000$\,\AA\ often lacked reliable experimental oscillator strengths
\citep{borreroAccurateAtomic2003, shiStatisticalEquilibrium2008,
    shchukinaNonlteDetermination2012}. \citet{shiSiliconAbundances2012}
investigate near-infrared lines in nearby stars with both LTE and NLTE,
and find that there is a larger departure from LTE for the infrared lines
than optical ones. They point out that weak lines are insensitive
to NLTE effects, whereas stronger lines show visible effects. The new
oscillator strengths also afford us the use of
near-infrared lines in our sample. These lines are not as affected by
blends as the optical lines \citep{shiStatisticalEquilibrium2008},
but some show strong NLTE cores. We do not include these lines to determine a
final abundance. During fitting, we clip points that are more than $0.5\%$ of
relative (normalised) flux further from the observations after an initial fit,
removing weak line blends and deviating line cores from the abundance
determination.

\begin{table*}[h!]
    \par
    \caption{
        Atomic data for spectral lines of \ion{Si}{i} and \ion{Si}{ii}.
    }
    \scalebox{0.82}{
        \begin{tabular}{lr@{ -- }llrlc|r@{ -- }lr@{ -- }l c c c c }
            \noalign{\smallskip}\hline\hline\noalign{\smallskip}
            $\lambda$    & \multicolumn{2}{c}{Transition} & \multicolumn{2}{c}{$\log gf$} & $\log gf$  & Weight    & \multicolumn{2}{c}{E$_{\text{low}}$~-~E$_{\text{upp}}$} & \multicolumn{2}{c}{E$_{\text{lim, low}}$~-~E$_{\text{lim, upp}}$} & n*$_{\text{low}}$ & n*$_{\text{upp}}$ & $\sigma_0$ & $\alpha$                                 \\
            \,[\AA]      & \multicolumn{2}{c}{ }          & \multicolumn{2}{c}{old}       & new        &           & [cm$^{-1}$]                                             & [cm$^{-1}$]                                                       & [cm$^{-1}$]       & [cm$^{-1}$]       &            &          & [a.u.] &                      \\
            \noalign{\smallskip}\hline\noalign{\smallskip}
            \ion{Si}{I}  &                                &                               &            &           &                                                         &                                                                   &                   &                   &            &          &        &       &      &       \\
            \noalign{\smallskip}\hline\noalign{\smallskip}
            5645.61*     & \Si{4{\rm s}}{3}{P}{o}{1}      & \Si{5\rm p}{3}{P}{}{2}        & -- 2.04(3) & G73+OBL91 & -- 2.067                                                & 1                                                                 & 39760             & 57468             & 65748      & 65748    & 2.054  & 3.640 & 1791 & 0.223 \\
            5665.55*     & \Si{4{\rm s}}{3}{P}{o}{0}      & \Si{5\rm p}{3}{P}{}{1}        & -- 1.94(3) & G73+OBL91 & -- 2.025                                                & 2                                                                 & 39683             & 57329             & 65748      & 65748    & 2.051  & 3.609 & 1772 & 0.222 \\
            5684.48*     & \Si{4{\rm s}}{3}{P}{o}{2}      & \Si{5\rm p}{3}{S}{}{1}        & -- 1.55(3) & G73+OBL91 & -- 1.606                                                & 2                                                                 & 39955             & 57542             & 65748      & 65748    & 2.062  & 3.656 & 1798 & 0.221 \\
            5690.43*     & \Si{4{\rm s}}{3}{P}{o}{1}      & \Si{5\rm p}{3}{P}{}{1}        & -- 1.77(3) & G73+OBL91 & -- 1.802                                                & 3                                                                 & 39760             & 57328             & 65748      & 65748    & 2.054  & 3.609 & 1772 & 0.222 \\
            5701.11*     & \Si{4{\rm s}}{3}{P}{o}{1}      & \Si{5\rm p}{3}{P}{}{0}        & -- 1.95(3) & G73+OBL91 & -- 1.981                                                & 3                                                                 & 39760             & 57296             & 65748      & 65748    & 2.054  & 3.602 & 1769 & 0.222 \\
            5708.40      & \Si{4{\rm s}}{3}{P}{o}{2}      & \Si{5\rm p}{3}{P}{}{2}        & -- 1.37(3) & G73+OBL91 & -- 1.388                                                & 0                                                                 & 39955             & 57468             & 65748      & 65748    & 2.062  & 3.640 & 1788 & 0.223 \\
            5772.15      & \Si{4{\rm s}}{1}{P}{o}{1}      & \Si{5\rm p}{1}{S}{}{0}        & -- 1.65(3) & G73+OBL91 & -- 1.643                                                & 0                                                                 & 40992             & 58312             & 65748      & 65748    & 2.105  & 3.841 & 2036 & 0.208 \\
            5780.38      & \Si{4{\rm s}}{3}{P}{o}{0}      & \Si{5{\rm p}}{3}{D}{}{1}      & -- 2.25(3) & G73+OBL91 & -- 2.156                                                & 0                                                                 & 39683             & 56978             & 65748      & 65748    & 2.051  & 3.536 & 1691 & 0.228 \\
            5793.07*     & \Si{4{\rm s}}{3}{P}{o}{1}      & \Si{5\rm p}{3}{D}{}{2}        & -- 1.96(3) & G73+OBL91 & -- 1.893                                                & 2                                                                 & 39760             & 57017             & 65748      & 65748    & 2.054  & 3.544 & 1704 & 0.228 \\
            5797.86      & \Si{4{\rm s}}{3}{P}{o}{2}      & \Si{5{\rm p}}{3}{D}{}{3}      & -- 1.95(3) & G73+OBL91 & -- 1.830                                                & 0                                                                 & 39955             & 57198             & 65748      & 65748    & 2.062  & 3.582 & 1755 & 0.223 \\
            5948.54      & \Si{4{\rm s}}{1}{P}{o}{1}      & \Si{5\rm p}{1}{D}{}{2}        & -- 1.13(3) & G73+OBL91 & -- 1.179                                                & 0                                                                 & 40992             & 57798             & 65748      & 65748    & 2.105  & 3.714 & 1845 & 0.222 \\
            6125.02      & \Si{3{\rm p^3}}{3}{D}{o}{1}    & \Si{5\rm f}{3}{D}{}{2}        & -- 1.465   & K07       & --                                                      & 0                                                                 & 45276             & 61598             & 114716     & 65748    & 1.257  & 5.141 & 3354 & 0.348 \\
            6142.49      & \Si{3{\rm p}}{3}{D}{o}{3}      & \Si{5\rm f}{3}{D}{}{3}        & -- 1.296   & K07       & --                                                      & 0                                                                 & 45321             & 61597             & 114716     & 65748    & 1.257  & 5.141 & 3354 & 0.348 \\
            6145.02      & \Si{3{\rm p}}{3}{D}{o}{2}      & \Si{5\rm f}{3}{G}{}{3}        & -- 1.311   & K07       & --                                                      & 0                                                                 & 45294             & 61562             & 114716     & 65748    & 1.257  & 5.119 & 3295 & 0.341 \\
            6237.32      & \Si{3{\rm p}}{3}{D}{o}{1}      & \Si{5\rm f}{3}{F}{}{2}        & -- 0.975   & K07       & --                                                      & 0                                                                 & 45294             & 61304             & 114716     & 65748    & 1.257  & 4.968 & 3081 & 0.351 \\
            6243.82      & \Si{3{\rm p}}{3}{D}{o}{2}      & \Si{5\rm f}{3}{F}{}{3}        & -- 1.244   & K07       & --                                                      & 0                                                                 & 45294             & 61305             & 114716     & 65748    & 1.257  & 4.968 & 3081 & 0.351 \\
            6244.47      & \Si{3{\rm p}}{3}{D}{o}{2}      & \Si{5\rm f}{1}{D}{}{2}        & -- 1.091   & K07       & --                                                      & 0                                                                 & 45294             & 61305             & 114716     & 65748    & 1.257  & 4.968 & 3081 & 0.351 \\
            6976.51      & \Si{4{\rm p}}{3}{D}{}{1}       & \Si{6{\rm d}}{3}{F}{o}{2}     & -- 1.07(3) & G73+OBL91 & --                                                      & 0                                                                 & 48020             & 62350             & 65748      & 65748    & 2.487  & 5.681 & 4600 & 0.530 \\
            7003.57      & \Si{4{\rm p}}{3}{D}{}{2}       & \Si{6{\rm d}}{3}{F}{o}{3}     & -- 0.793   & G73+OBL91 & --                                                      & 0                                                                 & 48102             & 62377             & 65748      & 65748    & 2.493  & 5.704 & 4700 & 0.531 \\
            7034.91      & \Si{3\rm d}{1}{D}{o}{2}        & \Si{5\rm f}{3}{F}{}{3}        & -- 0.78(3) & G73+OBL91 & --                                                      & 0                                                                 & 47352             & 61562             & 65748      & 65748    & 2.442  & 5.119 & 3232 & 0.338 \\
            7226.21      & \Si{3{\rm p}}{3}{D}{o}{1}      & \Si{4\rm f}{1}{D}{}{2}        & -- 1.41(3) & G73+OBL91 & --                                                      & 0                                                                 & 45276             & 59111             & 114716     & 65748    & 1.257  & 4.065 & 1745 & 0.307 \\
            7405.79      & \Si{3{\rm p}}{3}{D}{o}{1}      & \Si{4\rm f}{3}{F}{}{2}        & -- 0.72(3) & G73+OBL91 & --                                                      & 0                                                                 & 45276             & 58775             & 114716     & 65748    & 1.257  & 3.966 & 1585 & 0.304 \\
            7415.36      & \Si{3{\rm p}}{3}{D}{o}{2}      & \Si{4\rm f}{3}{F}{}{2}        & -- 0.65(3) & G73+OBL91 & --                                                      & 0                                                                 & 45294             & 58774             & 114716     & 65748    & 1.257  & 3.966 & 1585 & 0.304 \\
            7680.27*     & \Si{4{\rm p}}{1}{P}{}{1}       & \Si{5\rm d}{1}{D}{o}{2}       & -- 0.59(3) & G73+OBL91 & -- 0.678                                                & 2                                                                 & 47284             & 60301             & 65748      & 65748    & 2.437  & 4.487 & 2107 & 0.494 \\
            7918.38      & \Si{4{\rm p}}{3}{D}{}{1}       & \Si{5\rm d}{3}{F}{o}{2}       & -- 0.51(3) & G73+OBL91 & -- 0.666                                                & 0                                                                 & 48020             & 60645             & 65748      & 65748    & 2.487  & 4.636 & 2934 & 0.234 \\
            7932.35      & \Si{4{\rm p}}{3}{D}{}{2}       & \Si{5\rm d}{3}{F}{o}{3}       & -- 0.37(3) & G73+OBL91 & -- 0.472                                                & 0                                                                 & 48102             & 60705             & 65748      & 65748    & 2.493  & 4.664 & 2985 & 0.234 \\
            10288.94*    & \Si{4{\rm s}}{3}{P}{o}{0}      & \Si{4\rm p}{3}{S}{}{1}        & --         & --        & -- 1.622                                                & 2                                                                 & 39683             & 49400             & 65748      & 65748    & 2.493  & 4.664 & 739  & 0.230 \\
            10371.26     & \Si{4{\rm s}}{3}{P}{o}{1}      & \Si{4\rm p}{3}{S}{}{1}        & --         & --        & -- 0.789                                                & 0                                                                 & 39760             & 49400             & 65748      & 65748    & 2.493  & 4.664 & 739  & 0.230 \\
            10603.43     & \Si{4{\rm s}}{3}{P}{o}{1}      & \Si{4\rm p}{3}{P}{}{2}        & --         & --        & -- 0.394                                                & 0                                                                 & 39760             & 49188             & 65748      & 65748    & 2.493  & 4.664 & 727  & 0.231 \\
            10689.72     & \Si{4{\rm p}}{3}{D}{}{1}       & \Si{4\rm d}{3}{F}{o}{2}       & --         & --        & -- 0.017                                                & 0                                                                 & 48020             & 57372             & 65748      & 65748    & 2.493  & 4.664 & 1418 & 0.234 \\
            10694.25     & \Si{4{\rm p}}{3}{D}{}{2}       & \Si{4\rm d}{3}{F}{o}{3}       & --         & --        & +  0.155                                                & 0                                                                 & 48102             & 57450             & 65748      & 65748    & 2.493  & 4.664 & 1445 & 0.750 \\
            10749.37     & \Si{4{\rm s}}{3}{P}{o}{1}      & \Si{4\rm p}{3}{P}{}{1}        & --         & --        & -- 0.268                                                & 0                                                                 & 39760             & 49061             & 65748      & 65748    & 2.493  & 4.664 & 721  & 0.231 \\
            10784.56     & \Si{4{\rm p}}{3}{D}{}{2}       & \Si{4\rm d}{3}{F}{o}{2}       & --         & --        & -- 0.746                                                & 0                                                                 & 48102             & 57372             & 65748      & 65748    & 2.493  & 4.664 & 1417 & 0.296 \\
            10786.85     & \Si{4{\rm s}}{3}{P}{o}{1}      & \Si{4\rm p}{3}{P}{}{0}        & --         & --        & -- 0.380                                                & 0                                                                 & 39760             & 49028             & 65748      & 65748    & 2.493  & 4.664 & 719  & 0.231 \\
            10827.09     & \Si{4{\rm s}}{3}{P}{o}{2}      & \Si{4\rm p}{3}{P}{}{2}        & --         & --        & +  0.227                                                & 0                                                                 & 39955             & 49188             & 65748      & 65748    & 2.493  & 4.664 & 728  & 0.231 \\
            12390.15*    & \Si{4{\rm s}}{1}{P}{o}{1}      & \Si{4\rm p}{3}{P}{}{1}        & --         & --        & -- 1.805                                                & 2                                                                 & 40992             & 49061             & 65748      & 65748    & 2.493  & 4.664 & 730  & 0.234 \\
            12395.83*    & \Si{4{\rm s}}{3}{P}{o}{2}      & \Si{4\rm p}{3}{D}{}{1}        & --         & --        & -- 1.723                                                & 2                                                                 & 39955             & 48020             & 65748      & 65748    & 2.493  & 4.664 & 675  & 0.231 \\
            \noalign{\smallskip}\hline\noalign{\smallskip}
            \ion{Si}{II} &                                &                               &            &           &                                                         &                                                                   &                   &                   &            &          &        &       &              \\
            \noalign{\smallskip}\hline\noalign{\smallskip}
            6347.10      & \Si{4{\rm s}}{2}{S}{}{1/2}     & \Si{4{\rm p}}{2}{P}{o}{3/2}   & + 0.170    & K14       & + 0.182                                                 & 0                                                                 & 65500             & 81299             & 131838     & 131838   & 2.572  & 2.943 & 390  & 0.190 \\
            6371.36*     & \Si{4{\rm s}}{2}{S}{}{1/2}     & \Si{4{\rm p}}{2}{P}{o}{1/2}   & -- 0.040   & K14       & -- 0.120                                                & 1                                                                 & 65500             & 81299             & 131838     & 131838   & 2.572  & 2.943 & 390  & 0.190 \\
            \noalign{\smallskip}\hline\noalign{\smallskip}
        \end{tabular}}
    \tablefoot{
        An asterisk next to the wavelength signifies the line was in the chosen
        subsample.
        References to the ``old'' $gf$-values are
        G73+OBL91: \citet{garzAbsoluteOscillator1973}, renormalised by $+0.097$ dex
        according to \citet{obrianRadiativeLifetimes1991},
        K07: \citet{K07}, and
        K14: \citet{K14}. ``new'' $\log gf$
        values come from \citet{pehlivanrhodinExperimentalComputational2018}.
        The ``Weight'' column specifies our weighting for each line. The right
        portion of the table shows ABO theory parameters.
        ``E$_{\text{low}}$'' \& ``E$_{\text{upp}}$'' show the lower and upper energy
        levels of the	transitions in cm$^{-1}$, and ``E$_\mathrm{lim}$'' shows the
        series limit energy for that level (see e.g. \citet{heiterAtomicData2021}
        for details).
        ``n*$_{\text{low}}$'' \& ``n*$_{\text{upp}}$'', are the effective quantum
        numbers associated with the states of the transitions.
        ``$\sigma_0$'' is the line broadening cross section in atomic units and
        ``$\alpha$'' describes the power law velocity dependence.
    }
    \label{table:line_list}
\end{table*}

\subsection{Spectral synthesis}
In this study, a 3D hydrodynamical solar model atmosphere with an initial
silicon abundance of
$\lgesi = 7.55$ was used for the line synthesis.
Each line is synthesised with the \linfor\ code over 20 atmospheric
snapshots. The lines are synthesised in full 3D, including Doppler shifts,
and the profile is averaged in time and space (horizontally). We generate
a set of syntheses for each spectral line, each with a different $\gfe$
value. We alter $\log{gf}$ in this case, with a distinct stepping around a
central value for each line depending on the sensitivity of the lines to
oscillator strength changes.

Our investigation of the solar silicon abundance is mostly 3D LTE, but the
$-0.01$\,dex correction due to NLTE effects determined by
\citet{amarsiSolarSilicon2017} was used globally for our
calculated abundance. We investigate NLTE effects on broadening and
abundance in appendix \ref{appendix:matthias}.

\subsection{Line broadening}
Aside from the abundance, we also aim to investigate the effects of
broadening required to fit synthetic spectra to observations. The
broadening we see in the observed lines can be attributed to broadening by
macroscopic velocity fields, thermal
broadening, and atomic line broadening due to collisions with neutral particles
and electrons. For broadening due to macroscopic velocity
fields, though a tendency towards less turbulent flow is expected
in numerical models due to limited spatial resolution, our 3D syntheses are
broader than the observations, even prior to applying instrumental profile
broadening. The result is not unique to our work
\citep[see, e.g.][]{caffauPhotosphericSolar2015}, nor is it unique to
\cobold~ models (see Appendix~\ref{appendix:anish} for details).

\subsubsection{Collisional broadening}
Solar spectral lines are subject to broadening from atomic effects.
These are pressure broadening effects arising from van der Waals and other
interatomic forces. It turns out that the $7680$\,\AA\ line shows
sizeable Stark broadening, meaning the Stark effect is
non-negligible for some solar silicon lines and illustrating the importance
of accurately modelling these effects.
We use the theory by Anstee, Barklem and
O'Mara  \citep{ansteeInvestigationBrueckner1991,
    ansteeWidthCrosssections1995, barklemLineBroadening1998} to describe the
line broadening effects of neutral particles, predominantly neutral hydrogen
(hereafter ABO theory). Note that the broadening constants were specifically
calculated using existing tables from the above papers or through extended
calculations, and are shown in Table~\ref{table:line_list}. \linfor\
additionally takes the Stark broadening of lines into account, employing line
widths from from the Vienna Atomic Line Database (VALD)
\citep{kupkaVALD2Progress1999, kupkaVALD2New2000,ryabchikovaMajorUpgrade2015}
and assuming temperature dependence from the Uns\"old theory
\citep{unsoldPhysikSternatmospharen1955}.

The theory of collisional line broadening for neutral species
\citep{ansteeInvestigationBrueckner1991,ansteeWidthCrosssections1995} was
extended to singly ionised atoms in
\citet{barklemBroadeningStrong1998}.
For the lines corresponding to ionised Si in this study, we performed specific
calculations assuming $E_p = -4/9$\ atomic units (the average value of the
energy denominator of the second-order contribution to the interaction energy),
which is expected to give
reasonable estimates, though this assumption is less secure for ions than for
neutrals \citep[see][]{2018ApJ...857....2R, barklemBroadeningFe2005}.
Equation~1 of \citet{ansteeWidthCrosssections1995} describes the
line broadening cross section $\sigma$ (in atomic units) as a function of
the  relative collision velocity $v$, with respect to a reference value of
$v_0 = 10^4$\,\ms, with the parameter $\alpha$ giving the power law velocity
dependence:

\beq
\label{eq:abo}
\sigma(v) = \sigma_0\left(\frac{v}{v_0}\right)^{-\alpha} .\,
\eeq
The line width at a given temperature can then be obtained from this relation by
analytic integration over the Maxwellian velocity distribution (see Equation~3
of \citet{ansteeWidthCrosssections1995}).

\subsubsection{Rotational broadening}
For disk-integrated spectra, we consider fixed solid body rotation assuming
a synodic $v\sin{i} = 1.8$~\kms\ is present in the observations from the
Hamburg spectrum \citep{bartelsTwentysevenDay1934, grayObservationAnalysis2008}.
In total, $1$ vertical $16$ inclined rays were used ($4$ $\mu$-angles and $4$
$\phi$-angles) with Lobatto quadrature
\citep{abramowitzHandbookMathematical1965}. Rotational broadening is not
included for disk-centre spectra.

\subsection{Magnetic fields}
\subsubsection{Effects on model atmospheres}
It is known that weak magnetic fields are present in the solar surface
layers and there is evidence that the mean magnetic flux density
for these is of the order of $10^2$\,G
\citep{buenoSubstantialAmount2004, nordlundSolarSurface2009}. Numerical 3D
convection models predict that the granulation pattern is profoundly
affected in regions with high flux density
\citep{cattaneoInteractionConvection2003,
    voglerSimulationsMagnetoconvection2005, cheungSolarSurface2008}, in
agreement with observations. Additionally, photospheric material
becomes more transparent in magnetic concentrations due to their lower
density, allowing one to see into deeper, hence hotter, layers of the
solar atmosphere \citep{steinRealisticSolar2000}. In these regions, where
flux tubes are also heated through heat influx from the surrounding
material \citep{spruitPressureEquilibrium1976}, weak spectral lines will
experience a brightening of their core, meaning
magnetic fields can act on temperature-sensitive lines not only directly
through the Zeeman effect, but also indirectly due to temperature
stratification in line-forming regions \citep{fabbianSolarFe2012}.

\subsubsection{Effects on spectral synthesis}
In 1D model atmospheres, \citet{borreroRoleMagnetic2008} showed that magnetic
fields have non-negligible effects on spectral line synthesis, and
\citet{fabbianSolarAbundance2010} expanded this to 3D atmospheres,
finding an abundance correction for Fe of the order of $\sim +0.01$\,dex
when using magneto-convection models. \citet{fabbianSolarFe2012} used 28 iron
lines and found the average solar iron abundance to be $\sim0.03 - 0.11$\,dex
higher when using a magneto-convection model with
$\left\langle B_{\mathrm{vert}} \right\rangle = 200$\,G
and investigated models with average magnetic flux densities of
$\left\langle B_{\mathrm{vert}} \right\rangle = 0, 50, 100, 200$\,G.
They also showed that Zeeman broadening gains importance in the infrared,
and that the largest contribution to higher abundance is the indirect
effect of line-weakening caused by a warmer stratification as seen on an
optical depth scale.

Following this, \citet{fabbianContinuumIntensity2015}
were able to reproduce the observations of 2 \ion{O}{I} spectral lines
with blends with the use of a 3D MHD photospheric model with a uniform
vertical magnetic field of $200$\,G. They required an oxygen
abundance several centidexes higher than those suggested by
\citet{asplundChemicalComposition2009} to fit observations and again showed the
need to consider magneto-convection processes when considering problems
sensitive to the  shape of spectral features, such as spectral synthesis and
fitting.

As shown in Table~\ref{table:model_atm}, we use two magnetic model
atmospheres to investigate the effects of the magnetic field strength on
the fitted abundance and broadening values.
Zeeman broadening is not taken into account in the spectral synthesis. Note that
\citet{shchukinaImpactSurface2015} and \cite{shchukinaImpactSurface2016} also
show that such vertical fields as in \citet{fabbianSolarFe2012}
overestimate the effects when compared with a 3D MHD model with a
self-consistent small-scale dynamo and more realistic magnetic fields. With this
in mind, and with the fact that the magnetic models used here are not of a high
enough resolution to quantitatively investigate the effects of magnetic fields,
we consider only differential effects.
Section~\ref{sec:results} shows these differences
and concludes that the extra broadening present in the msc600 model
syntheses could be partly attributed to the lack of magnetic fields.
\section{Spectral fitting routine}
\label{sec:fitting}
\noindent
The synthesised lines are fitted to the observations from the Hamburg
spectrum via \chisq\ minimisation, where we define \chisq\ as

\begin{equation}
    \chi^2 = \Delta x^\mathrm{T} \mathbf{C}^{-1} \Delta x ,\,
    \label{eq:chi2}
\end{equation}
where $\Delta x$ is the vector of residuals and $\mathbf{C}$
is the covariance matrix of the data. We use maximum likelihood estimation
using \chisq\ rather than the conventional least-squares minimisation in
order to incorporate the errors due to covariance in the spectrum.

\begin{figure}[h]
    \centering
    \includegraphics[width=90mm]{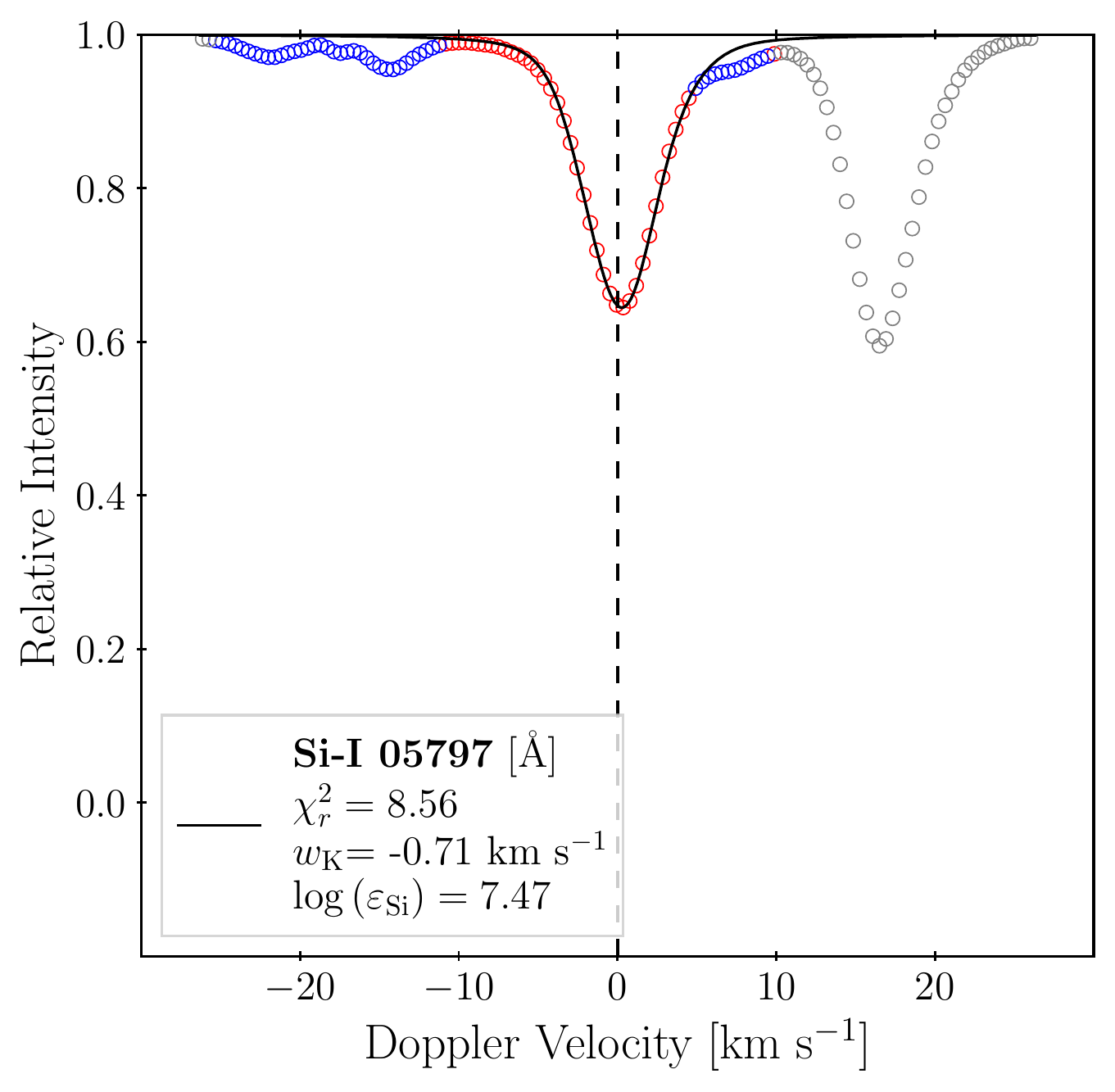}
    \caption{Synthesised fitted line profile (black line) against the
        observations from the Hamburg atlas (points). Grey points were cut from
        the initial fit; blue points show areas where the deviation
        between initial fit and data is too high, and so the subsequent fits
        do not use the offending points; red points are used for computing
        all fitted quantities. Removing the points on right hand side removes
        the strong blend, and sigma-clipping handles weaker ones during fitting.
        $w_\mathrm{K}$ is the width of the broadening kernel used. Note that
        this line is \textit{not} used in the final determination of the
        silicon abundance because of the large line blends.}
    \label{fig:line_example}
\end{figure}

In this analysis, we fit disk-centre intensities for abundance
determination,  but disk-integrated spectra are also used when choosing
the line sample. To mask weak blends, a window is chosen around each line for
the fitting procedure. An example is shown in Fig.~\ref{fig:line_example}.

The routine involves first loading the observations and determining the
covariance between pixels for a given line
(see Section~\ref{subsec:noise} for details). Then, during the fitting,
instrumental profile broadening and rotational broadening
(for disk-integrated spectra) are applied
to the syntheses. A monotonic cubic interpolation across abundance is used to
find the closest synthesis matching the observations,
now also fitting the wavelength shift, to create an array of syntheses.
Finally, for the nine abundances in steps of
$0.05$~dex that were synthesised, the abundance is fit by interpolating
through the array of syntheses and constructing a spectrum from the overall best
fitting points. Sigma-clipping is employed to improve subsequent fits.
The tight window is necessary with our particular combination of syntheses
and observations to minimise the effects that weak line blends have on fitting.

Some infrared lines exhibited strong NLTE cores, and we do not include
these lines in our final sample used to
determine abundance. However, we did attempt to fit
these lines by masking the core and fitting the line up to $70\%$ residual
intensity, inspired by the work of
\citet{shiStatisticalEquilibrium2008}. Their work revolves
around accurately treating NLTE effects, and we noticed that infrared lines
shown in this work could be fit in LTE up to this $70\%$ intensity. In the end,
clipping points at the $0.5\%$ level was more versatile and useful than simply
masking the line cores. This choice of $0.5\%$
eliminates stronger offending blends while still retaining important
characteristics in the line profile. We fitted some infrared lines with strong
NLTE cores with this method, but ultimately chose to leave them out of the
subsample used to derive an abundance since they
require excess negative broadening (see Sec.~\ref{sec:broadening}) to
fit, which could be indicative of smaller
NLTE effects present in the fitted parts of the line profile.

\subsection{Profile (de)broadening}
\label{sec:broadening}
In the present investigation, we found that in many cases the synthetic
line profiles were already broader than
the observations, even prior to the application of instrumental broadening.
Such a problem is specific to spectral synthesis calculations based on
3D model atmospheres \citep{caffauPhotosphericSolar2015},
as broadening effects of the stellar
velocity field in 1D hydrostatic models are added in an ad-hoc fashion
via fitting micro- and
macro-turbulent broadening to the observations. Hence, mismatches of the
overall broadening between model and observations cannot occur.

In our case, we could either broaden the observations or de-broaden the
syntheses. Unsurprisingly, broadening the
observations a priori by $2.5$\,\kms\ (the value that allowed our syntheses
to fit the observations well) resulted
in better statistical fits, as the lines appeared more Gaussian-like.
However, this removes information about the
line, such as weak blends and the overall shape. Instead, we chose to
implement the capability to de-broaden our syntheses, rendering them
narrower to better fit the observations as opposed to
broadening observations instead. This is stated as equivalent to a
negative broadening, or broadening with a
kernel that has a negative full width at half-maximum (FWHM).  We hoped
to maintain the ability of 3D
line profiles that allows one to identify weak blends that can degrade
fits without being immediately recognised. This provides us a fully
invertible method when it comes to fitting the broadening value of
syntheses. Note that the broadening value we fit is still much smaller
than the values of micro- and  macro-turbulence fitted in 1D models, and
our 3D model still reproduces line shapes and asymmetries.

We formally associate broadening with convolution, and de-broadening
(negative broadening) as deconvolution and use a kernel inspired by
\citet{dorfiSimpleAdaptive1987}. This kernel is composed of a double
exponential decay centred on zero (see Appendix~\ref{appendix:ddsmooth} for
details of the implementation), and is referred to throughout this work as the
$G_n$ kernel, where $n$ is an integer representing the 'order' of the
kernel. We can convolve the $G_1$ kernel (Equation \ref{eq:g1}) with itself
to produce more Gaussian-like kernels, but note that this brings back
the noise amplification that we aimed to mitigate with the use of the
$G_1$ kernel:

\begin{equation}
    \label{eq:g1}
    g_1(x, x') = G_1(x-x') = \frac{\alpha}{2}\exp(-\alpha \left\lvert x-x'\right\rvert).
\end{equation}
Here, $x$ is position (in velocity space) and $\alpha$ controls the width of the
kernel.

Gaussian and sinc functions were both considered as instrumental profile
broadening kernels, but these do not allow for efficient and optimal
computation of a de-broadened profile, since as the Fourier transform of these
kernel functions rapidly goes to zero precluding a deconvolution, small
disturbances cause large spikes in the wings of debroadened spectra. Across
all lines (and across the downsample), the choice between the $G_1$ and $G_3$
kernels does not affect the fitted abundance. The $G_1$ kernel does result
in slightly better statistical fits, and so we favour it in this study.
\subsection{Photometric noise model}
\label{subsec:noise}
Often, the pixel-to-pixel correlation of the signal in the spectrum is
ignored, despite being rather commonly present due to instrumental
imperfections during detection as well as steps during data reduction.
The Hamburg atlas was rebinned at steps of $3.8$ m\AA.
This rebinning introduced a correlation between pixels.
We implement a photometric noise model that considers
the relation between neighbouring pixels in the observations, whose
correlated signal introduces correlation in
the noise values in each pixel (given by the root mean-squared error
(RMSE) of a window). In order to model this, we require a representation of the
covariance matrix of the noise. This covariance matrix,
or correlated noise matrix, can then be applied to the set of observations
in the fitting routine to describe the
pixel-to-pixel correlation of the noise in the spectrum. Due to the nature
of the covariance matrix, it must be positive semi-definite. To accurately
estimate the matrix of correlated noise, we fit the autocorrelation
function of continuum regions of the spectrum with an exponential decay.

\section{Results}
\label{sec:results}
\subsection{Line syntheses}

The spectral line syntheses for the chosen $11$ lines are shown in
Fig. \ref{fig:line_syn}. The lines are synthesised with the non-magnetic
msc600 model atmosphere and use the $G_1$ broadening kernel during fitting.
We also use a covariance matrix for pixel-correlated noise, which results
in an increase in mean abundance of $0.001$ dex compared to assuming
uncorrelated noise. Our final derived photospheric solar silicon abundance is
$\lgesi$ = \abufinal, including the $-0.01$ dex correction from NLTE effects
\citep{amarsiSolarSilicon2017}.

\begin{figure}
    \centering
    \includegraphics[width=90mm]{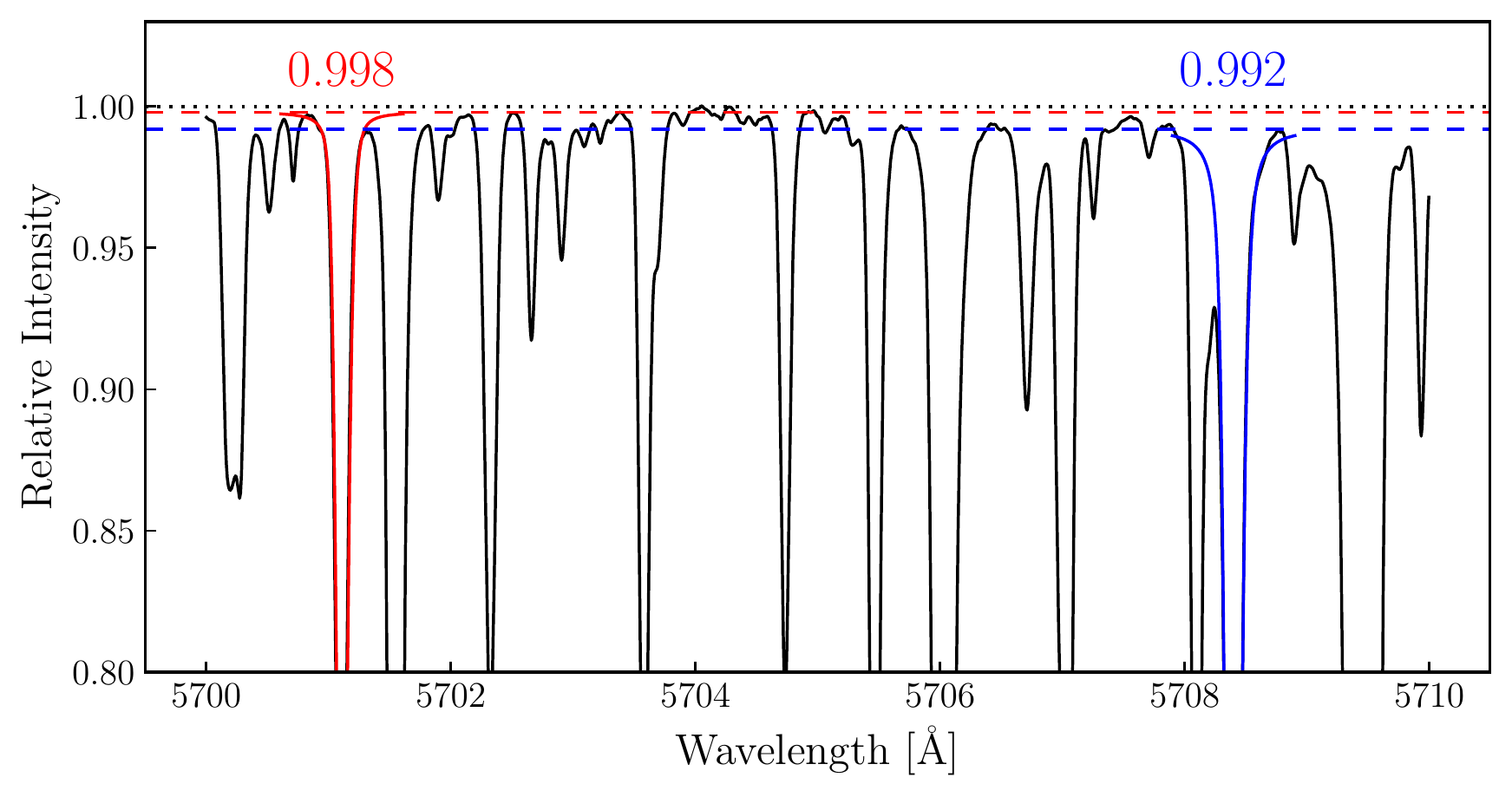}
    \caption{
        Comparison of two silicon line profiles while locally fitting the
        continuum.
        Observations are shown in black and the fitted syntheses at $5701.11$
        and \lineAng{5708.40} are shown in red and blue, respectively, along
        with the fitted continuum value. Across a wider wavelength range,
        the various fitted continua are not consistent with one another.
    }
    \label{fig:continuum}
\end{figure}

\begin{figure*}
    \centering
    \includegraphics[width=140mm, height=195mm]{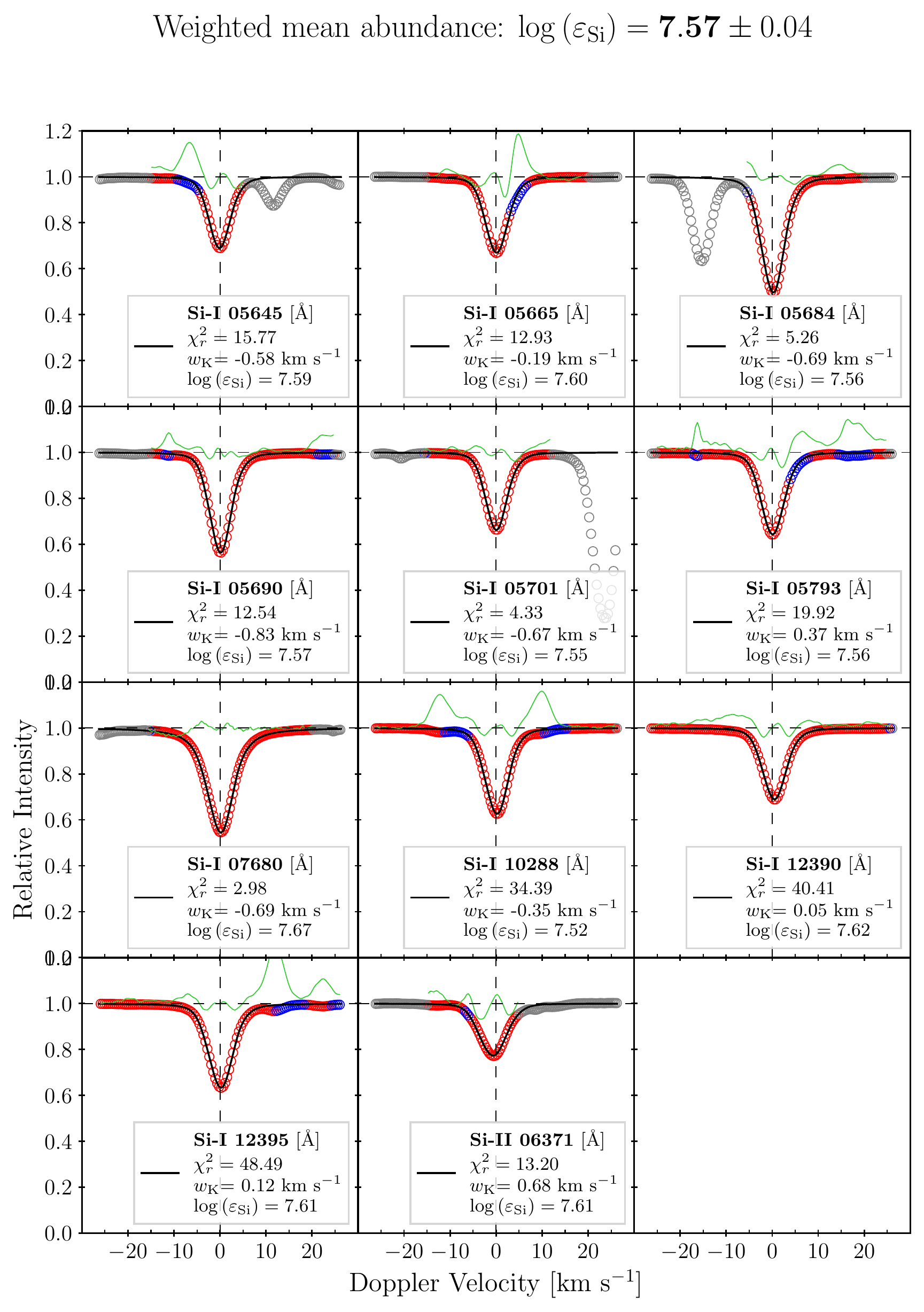}
    \caption{Synthesised fitted line shapes (black solid line) against
        observations from the Hamburg spectrum (points) for 11
        silicon lines for the msc600 model. Grey points were removed before
        fitting, blue points show areas where sigma-clipping was employed to
        remove poorly fitting points, and red points were used for determining
        final fitted quantities. Green lines show the residuals
        increased by a factor of 10 for readability. In each panel, the name of
        the line, the reduced \chisq\, value, the width of the broadening kernel
        $w_\mathrm{K}$, and the LTE abundance $\lgesi$ are shown. The final
        weighted mean abundance of \abufinal\, includes the $-0.01$ dex NLTE
        correction, a fixed continuum, no pre-broadening, and de-broadening.}
    \label{fig:line_syn}
\end{figure*}

We find that simultaneously fitting the continuum for the entire selection
of spectral lines systematically lowers the abundance by $0.01$\,dex. A
local fit of the continuum is therefore not representative of the spectrum
on a larger scale, and we rely on the normalisation provided by the Hamburg
atlas. This comparison is shown for two lines in Fig.~\ref{fig:continuum}.
Broadening the observations a priori by $2.5$\,\kms
(to counter the overly broadened syntheses)
increases the abundance across all configurations by $0.02$\, dex as this
convolves weak line blends into the primary line shape, rendering the
observations more Gaussian-like and removing information (such as the
observational line shape) in the process.

\subsection{Magnetic field effects}
\begin{figure*}
    \centering
    \includegraphics[width=175mm]{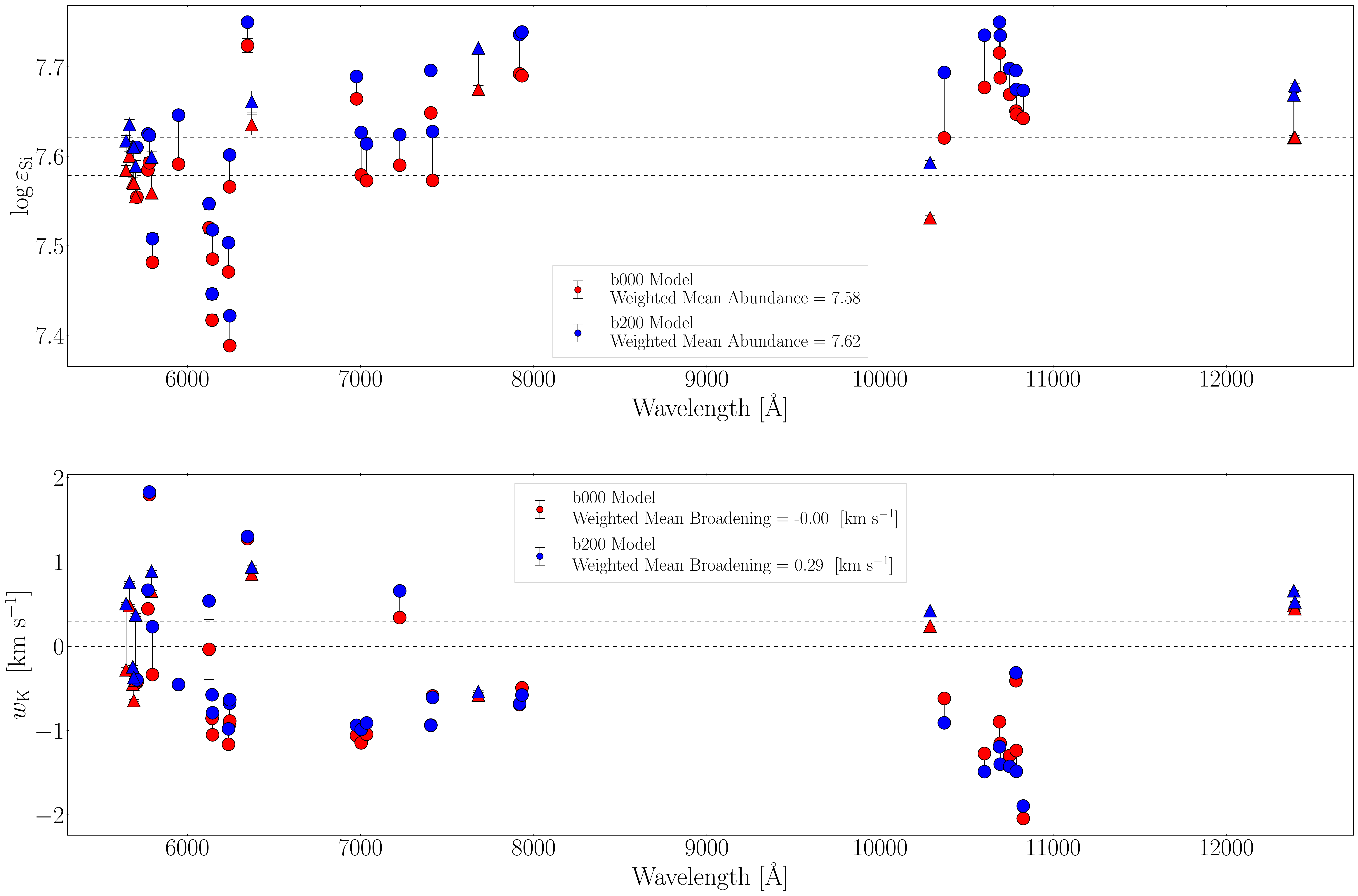}
    \caption{Comparison of abundance and broadening between the b000 (red
        points), and b200 (blue points) models. The red and blue horizontal
        dashed lines follow the same colour scheme and show the
        weighted average values. Triangles indicate the lines used in the
        subsample. Increasing the magnetic field strength increases fitted
        abundance and decreases negative broadening.
    }
    \label{fig:mag_parameter_comparison}
\end{figure*}

Figure \ref{fig:mag_parameter_comparison} shows the difference in
abundance and broadening fitted when comparing different magnetic model
atmospheres: one without a magnetic field and one with a magnetic field of
200 G (b200).

\begin{figure}
    \centering
    \includegraphics[width=80mm]{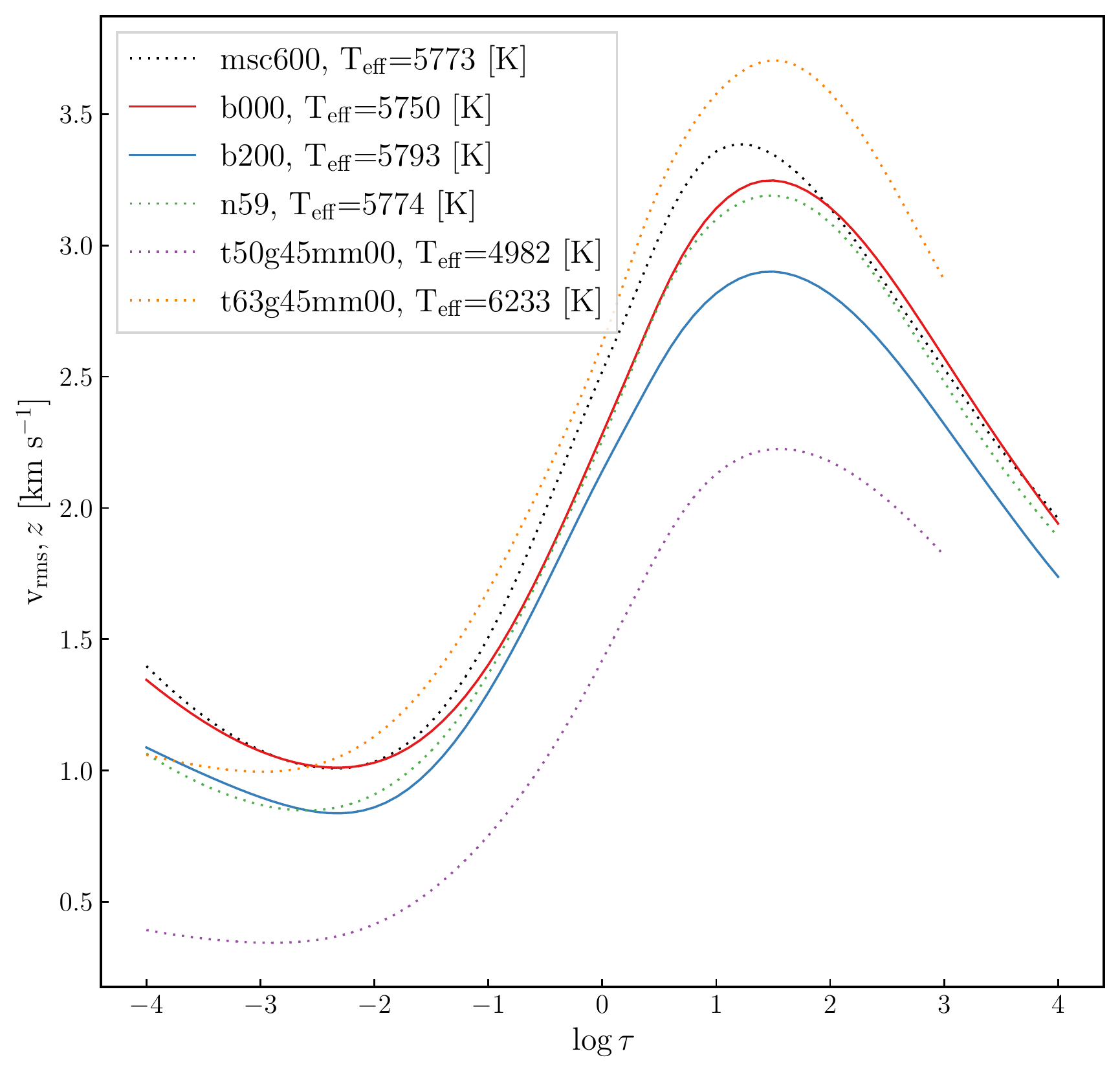}
    \caption{
        Vertical RMS velocity profiles in and around line-forming regions for
        the b000, b200 models (solid lines) with the msc600, n59, and 2 other
        models at much lower and higher effective temperatures shown for
        comparison (dotted lines). The b200 model clearly has lower RMS
        velocity than the other solar-type models, even though its effective
        temperature is higher.
    }
    \label{fig:v_z}
\end{figure}

The fitted results of the magnetic models should not be used
on an absolute scale, as the model atmosphere grids are too coarse to
resolve the detailed structure of small magnetic flux concentrations.
Additionally, as the effective temperatures of these models are further from
the nominal $\Teff=5772$ K \citep{prsaNominalValues2016}, we apply a correction
to each line's fitted
abundance to account for this change in temperature. The correction was
derived from the snapshot-to-snapshot variation of $\Teff$ and equivalent width.
The highest correction, for the b000 model, was $+0.015$ dex for the
\ion{Si}{II} lines, while the \ion{Si}{I} lines averaged a $-0.005$ dex
correction. \ion{Si}{I} and \ion{Si}{II} lines show opposite trends in each
model. The models are used for differential comparison, noting that
increasing the magnetic field strength increases both the fitted abundance and
broadening values. A possible implication is that the over-broadening of
synthesised lines is caused by a lack of magnetic fields in the model
atmosphere, as magnetic field lines constrain the flow of
material in the solar atmosphere, reducing turbulence and thereby the line
broadening. Fig.~\ref{fig:v_z} illustrates this point for all $4$ models
alongside $2$ models with much higher and lower effective temperatures. Though
the b200 model has a higher effective temperature than the other solar-type
models, its still has a lower vertical RMS velocity. Comparing to the t63g45mm00
model at $6233$ K, the increase in effective temperature in the b200 model to
overcome the magnetic field effects would need to be much greater than the
current model's value.

Additionally, a magnetic field strength of $200$ G still does not give the
full amount of de-broadening required to fit the observations and is higher
than the value of up to $75$ G expected in the majority of the quiet
photosphere \citep{ramirezvelezStrengthDistribution2008}. Again, as shown by
\citet{shchukinaImpactSurface2015} \& \citet{shchukinaImpactSurface2016}, a
vertical field of $200$ G would overestimate the effects when compared with a
self-consistent 3D MHD model with a small-scale dynamo.
The syntheses do not include the ~$1$~\kms\ instrumental broadening, meaning the
nominal $w_\mathrm{K} \approx 1$~\kms\ -- hence even further de-broadening from
magnetic fields would be required. Therefore, our results are suggestive but not
definitive that the lack of magnetic fields contributes to the over-broadening
of spectral lines, and the high magnetic field strengths required to produce the
required broadening would not be consistent with 3D MHD models with small-scale
dynamos.

\subsection{Effects of model differences}
\begin{figure*}
    \centering
    \includegraphics[width=175mm]{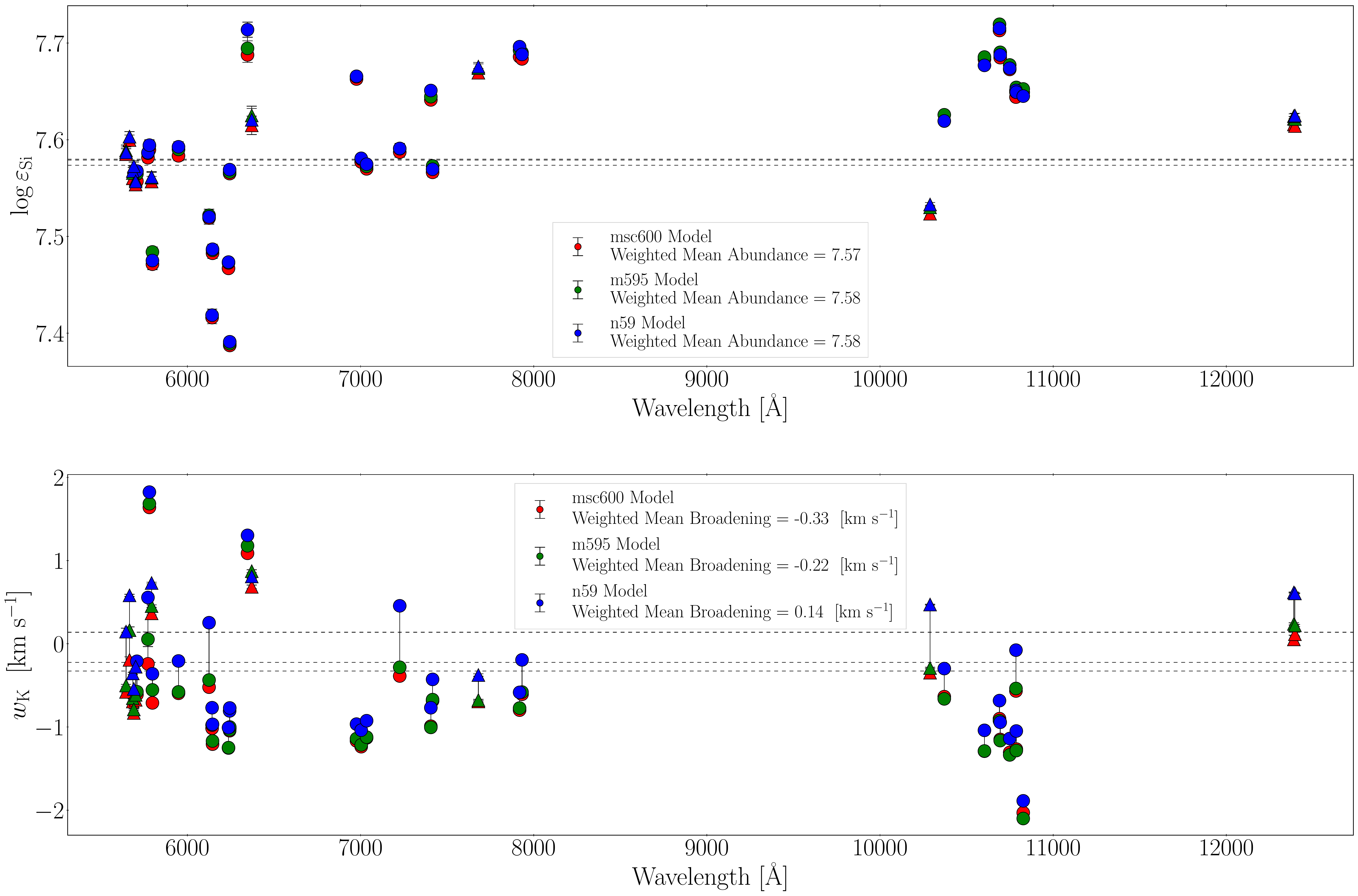}
    \caption{Comparison of abundance and broadening between the msc600
        (red points), m595 (green points) and n59 (blue points) models.
        Triangles indicate the lines used in the subsample.
        The red, green and blue horizontal dashed lines follow the same colour
        scheme and show the weighted average values.}
    \label{fig:model_parameter_comparison}
\end{figure*}

\begin{figure}
    \centering
    \includegraphics[width=80mm]{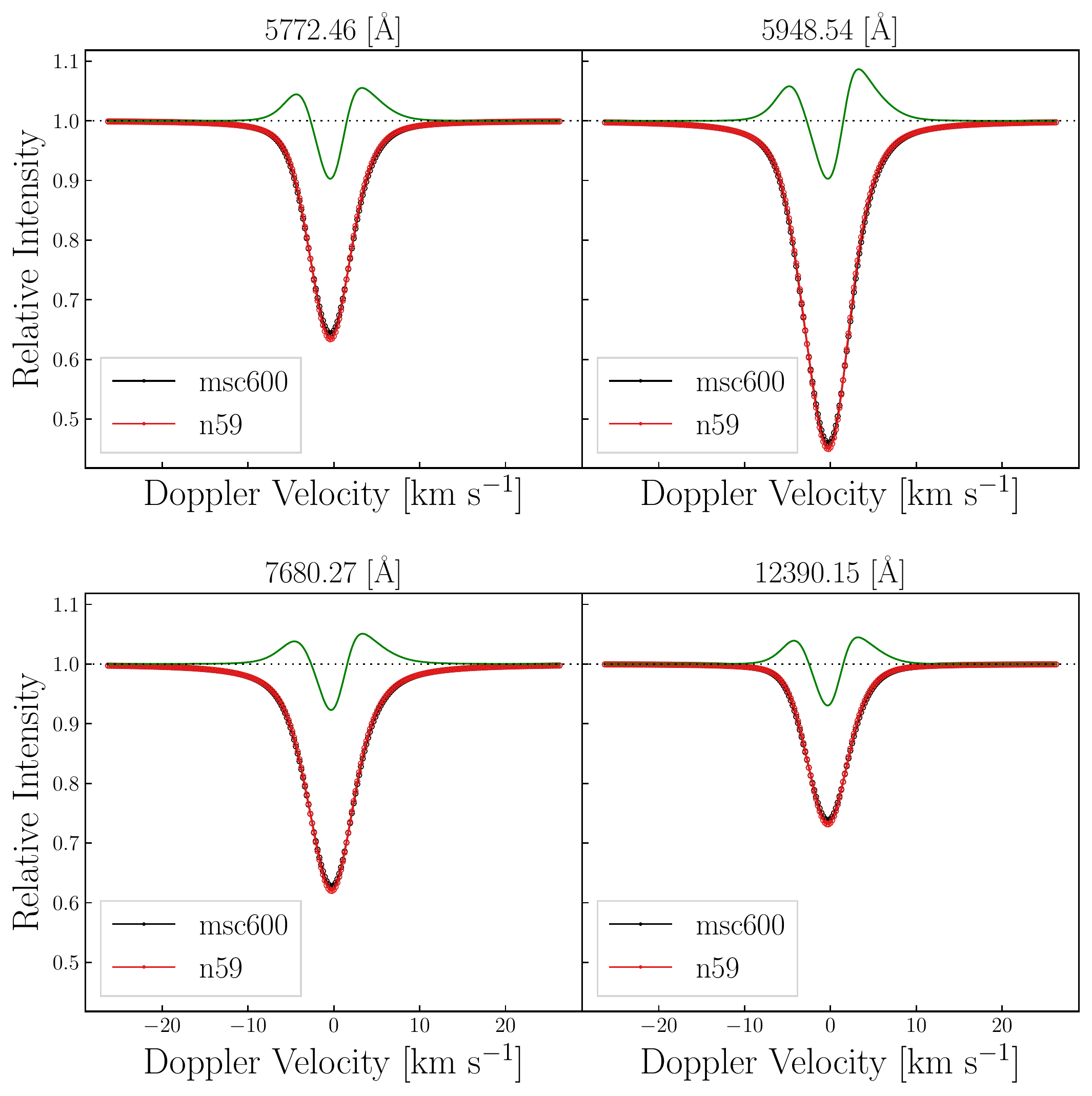}
    \caption{
        Line profiles from msc600 (black) and n59 (red) models with the
        residuals magnified by a factor of $10$ in green. The difference in the
        line profiles is greatest in the line core.
    }
    \label{fig:resolution}
\end{figure}

The models presented in this work have different spatial resolutions and utilise
different radiation transport (RT) schemes. As a comparison,
Fig.~\ref{fig:model_parameter_comparison} shows the difference between
the fitted abundance and broadening values in the msc600, m595 and n59 models.
We find that using the coarser models predicts a slightly higher abundance,
and the lines are not synthesised as broad as when using the finer msc600 model,
which is shown in Fig.~\ref{fig:resolution}. This is in line with the reduced
level of turbulence in the n59 model due to its lower resolution. The n59 model
also has a significantly higher extra viscosity relative to the msc600 model,
and utilises a long characteristic RT scheme while the msc600
model uses a newer, multiple short characteristic scheme. The m595 model has
the same parameters as msc600, except the spatial resolution, which is that of
n59. Its fitted broadening lies between n59 and msc600, but is closer to the
latter model. This suggests that, rather than the spatial resolution, it is the
RT scheme and viscosity parameters that primarily affect the line broadening;
though the spatial resolution does have a small effect.
All models use the same opacity table and equation of state. With the
chosen line sample, generally positive broadening is required for the n59 model;
however, when considering all lines, the majority still require negative
broadening to fit the observations.

\subsection{Disk-centre and disk-integrated spectrum differences}
\begin{figure*}
    \centering
    \includegraphics[width=160mm]{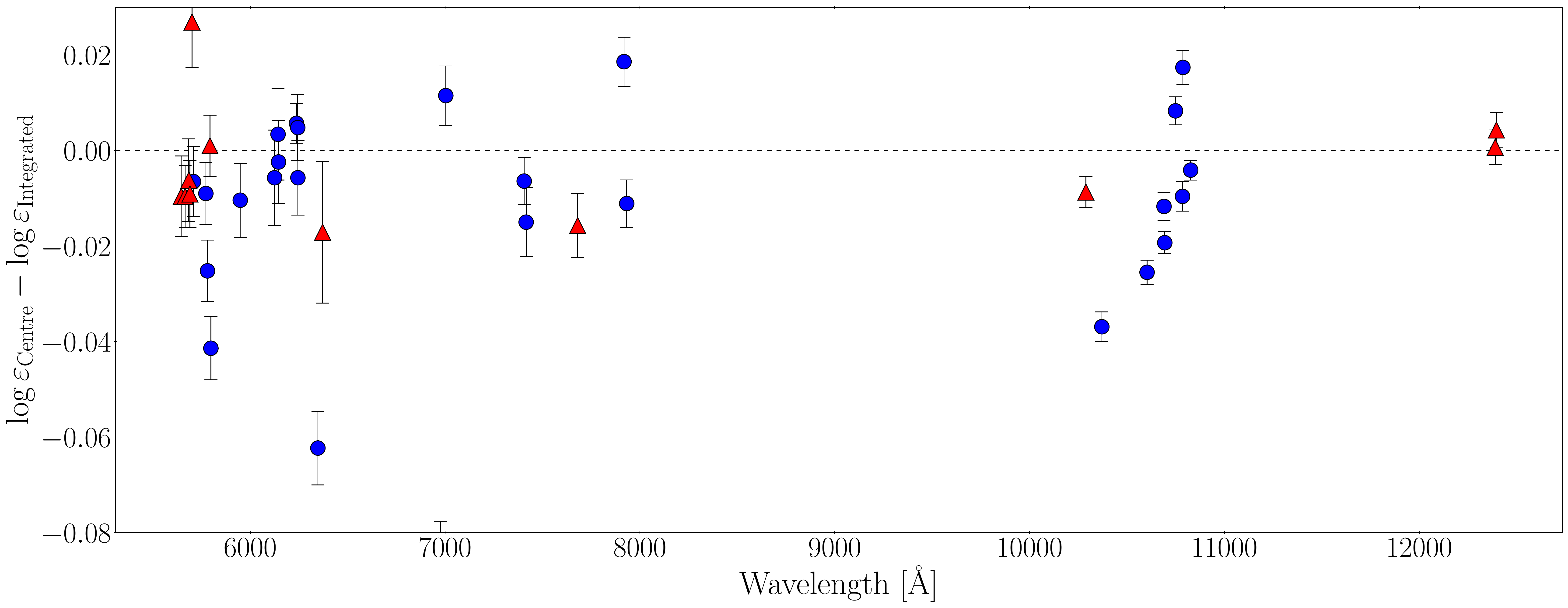}
    \caption{Differences between disk-centre and disk-integrated fitted
        abundances for the msc600 model. Red triangles indicate the lines in the
        subsample used for the final abundance calculation.}
    \label{fig:clv_parameter_comparison}
\end{figure*}

Comparisons between disk-centre and disk-integrated spectra for the msc600
model are shown in Fig.~\ref{fig:clv_parameter_comparison}. Disk-integrated
spectra show a $0.01$ dex higher abundance on average, and a
broadening $0.03$~\kms\ more negative than the disk-centre spectra across the
full sample of $39$ lines. The correspondence between disk-centre and
disk-integrated fitted abundance and broadening is hence quite satisfactory.
The infrared lines not chosen in the subsample show higher
deviation than most optical lines, perhaps due to NLTE core effects that
are more prominent in the disk-integrated spectrum. Additionally, the largest
deviation is given by the \ion{Si}{II} \lineAng{6347.10} line, which
is not used in the subsample because of this large deviation.
The other \ion{Si}{II} at \lineAng{6371.36} line
is used in the subsample and shows a large uncertainty in the fitted abundance.

The fitted abundance and broadening values for the disk-centre and
disk-integrated syntheses, as well as the old and new oscillator strengths used
in this study are given in Table~\ref{table:clv_abu}. Only the disk-centre
spectra and new $gf$-values are used for determining the final abundance.
\begin{table*}
    \centering
    \caption{Fitted 3D LTE abundance and broadening values for disk-centre (DC)
        and disk-integrated (DI) spectra with their formal statistical
        uncertainties.
    }
    \begin{tabular}{lrrllll}
        \noalign{\smallskip}\hline\hline\noalign{\smallskip}
        Wavelength         & W$_\lambda$ (DC)    & W$_\lambda$ (DI)    & $\lgesi$ (DC)     & $\lgesi$ (DI)     & $w_\mathrm{K}$ (DC) & $w_\mathrm{K}$ (DI) \\
        \,[$\mathrm{\AA}$] & [m$\mathrm{\AA}$]   & [m$\mathrm{\AA}$]   &                   &                   & [km s$^{-1}$]       & [km s$^{-1}$]       \\
        \noalign{\smallskip}\hline\noalign{\smallskip}
        \ion{Si}{I}        &                     &                     &                   &                   &                     &                     \\
        \noalign{\smallskip}\hline\noalign{\smallskip}
        5645.6128*         & $37.285 \pm 0.004$  & $38.465 \pm 0.004$  & $7.583 \pm 0.005$ & $7.590 \pm 0.006$ & $-0.62 \pm 0.01$    & $-0.60 \pm 0.02$    \\
        5665.5545*         & $40.850 \pm 0.003$  & $42.100 \pm 0.003$  & $7.589 \pm 0.004$ & $7.600 \pm 0.005$ & $-0.36 \pm 0.02$    & $-0.40 \pm 0.04$    \\
        5684.4840*         & $66.447 \pm 0.003$  & $65.793 \pm 0.003$  & $7.560 \pm 0.006$ & $7.566 \pm 0.007$ & $-0.70 \pm 0.01$    & $-0.89 \pm 0.01$    \\
        5690.4250*         & $54.123 \pm 0.003$  & $54.875 \pm 0.003$  & $7.566 \pm 0.006$ & $7.581 \pm 0.005$ & $-0.83 \pm 0.01$    & $-1.12 \pm 0.01$    \\
        5701.1040*         & $40.941 \pm 0.004$  & $41.240 \pm 0.005$  & $7.554 \pm 0.006$ & $7.554 \pm 0.007$ & $-0.67 \pm 0.01$    & $-0.81 \pm 0.02$    \\
        5708.3995          & $83.932 \pm 0.001$  & $81.971 \pm 0.002$  & $7.547 \pm 0.003$ & $7.557 \pm 0.004$ & $-0.61 \pm 0.01$    & $-0.79 \pm 0.01$    \\
        5772.1460          & $58.461 \pm 0.003$  & $58.254 \pm 0.003$  & $7.578 \pm 0.005$ & $7.590 \pm 0.004$ & $-0.31 \pm 0.02$    & $-0.26 \pm 0.25$    \\
        5780.3838          & $34.914 \pm 0.004$  & $38.252 \pm 0.005$  & $7.618 \pm 0.007$ & $7.664 \pm 0.007$ & $+1.94 \pm 0.02$    & $+2.49 \pm 0.00$    \\
        5793.0726*         & $47.079 \pm 0.003$  & $47.932 \pm 0.004$  & $7.553 \pm 0.005$ & $7.565 \pm 0.006$ & $+0.27 \pm 0.02$    & $+0.49 \pm 0.02$    \\
        5797.8559          & $44.799 \pm 0.003$  & $46.259 \pm 0.002$  & $7.476 \pm 0.005$ & $7.497 \pm 0.003$ & $-0.67 \pm 0.01$    & $-0.48 \pm 0.03$    \\
        5948.5410          & $97.658 \pm 0.002$  & $94.248 \pm 0.003$  & $7.581 \pm 0.005$ & $7.598 \pm 0.006$ & $-0.60 \pm 0.01$    & $-0.82 \pm 0.01$    \\
        6125.0209          & $34.496 \pm 0.005$  & $34.737 \pm 0.006$  & $7.517 \pm 0.006$ & $7.529 \pm 0.007$ & $-0.58 \pm 0.02$    & $-0.53 \pm 0.03$    \\
        6142.4832          & $39.323 \pm 0.004$  & $38.765 \pm 0.006$  & $7.408 \pm 0.006$ & $7.413 \pm 0.007$ & $-1.10 \pm 0.01$    & $-1.12 \pm 0.02$    \\
        6145.0159          & $44.040 \pm 0.004$  & $42.187 \pm 0.005$  & $7.487 \pm 0.006$ & $7.479 \pm 0.006$ & $-1.17 \pm 0.01$    & $-1.36 \pm 0.01$    \\
        6237.3191          & $81.782 \pm 0.002$  & $76.779 \pm 0.003$  & $7.468 \pm 0.003$ & $7.462 \pm 0.003$ & $-1.23 \pm 0.01$    & $-1.39 \pm 0.01$    \\
        6243.8146          & $57.363 \pm 0.003$  & $54.134 \pm 0.004$  & $7.574 \pm 0.004$ & $7.567 \pm 0.005$ & $-0.96 \pm 0.01$    & $-1.13 \pm 0.01$    \\
        6244.4655          & $54.674 \pm 0.003$  & $52.042 \pm 0.004$  & $7.390 \pm 0.005$ & $7.388 \pm 0.005$ & $-0.98 \pm 0.01$    & $-1.15 \pm 0.01$    \\
        6976.5129          & $56.438 \pm 0.003$  & $52.806 \pm 0.003$  & $7.654 \pm 0.003$ & $7.648 \pm 0.004$ & $-1.22 \pm 0.01$    & $-1.22 \pm 0.01$    \\
        7003.5690          & $75.507 \pm 0.002$  & $68.469 \pm 0.003$  & $7.583 \pm 0.003$ & $7.569 \pm 0.005$ & $-1.43 \pm 0.01$    & $-1.39 \pm 0.01$    \\
        7034.9006          & $83.976 \pm 0.002$  & $76.443 \pm 0.003$  & $7.575 \pm 0.003$ & $7.565 \pm 0.004$ & $-1.13 \pm 0.01$    & $-1.28 \pm 0.01$    \\
        7226.2079          & $43.345 \pm 0.004$  & $42.227 \pm 0.004$  & $7.587 \pm 0.005$ & $7.597 \pm 0.006$ & $-0.24 \pm 0.02$    & $-1.47 \pm 0.05$    \\
        7405.7718          & $110.020 \pm 0.002$ & $101.650\pm 0.002$  & $7.634 \pm 0.004$ & $7.655 \pm 0.004$ & $-1.31 \pm 0.00$    & $-1.27 \pm 0.01$    \\
        7415.9480          & $105.090 \pm 0.002$ & $99.125 \pm 0.002$  & $7.558 \pm 0.004$ & $7.590 \pm 0.005$ & $-0.83 \pm 0.01$    & $-0.87 \pm 0.01$    \\
        7680.2660*         & $102.790 \pm 0.002$ & $95.068 \pm 0.002$  & $7.669 \pm 0.004$ & $7.698 \pm 0.005$ & $-0.93 \pm 0.01$    & $-0.92 \pm 0.01$    \\
        7918.3835          & $103.440 \pm 0.002$ & $97.816 \pm 0.002$  & $7.690 \pm 0.003$ & $7.723 \pm 0.004$ & $-0.82 \pm 0.01$    & $-0.85 \pm 0.01$    \\
        7932.3479          & $127.030 \pm 0.002$ & $117.070 \pm 0.002$ & $7.689 \pm 0.004$ & $7.706 \pm 0.004$ & $-0.61 \pm 0.01$    & $-0.81 \pm 0.01$    \\
        10288.9440*        & $88.692 \pm 0.001$  & $85.213 \pm 0.001$  & $7.522 \pm 0.002$ & $7.538 \pm 0.003$ & $-0.37 \pm 0.01$    & $-0.39 \pm 0.01$    \\
        10371.2630         & $199.010 \pm 0.001$ & $189.420 \pm 0.001$ & $7.621 \pm 0.002$ & $7.668 \pm 0.003$ & $-0.63 \pm 0.00$    & $-1.08 \pm 0.00$    \\
        10603.4250         & $295.370 \pm 0.001$ & $281.120 \pm 0.001$ & $7.686 \pm 0.002$ & $7.742 \pm 0.002$ & $-1.24 \pm 0.00$    & $-1.88 \pm 0.00$    \\
        10689.7160         & $228.360 \pm 0.001$ & $205.700 \pm 0.001$ & $7.712 \pm 0.003$ & $7.741 \pm 0.002$ & $-0.91 \pm 0.00$    & $-1.28 \pm 0.00$    \\
        10694.2510         & $251.900 \pm 0.001$ & $230.870 \pm 0.001$ & $7.679 \pm 0.002$ & $7.727 \pm 0.002$ & $-1.15 \pm 0.00$    & $-1.51 \pm 0.00$    \\
        10749.3780         & $330.400 \pm 0.000$ & $308.620 \pm 0.001$ & $7.677 \pm 0.001$ & $7.713 \pm 0.002$ & $-1.44 \pm 0.00$    & $-2.08 \pm 0.03$    \\
        10784.5620         & $106.910 \pm 0.001$ & $98.584 \pm 0.001$  & $7.645 \pm 0.002$ & $7.664 \pm 0.002$ & $-0.54 \pm 0.01$    & $-0.72 \pm 0.01$    \\
        10786.8490         & $296.440 \pm 0.001$ & $278.200 \pm 0.001$ & $7.651 \pm 0.001$ & $7.687 \pm 0.002$ & $-1.29 \pm 0.00$    & $-1.98 \pm 0.03$    \\
        10827.0880         & $473.350 \pm 0.000$ & $433.190 \pm 0.001$ & $7.662 \pm 0.001$ & $7.678 \pm 0.002$ & $-2.32 \pm 0.01$    & $-1.97 \pm 0.02$    \\
        12390.1540*        & $89.770 \pm 0.001$  & $87.503 \pm 0.001$  & $7.615 \pm 0.002$ & $7.613 \pm 0.003$ & $-0.11 \pm 0.03$    & $-0.29 \pm 0.83$    \\
        12395.8320*        & $111.000 \pm 0.001$ & $106.810 \pm 0.001$ & $7.613 \pm 0.002$ & $7.609 \pm 0.003$ & $+0.08 \pm 0.03$    & $-0.38 \pm 0.02$    \\
        \noalign{\smallskip}\hline\hline\noalign{\smallskip}
        \ion{Si}{II}       &                     &                     &                   &                   &                     &                     \\
        \noalign{\smallskip}\hline\noalign{\smallskip}
        6347.1087          & $55.354 \pm 0.002$  & $48.835 \pm 0.002$  & $7.675 \pm 0.005$ & $7.547 \pm 0.005$ & $+0.91 \pm 0.01$    & $+1.43 \pm 0.01$    \\
        6371.3714*         & $37.936 \pm 0.004$  & $32.226 \pm 0.006$  & $7.610 \pm 0.010$ & $7.469 \pm 0.011$ & $+0.59 \pm 0.02$    & $+1.25 \pm 0.03$    \\
        \noalign{\smallskip}\hline\noalign{\smallskip}
    \end{tabular}
    \label{table:clv_abu}
    \tablefoot{
        $w_\mathrm{K}$ is the width of the broadening kernel. The old and new
        $\log{gf}$ values are also provided. An asterisk next to the wavelength
        signifies the line was in the chosen subsample.
    }
\end{table*}

\subsection{Comparisons with meteoritic abundances}

\begin{figure}
    \centering
    \includegraphics[width=90mm]{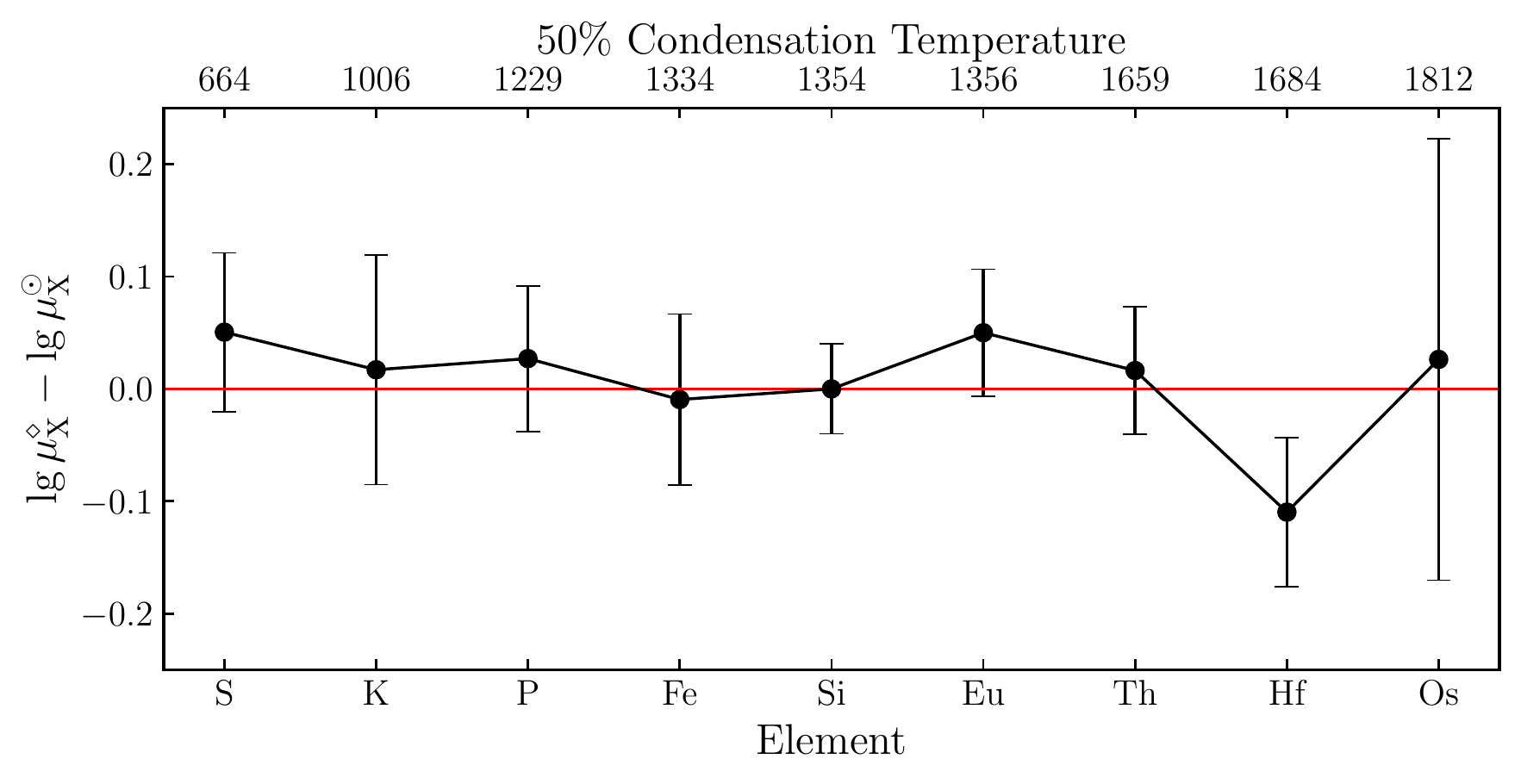}
    \caption{Elemental abundance differences between CI chondrites
        ($\mu_X^\diamond$) and solar photospheric composition ($\mu_X^\odot$)
        according to \cobold\ in ascending order of the condensation
        temperatures of the elements. The silicon abundance difference is zero
        by definition, but shown to illustrate the scale of the error.}
    \label{fig:meteoric_comparison}
\end{figure}

Type-I carbonaceous chondrites (CI chondrites) constitute a special class of
meteorites. Their chemical composition of refractory elements is believed to
reflect the composition of the early solar system
\citep[e.g.,][]{loddersAbundancesElements2009}. Conventionally, meteoritic
abundances are given relative to silicon on the so-called cosmochemical scale,
here for an element X written as
\begin{equation}
    \Ac{X} \equiv 10^6\times \frac{\nd{X}}{\nd{Si}},
\end{equation}
where \nd{X}\ denotes the number density per volume of element~X. In this
paper, we gave abundances on the astronomical scale defined by
\begin{equation}
    \Aa{X} \equiv 10^{12}\times \frac{\nd{X}}{\nd{H}}\,.
    \label{e:AaX}
\end{equation}
With these definitions we have\footnote{$\lg\equiv\log_{10}$} $\lg\Ac{Si}=6$,
and $\lg\Aa{H}=12$. Conversion from the abundance of an element~X from the
astronomical to the cosmochemical scale reads
\begin{equation}
    \lg\Ac{X} = \lg\Aa{X} - \lg\Aa{Si} + 6
    \label{e:atoc}
\end{equation}
which necessitates the knowledge of the silicon abundance on the astronomical
scale. Equation~\eref{e:atoc} shows that an increase of the silicon abundance
-- as found in this work in comparison to earlier results -- leads to a
corresponding decrease of an abundance given on the cosmochemical scale. Since
the abundance on the cosmochemical scale involves two abundances on the
astronomical scale the conversion generally leads to an increase of the
resulting uncertainty for an abundance -- assuming that there are no
significant correlations among the individual input uncertainties. Dealing
with a ratio of two metal abundances here has the advantage that settling
effects over the lifetime of the Sun should cancel out to first order
\citep{loddersAbundancesElements2009}, and we can directly compare to the
corresponding meteoritic ratios. In Fig.~\ref{fig:meteoric_comparison} we
compare these abundance ratios on the cosmochemical scale in ascending order
of the $50\%$ condensation temperatures of the elements.  All condensation
temperatures are from \citet{palmeSolarSystem2014} and meteoritic abundances
are taken from \citet[][her Tab.~3, present solar
    system]{loddersRelativeAtomic2021}. We use the solar photospheric abundances
of non-volatile elements based on \cobold\ models presented in
\citet{caffauSolarChemical2011}. The uncertainties indicated in
Fig.~\ref{fig:meteoric_comparison} are dominated by the uncertainties of the
spectroscopically determined photospheric abundances. The uncertainty of the
silicon abundance noticeably contributes here. The error bars are perhaps
over-estimated since some compensatory effects due to error correlations might
be present. With the exception of hafnium (a well-known problem case)
meteoritic and photospheric abundances are consistent with each other on the
$1\,\sigma$ level. However, for being able to identify possible differences a
reduction of the uncertainties appears desirable.

\subsection{Mass fractions of hydrogen, helium, and metals}

In this section we calculate the mass fractions of hydrogen~\mfX, helium~\mfY,
and metals~\mfZ\ which are of particular interest for stellar structure. Our
intention is not so much to provide the absolute numbers but rather to
demonstrate the involved uncertainties.
\citet{serenelliImplicationsSolar2016} and \citet{vinyolesNewGeneration2017}
both advocate the use of meteoritic abundances for elements heavier than C, N,
O in the Sun. We follow this idea, and augment the 12 photospheric abundances
from \citet{caffauSolarChemical2011} (Li, C, N, O, P, S, K, Fe, Eu, Os, Hf,
Th) and the newly derived silicon abundance by the meteoritic abundances given
by \citet{loddersRelativeAtomic2021}. To this end, we need to bring
photospheric and meteoritic abundances onto the same scale. For relating the
abundances we use the abundance of \textit{silicon only}. Using
Eq.~\eref{e:AaX} we obtain
\begin{equation}
    \lg\Aam{X}  = \lg\Acm{X} + \lg\Aas{Si} - 6
    \hspace{2em}\mbox{or}\hspace{2em}
    \Aam{X}  = \frac{\ndm{X}}{\ndm{Si}} \, \Aas{Si},
    \label{e:ctoa}
\end{equation}
where diamonds ($\diamond$) indicate meteoritic values, the Sun symbol
($\odot$) solar photospheric values. The basic assumption underlying
Eq.~\eref{e:ctoa} is that $\Aam{Si}=\Aas{Si}$, that is to say, silicon is not
subject to differentiation effects, neither in the solar photosphere, nor in the
meteorites.

Equation~\eref{e:ctoa} shows that in the conversion of the meteoritic
abundances we have to take into account the uncertainties of the individual
species including the silicon abundance as measured in meteorites. The
normalisation puts the meteoritic silicon abundance on a value of $10^6$,
however, with an uncertainty of 3.4\,\%. The uncertainty of the solar silicon
abundance contributes to the meteoritic uncertainties in the
conversion to the astronomical scale. An input neither directly obtained
from meteorites nor from photospheric spectroscopy is the helium
abundance. Here, we assume a ratio
$\nds{He}/\nds{H}=(8.38\pm0.39)\times10^{-2}$ as given by
\citet{loddersRelativeAtomic2021} for the present Sun which is
motivated from helioseismic measurements. With these ingredients and atomic
weights assumed to be precisely known we ran Monte-Carlo error propagations
and obtained $\mfX=0.7382\pm0.0084$, $\mfY= 0.2456\pm0.0086$,
$\mfZ=0.0162\pm0.0015$, and $\mfZ/\mfX=0.0220\pm 0.0020$ (independent of the
abundance of helium). The uncertainties of \mfX\ and \mfY\ are dominated by
the uncertainty of the helium abundance. In order to further quantify the
uncertainties of~\mfZ\ we present the build-up of the overall uncertainty of
\mfZ\ by sequentially adding sources of uncertainty. We obtain the sequence
$\sigma_\mfZ=(65, 77, 79, 148)\times 10^{-5}$ when adding the uncertainties of
the meteoritic abundances, of the photospheric abundance of silicon, of all
photospheric abundances except CNO, and of all contributions including from CNO,
respectively. This perhaps odd-appearing procedure is motivated by the fact that the
individual sources of uncertainty are not simply additive (rather additive in
quadrature), and there is some coupling emergent from the presence of the
photospheric silicon abundance in the conversion given by
Eq.~\eref{e:ctoa}. Keeping this limitation in mind, we see that the
contribution to the overall uncertainty by the meteorites is not
insignificant, and noticeably increased by the uncertainty of the
photospheric silicon abundance. We emphasise that the ``meteoritic'' abundances
include the noble gases (particularly neon). Their assumed values are coming
from measurements other than those in meteorites. The photospheric
abundances other than of CNO contribute little to the overall uncertainty, the
most important contribution stems -- perhaps unsurprisingly -- from CNO.
We conclude that the procedure of using the average of several refractory
elements \citep[e.g.,][]{loddersAbundancesElements2009} is a good way to
reduce the overall uncertainty on the mass fraction of metals.
The precision to which the CNO elements are measured in the solar
photosphere is most important. Beyond that, any reduction of the uncertainty of
photospheric or meteoritic abundances is helpful.


\section{Discussion}
\label{sec:discussion}
\subsection{Comparisons between models and line samples}
Both the choice of model atmosphere and the line sample affect the final
abundance (and the level of broadening or debroadening required).
Across all configurations, prior broadening of the observations increases
the fitted abundance. Moreover, there is a spread of fitted abundances for
even a single model based on the chosen line sample and fitting method.
Overall, it is the combination of more or less judicious decisions taken
during the fitting process that ultimately determines the final fitted
abundance value.

\subsection{Comparisons with previous works}

\begin{figure}
    \centering
    \includegraphics[width=90mm]{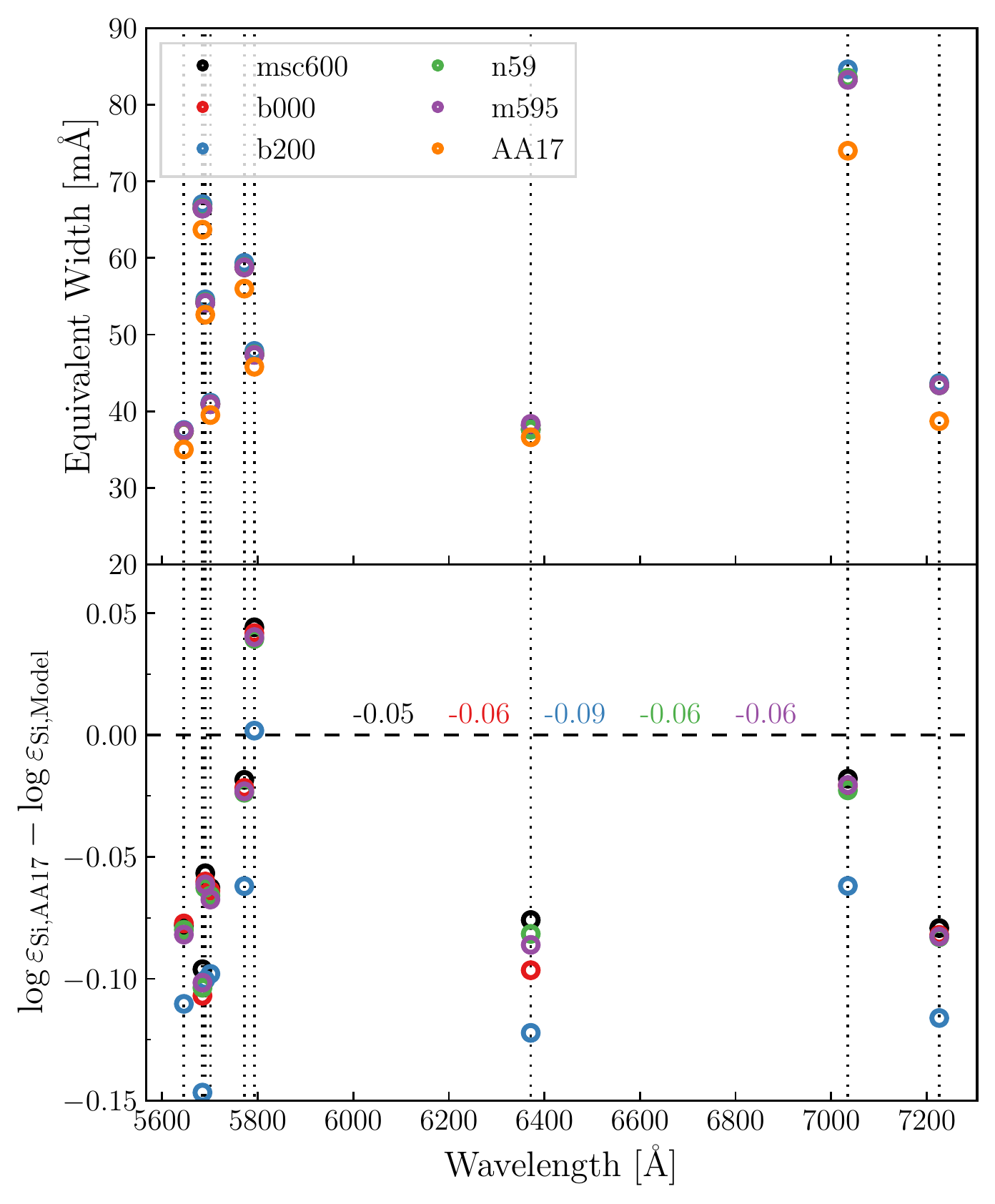}
    \caption{
        Differences in equivalent width (top) and fitted abundance (bottom) in
        comparison to the values in \citet{amarsiSolarSilicon2017}. The numbers
        in the bottom panel give the mean abundance difference using the
        weighting scheme of \citet{amarsiSolarSilicon2017}. The difference
        in abundance includes all sources, not just the difference in equivalent
        width.
    }
    \label{fig:anish_ew}
\end{figure}

Our derived abundance of \abufinal\ is $0.06$ dex higher than the abundance
recently presented in \citet{asplundChemicalMakeup2021c}.
As a comparison to their work, we calculated a mean abundance using the
line sample they presented (leaving out the $6741.61$ \AA\ line) alongside their
weights, and find an abundance of $7.55 \pm 0.02$, which is $0.04$ dex
higher than that presented in their work. Part of the difference in fitted
abundance stems from the new oscillator strength data used; when using the same
oscillator strength data, we find an abundance of $7.54 \pm 0.03$. This is
consistent with the average difference of the oscillator strength data used for
these lines. The
equivalent widths obtained by our line profile fitting for all models are higher
than those given in \citet{amarsiSolarSilicon2017}, and are shown in
Table~\ref{table:eqw_aa17}.
The corresponding differences in abundance are shown in Fig.~\ref{fig:anish_ew}.
Altogether, the dominant difference comes from the new oscillator strength data
(~$0.04$ dex) and equivalent widths (~$0.02$ dex), while line selection,
weighting and 3D model are of secondary importance (~$0.01$ dex).

\begin{table}
    \caption{
        Equivalent widths for the four models presented in this study
        alongside those given in \citet{amarsiSolarSilicon2017} (AA17).
    }
    \scalebox{0.90}{
        \begin{tabular}{rrrrrrr}
            \toprule
            Wavelength   & msc600   & b000     & b200     & n59      & m595     & AA17 \\
            \,[$\AA$]    & [m$\AA$] & [m$\AA$] & [m$\AA$] & [m$\AA$] & [m$\AA$]        \\
            \midrule
            \ion{Si}{I}  &          &          &          &          &                 \\
            \midrule
            5645.6128    & 37.4     & 37.4     & 37.5     & 37.4     & 37.4     & 35.0 \\
            5684.4840    & 66.5     & 67.0     & 67.0     & 66.5     & 66.5     & 63.7 \\
            5690.4250    & 54.2     & 54.2     & 54.6     & 54.2     & 54.1     & 52.6 \\
            5701.1040    & 40.9     & 40.9     & 41.1     & 40.9     & 40.9     & 39.5 \\
            5772.1460    & 58.8     & 58.8     & 59.4     & 58.8     & 58.8     & 56.0 \\
            5793.0726    & 47.4     & 47.4     & 47.9     & 47.4     & 47.4     & 45.8 \\
            7034.9006    & 83.4     & 83.4     & 84.6     & 83.5     & 83.2     & 74.0 \\
            7226.2079    & 43.4     & 43.4     & 43.7     & 43.4     & 43.4     & 38.7 \\
            \midrule
            \ion{Si}{II} &          &          &          &          &                 \\
            \midrule
            6371.3714    & 38.1     & 37.6     & 37.6     & 37.6     & 38.3     & 36.6 \\
            \bottomrule
        \end{tabular}
    }
    \tablefoot{
        Our
        higher equivalent widths lead to a $+0.02$ dex increase in abundance across
        all cases (using the weighting of AA17).
    }
    \label{table:eqw_aa17}
\end{table}

\subsection{Uncertainties}
\label{subsec:uncertainties}

We use the root-mean-squared error (RMSE) of the selected sample
to capture the final uncertainty on
the fitted abundance. This uncertainty represents the uncertainties through
the entire fitting procedure, including uncertainties in oscillator strengths
as well as statistical uncertainties. The RMSE is given by

\begin{equation}
    \mathrm{RMSE} = \sqrt{\frac{1}{N}\sum_i^L{(y_i - \hat{y})^2}},
    \label{eq:rmse}
\end{equation}

where $\hat{y}$ is the weighted mean abundance, $y_i$ is the weighted
abundance calculated for a single line, $L$ is the number of lines used to
determine the weighted mean and $N$ is the sum of the weights. We chose this
estimator since each line
yields a different fitted abundance, and the RMSE naturally incorporates this
scatter. Additionally, since we compute the RMSE on the final fitted abundances,
the uncertainties in fitting and in the oscillator strengths are already
represented in the scatter. On average, the oscillator strength uncertainty is
$8.57\%$, which is compatible with our RMSE uncertainty in abundance. For other
fitted quantities, we use the statistical uncertainties from the fitting
procedure and other relevant sources of error, such as uncertainties in
ABO theory parameters for broadening.

Despite careful line selection, new oscillator strength values and an
improved broadening theory, there is still a
substantial scatter in individual fitted abundance values. The scatter is
included in the uncertainties by means of the RMSE.
The (maximum - minimum) scatter in abundance values is $0.15$ dex for our main
configuration, with an RMSE of $0.04$. Using the line list of
\citet{amarsiSolarSilicon2017} decreases the scatter to $0.06$ dex and the RMSE
to $0.02$. Again, this shows that the choice of lines to use has a
non-negligible impact on the final derived abundance.

\begin{figure}
    \centering
    \includegraphics[width=80mm]{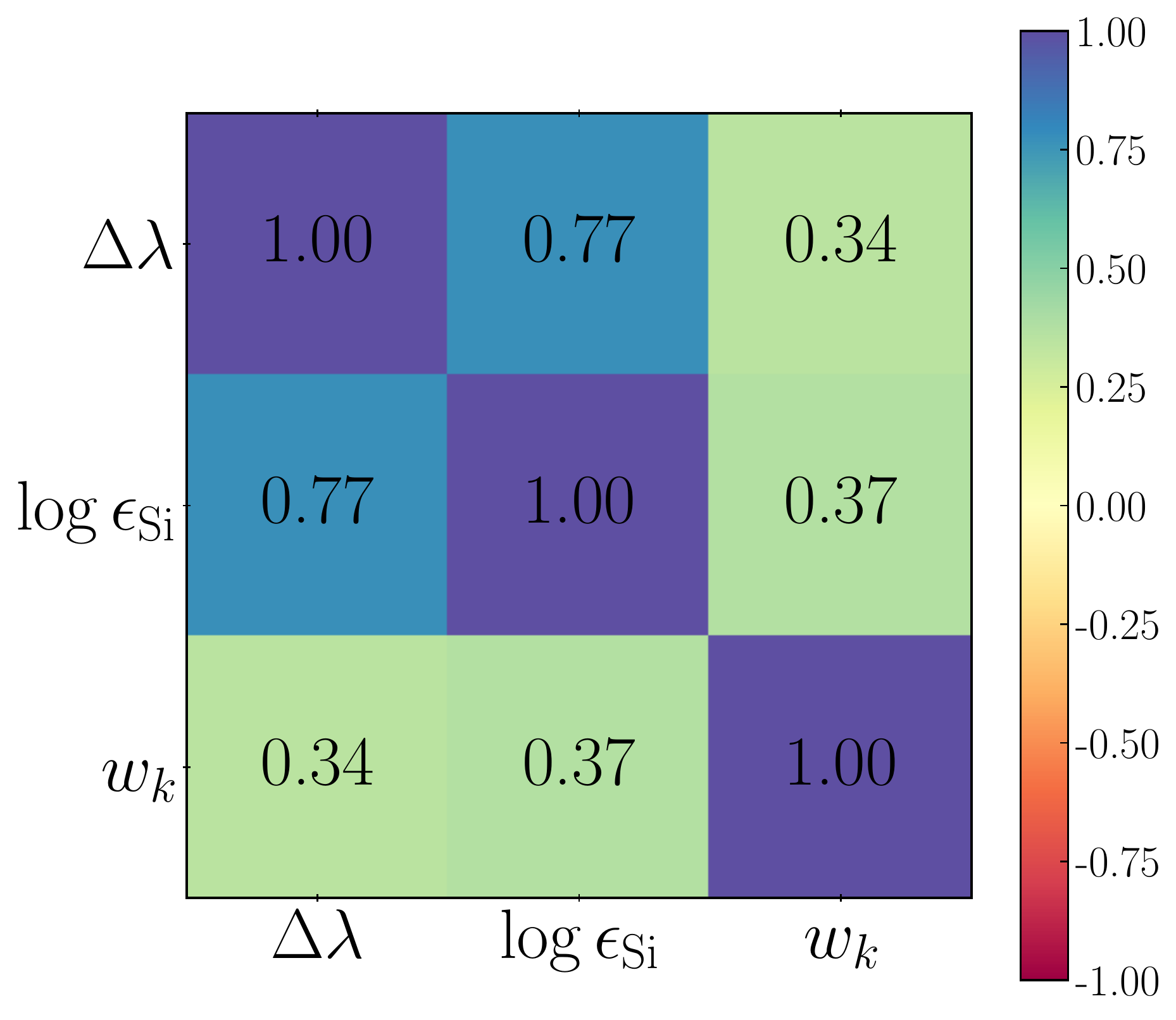}
    \caption{Correlation coefficient between fitted parameters for the msc600 model.
        $\Delta \lambda$ is the wavelength shift, $\lgesi$ is the abundance
        and $w_\mathrm{K}$ is the width of the broadening kernel.}
    \label{fig:correlation}
\end{figure}

We make use of the `mpfitcovar' routine
\citep{markwardtNonlinearLeastsquares2009} which takes in the spectrum and
correlated noise model and fits the free parameters of our routine
(normally abundance, broadening and wavelength shift). We use \chisq\
statistics to find the best fitting parameter as
well as the errors on those parameters. There is some correlation between
the quantities themselves, shown in Fig.~\ref{fig:correlation} for the
msc600 model.

\section{Conclusions}
\label{sec:conclusion}
\noindent
We have presented a 3D LTE analysis of 39 silicon lines using \cobold\
model atmospheres and the \linfor\
spectral synthesis code. Of these, a total of $11$ were selected for the
abundance analysis, comprising of 7 optical \ion{Si}{I} lines, 3 near-infrared
\ion{Si}{I} lines and 1 \ion{Si}{II} line. New oscillator strengths from
\citet{pehlivanrhodinExperimentalComputational2018} were used,
enabling the use of infrared lines alongside optical ones and
providing smaller uncertainties for oscillator strengths. Compared to the
previous experimental strengths from \citet{garzAbsoluteOscillator1973},
the new $\log(gf)$ values and weighting scheme decrease the formal statistical
uncertainty across the relevant lines from $0.07$ dex to $0.04$ dex.
An improved broadening theory also helped to constrain statistical
uncertainties further.

Our main conclusions are as follows:

\begin{itemize}
    \item {
          We find a photospheric solar silicon abundance of
          $\lgesi$ = \abufinal, including the $-0.01$ dex correction from NLTE
          effects investigated in \citet{amarsiSolarSilicon2017}.
          The $0.06$ dex increase with
          respect to the recent studies by \citet{asplundChemicalMakeup2021c},
          \citet{amarsiSolarSilicon2017} suggests that the
          determination of the solar silicon abundance is not yet a firmly
          solved problem. Our advocated configuration uses the $G_1$
          broadening kernel, the higher resolution msc600 model and our chosen
          subsample of lines.
          }

    \item{
          Several factors affect the fitted abundance and broadening, but the
          line selection plays the primary role. We focus on lines that are
          devoid of major blends, have updated oscillator strengths,
          and also where syntheses match well with observed line shapes.
          Notably, the near-infrared lines give higher abundances than
          optical lines on average.
          }

    \item{
          The over-broadened line syntheses we see in this work are not
          specific to the \cobold\ atmospheres. Comparisons were made with
          \stagger\ + \balder\ (see Appendix~\ref{appendix:anish} for
          details), and we are able to refit their abundances and
          broadening values using both the msc600 and n59 models.
          Broadening-wise, their syntheses lie between the msc600 and n59
          models described here. Additionally, the overly broadened line
          syntheses are caused by the combination of various effects,
          including velocity fields, atomic broadening and neglect of magnetic
          field effects -- we did not find a single definite cause for
          over-broadening.
          }

    \item{
          Using a magnetic model with a magnetic field strength of $200$\,G
          increases the fitted abundance and reduces the Doppler broadening
          when compared to the same model with a magnetic field strength of
          $0$\,G. These differential results could point
          towards over-broadened syntheses resulting from a lack of
          consideration of magnetic field effects, particularly that strong
          magnetic fields impede turbulent flow in the atmosphere
          \citep{cattaneoInteractionConvection2003}, resulting in narrower
          lines. However, the vertical magnetic fields that would be required
          to offset the negative broadening are much too large to be
          consistent with 3D MHD models with small-scale dynamos
          \citep{shchukinaImpactSurface2015,shchukinaImpactSurface2016},
          meaning magnetic field effects likely do not play a major role in
          the overbroadening of the line syntheses.
          }

    \item{
          The atomic broadening cross-sections for the silicon lines presented
          in this work are remarkably large, which subsequently increases
          the effect of collisional broadening. Alongside the effects of
          magnetic fields, the collisional broadening may then also contribute
          to the over-broadening in the line syntheses.
          }

    \item{
          Meteoritic abundances when transformed onto the astronomical scale are
          increased with respect to the previous study by
          \citet{palmeSolarSystem2014} due to the increase in our photospheric
          silicon abundance. The differences of the non-volatile elements
          available from \cobold-based analyses are -- except for hafnium --
          consistent with zero, however, with significant uncertainties.
          \citet{serenelliImplicationsSolar2016} and
          \citet{vinyolesNewGeneration2017} both advocate the use of meteoritic
          abundances for elements heavier than C, N, O in the Sun, but using only
          the silicon abundance for the conversion between cosmochemical and
          astronomical scale \citep[e.g., ][]{asplundSolarChemical2005} would
          give a $0.05$\,dex increase for non-volatile metals. The sizeable
          uncertainty of the silicon abundance found in this study lets the use of
          multiple elements for referencing meteoritic and photospheric abundances
          appear attractive, such as done in
          \citet{loddersAbundancesElements2009}.
          }

    \item{
          A local fit of the continuum level is clearly at odds when looking
          at the spectrum on a large scale.
          Performing such a fit results in an artificial lowering of the
          equivalent width, and systematically lowers
          the abundance by $0.01$\,dex.
          }

    \item{
          We find no strong evidence of NLTE effects severely affecting
          abundance calculations in optical lines and the \textit{chosen}
          infrared
          lines beyond the $-0.01$ dex correction included, though the
          negative broadening required could be indicative
          of minor NLTE effects being present. Appendix
          \ref{appendix:matthias} explains the observed
          NLTE effects in 1D in more detail.
          }

    \item{
          An NLTE solar silicon abundance of \abufinal\ could improve the
          differences for solar neutrino fluxes, sound
          speed profiles and the surface helium fraction.
          \citet{vinyolesNewGeneration2017} show that a solar model
          with the composition proposed in \citet{grevesseStandardSolar1998}
          statistically performs better in regard to the solar sound speed
          profile than a model with a solar composition proposed in
          \citep{asplundChemicalComposition2009}. The former composition
          uses $\lgesi = 7.56\pm0.01$, while the latter uses
          $\lgesi = 7.51\pm0.01$. \citet{serenelliImplicationsSolar2016}
          show that the silicon abundance of $\lgesi = 7.82$ from
          \citet{vonsteigerSolarMetallicity2016} gives worse fits overall for
          solar neutrino fluxes, sound speed profiles, and the surface
          helium fraction, so a silicon abundance of \abufinal , closer to
          the abundance derived from the \citet{grevesseStandardSolar1998}
          composition, clearly results in an improvement relative to
          \citep{asplundChemicalComposition2009}.
          }
\end{itemize}

All in all, our analysis suggests that the photospheric solar Si abundance is
not yet a definitively solved problem. The use of state-of-the-art 3D model
atmospheres and an improved broadening theory is essentially a requirement to
accurately reproduce line shapes. Even with these improvements, our synthetic
line profiles were generally overbroadened with respect to the observations, and
it is unlikely that this feature is unique to \cobold\ model atmospheres or to
Si lines.

\begin{acknowledgements}
    S.A.D. and H.G.L. acknowledge financial support by the Deutsche
    Forschungsgemeinschaft (DFG, German Research Foundation) -- Project-ID
    138713538 -- SFB 881 (``The Milky Way System'', subproject A04).
    A.K. acknowledges support of the "ChETEC" COST Action (CA16117) and from the
    European Union's Horizon 2020 research and innovation programme under grant
    agreement No 101008324 (ChETEC-INFRA).
    P.S.B acknowledges support from the Swedish Research Council through
    individual project grants with contract Nos. 2016-03765 and 2020-03404.
    This work has made use of the VALD database, operated at Uppsala University,
    the Institute of Astronomy RAS in Moscow, and the University of Vienna.
    The authors are grateful for the assistance of A. M. Amarsi
    in this work, especially for comparing differences in model atmospheres, to
    L. Mashonkina for a discussion on NLTE effects, and to H. Hartman for
    providing assistance related to the new oscillator strengths.
\end{acknowledgements}

\bibpunct{(}{)}{;}{a}{}{,} 
\bibliographystyle{aa}
\bibliography{Silicon.bib}

\begin{appendix}
    %
    \section{Comparisons between \linfor/\cobold\ and \balder/\stagger}
    \label{appendix:anish}
    \noindent


    There was motivation to investigate whether the extra broadening present in the
    syntheses produced by \linfor\ using the \cobold\ model atmospheres was unique
    to these codes. To test this, we compared six LTE \ion{Si}{I}
    line syntheses using our models against those
    produced by \balder\ \citep{amarsiEffectiveTemperature2018}, a custom version of
    \texttt{Multi3D} \citep{leenaartsMULTI3DDomainDecomposed2009} (data provided by
    A. Amarsi, priv. comm.). These data were calculated on a 3D hydrodynamic
    \stagger\ model solar atmosphere \citep{colletStaggerGridProject2011}.
    These lines were: $5690$, $5780$, $5793$, $6244$, $6976$ and $7680$\,\AA.

\begin{figure}[h]
        \includegraphics[width=90mm]{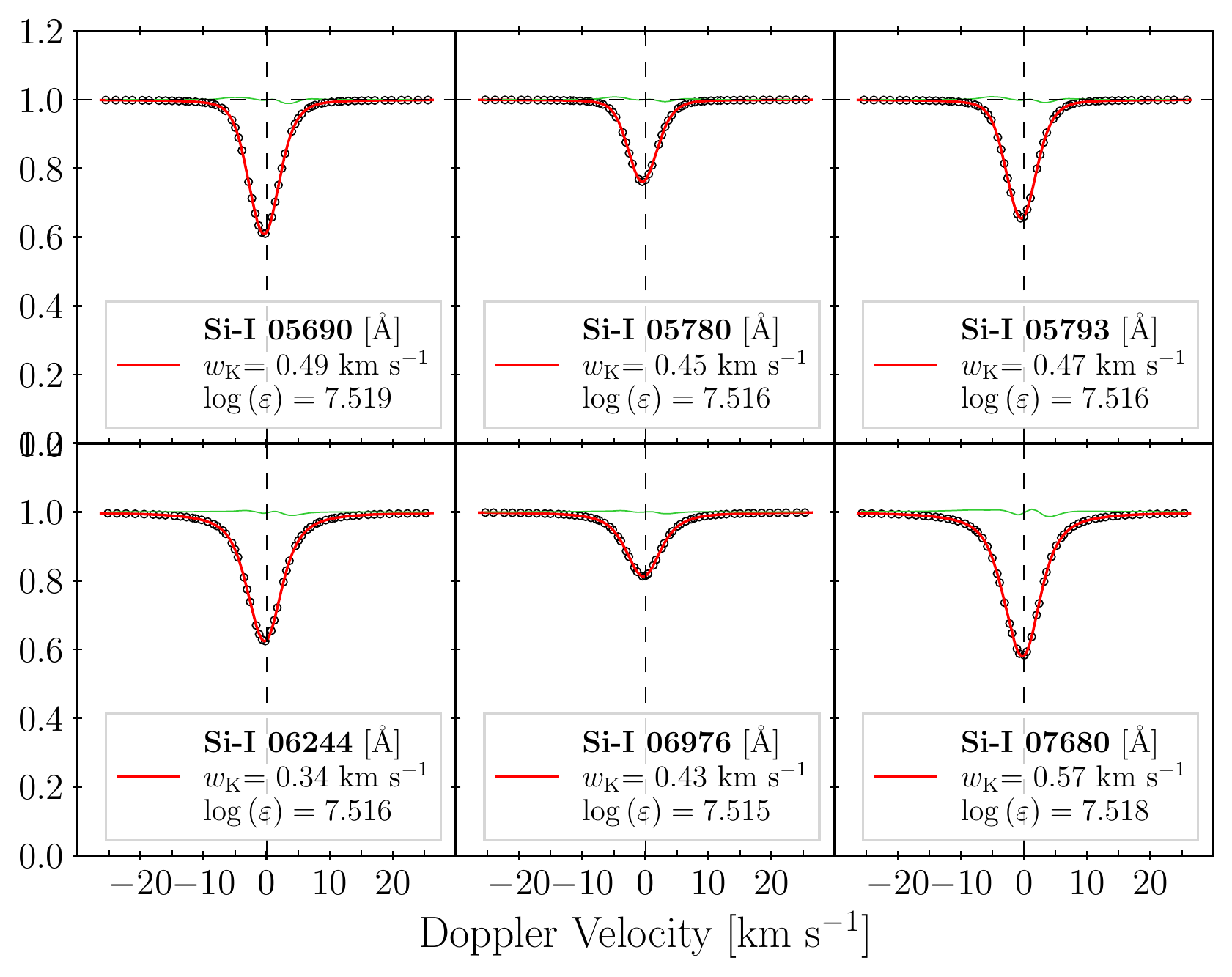}
        \caption{
            \linfor\ + \cobold\ n59 model syntheses (red lines) fit
            to \balder\ + \stagger\ (black points). Green lines show residuals
            increased by a factor of 10. The mean abundance as derived by the
            \cobold\ fit is $7.52\pm0.01$.}
        \label{fig:stagger_n59}
    \end{figure}

    We use the same $\log{gf}$ values, ABO parameters and central wavelengths
    for the lines in both sets of syntheses.
    After fitting for the abundance ($\lgesi=7.51$ was assumed in the BALDER
    syntheses), the results are near-identical, and the fits shown in
    Figs.~\ref{fig:stagger_n59} \& \ref{fig:stagger_msc600} using
    the n59 and 600 models are excellent. We are able to refit the abundance and
    broadening of the \balder\ syntheses. The black points represent the \balder\
    synthesis, and the red lines are the \linfor\ syntheses after fitting the lines
    synthesised with \balder.

    \begin{table}[h]
        \centering
        \caption{
            Fitted broadening values for six lines chosen as a comparison
            to \stagger.
        }
        \begin{tabular}{rllrrrr}
            \noalign{\smallskip}\hline\hline\noalign{\smallskip}
                      & n59 --        & msc600 --     & n59 --        & msc600 --     \\
                      & Stagger       & Stagger       & Hamburg       & Hamburg       \\
            \noalign{\smallskip}\hline\noalign{\smallskip}
            $\lambda$ & \wk           & \wk           & \wk           & \wk           \\
            \noalign{\smallskip}\hline\noalign{\smallskip}
            \,[$\AA$] & [km s$^{-1}$] & [km s$^{-1}$] & [km s$^{-1}$] & [km s$^{-1}$] \\
            \noalign{\smallskip}\hline\noalign{\smallskip}
            5690.4250 & $0.49$        & $-0.33$       & $-0.54$       & $-0.83$       \\
            5780.3838 & $0.45$        & $-0.40$       & $1.47$        & $1.40$        \\
            5793.0726 & $0.47$        & $-0.36$       & $0.72$        & $0.27$        \\
            6244.4655 & $0.34$        & $-0.47$       & $-0.79$       & $-0.98$       \\
            6976.5129 & $0.43$        & $-0.38$       & $-1.05$       & $-1.23$       \\
            7680.2660 & $0.57$        & $-0.08$       & $-0.38$       & $-0.69$       \\
            \noalign{\smallskip}\hline\noalign{\smallskip}
        \end{tabular}
        \tablefoot{
            The values come from fitting the
            n59 and msc600 models fitting to the \stagger\ models (first two columns)
            and to
            the Hamburg atlas observations (last two columns). In all cases, the n59
            model is less broad than the \stagger\
            model syntheses while the msc600 model is broader. However, since the n59
            model generally also requires significant negative broadening to fit the
            observations, the \stagger\ models must also yield too broad lines compared to
            the observations.
        }
        \label{table:stagger_comparison}
    \end{table}

    Two \cobold\ model atmospheres were used for the analysis. The msc600 solar
    model is a newer model that has a higher spatial resolution and larger box size
    than the older n59 model (see Table~\ref{table:model_atm} for details). The
    n59 model requires positive broadening to fit the \balder\ syntheses,
    while the msc600 requires negative broadening. Broadening-wise, then, the
    \balder\ syntheses lie somewhere between our two chosen models - they are
    broader than the n59 model, but not as broad as the msc600
    model. Table \ref{table:stagger_comparison} shows the broadening values required
    by our models to reproduce the \stagger\ syntheses and the observations from
    the Hamburg atlas. While the n59 model is less broad than the Stagger
    model in all cases and the msc600 model is broader, both of these models still
    generally require negative broadening to fit a majority of the lines.
    Again, no instrumental broadening (typically $\sim1$~\kms\ FWHM) was
    applied to the synthetic spectra, as should be possible without the presence of
    over-broadening.
    This shows the issue of the syntheses being over-broadened is not unique
    to \cobold\ + \linfor\, but is also present in the syntheses produced by
    \balder\ + \stagger.

    \begin{figure}[h]
        \includegraphics[width=90mm]{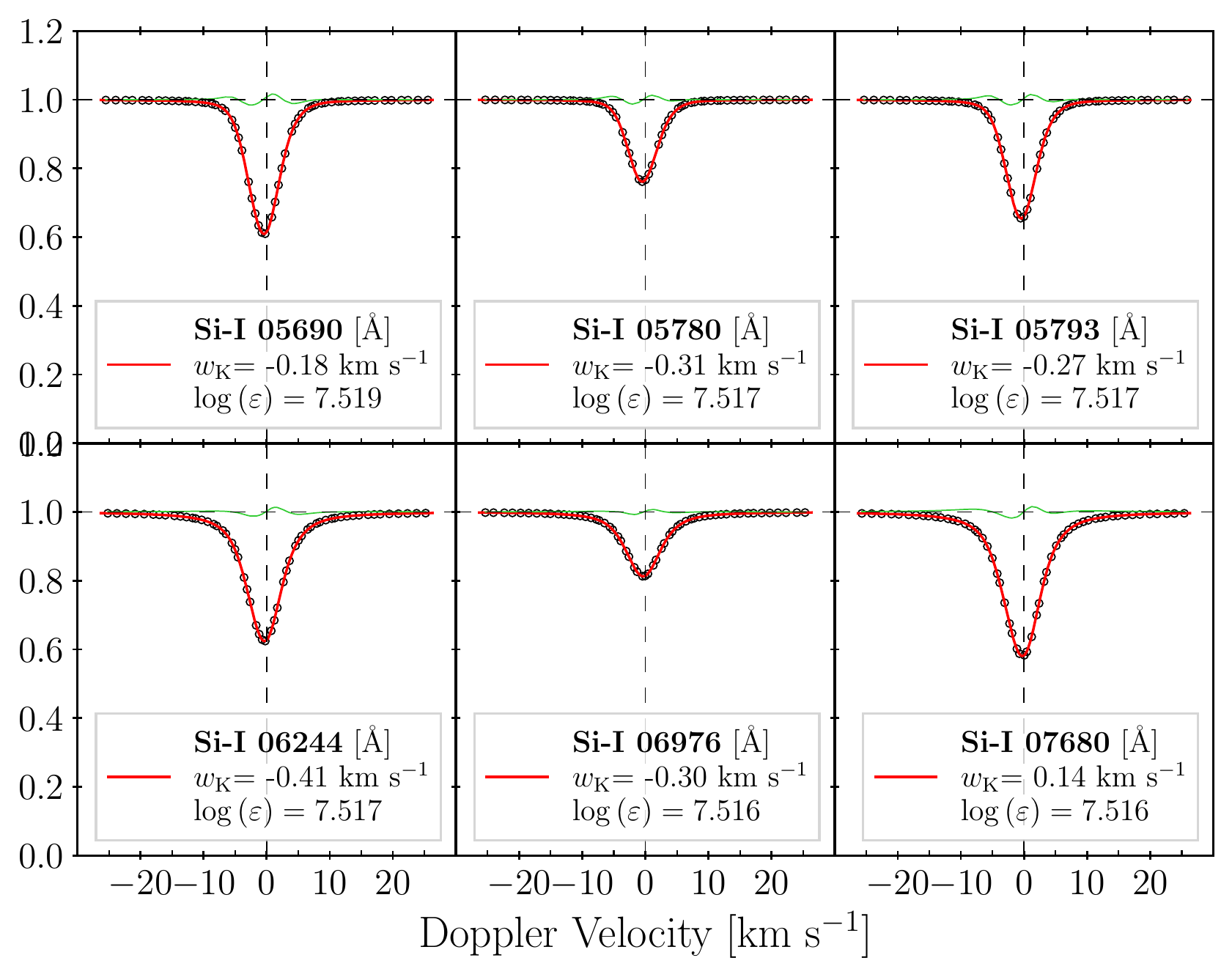}
        \caption{
            \linfor\ + \cobold\ m595 model model syntheses (red lines) fit
            to \balder\ + \stagger\ (black points). Green lines show residuals
            increased by a factor of 10. The mean abundance as derived by the
            \cobold\ fit is $7.52\pm0.01$.}
        \label{fig:stagger_msc600}
    \end{figure}

    \begin{figure}[h]
        \includegraphics[width=90mm]{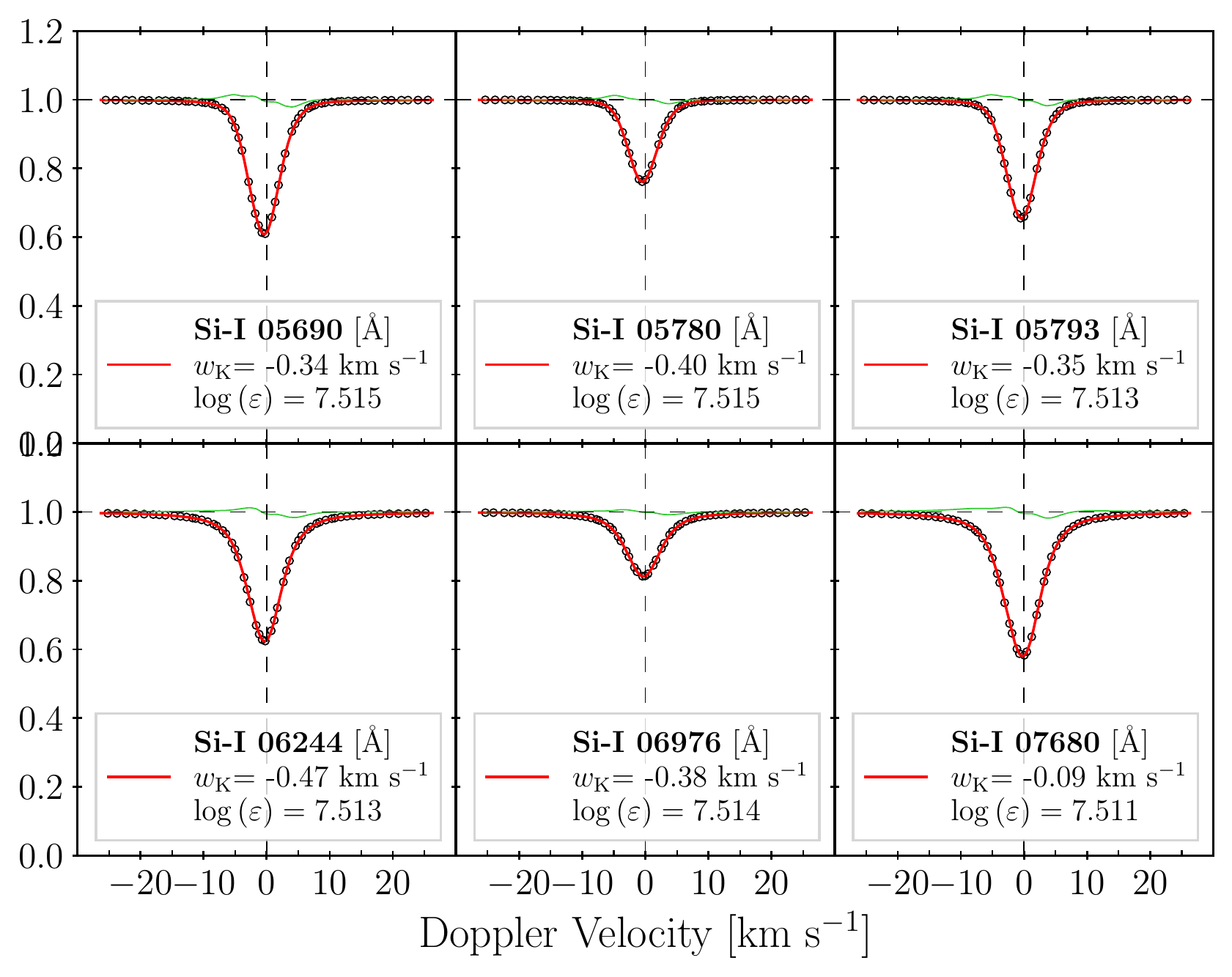}
        \caption{
            \linfor\ + \cobold\ msc600 model model syntheses (red lines) fit
            to \balder\ + \stagger\ (black points). Green lines show residuals
            increased by a factor of 10. The mean abundance as derived by the
            \cobold\ fit is $7.51\pm0.01$.}
        \label{fig:stagger_msc600}
    \end{figure}
    \clearpage
    \newpage
    %
    \section{1D LTE versus NLTE comparisons in \linfor}
    \label{appendix:matthias}
\begin{figure}[h]
        \centering
        \includegraphics[width=90mm]{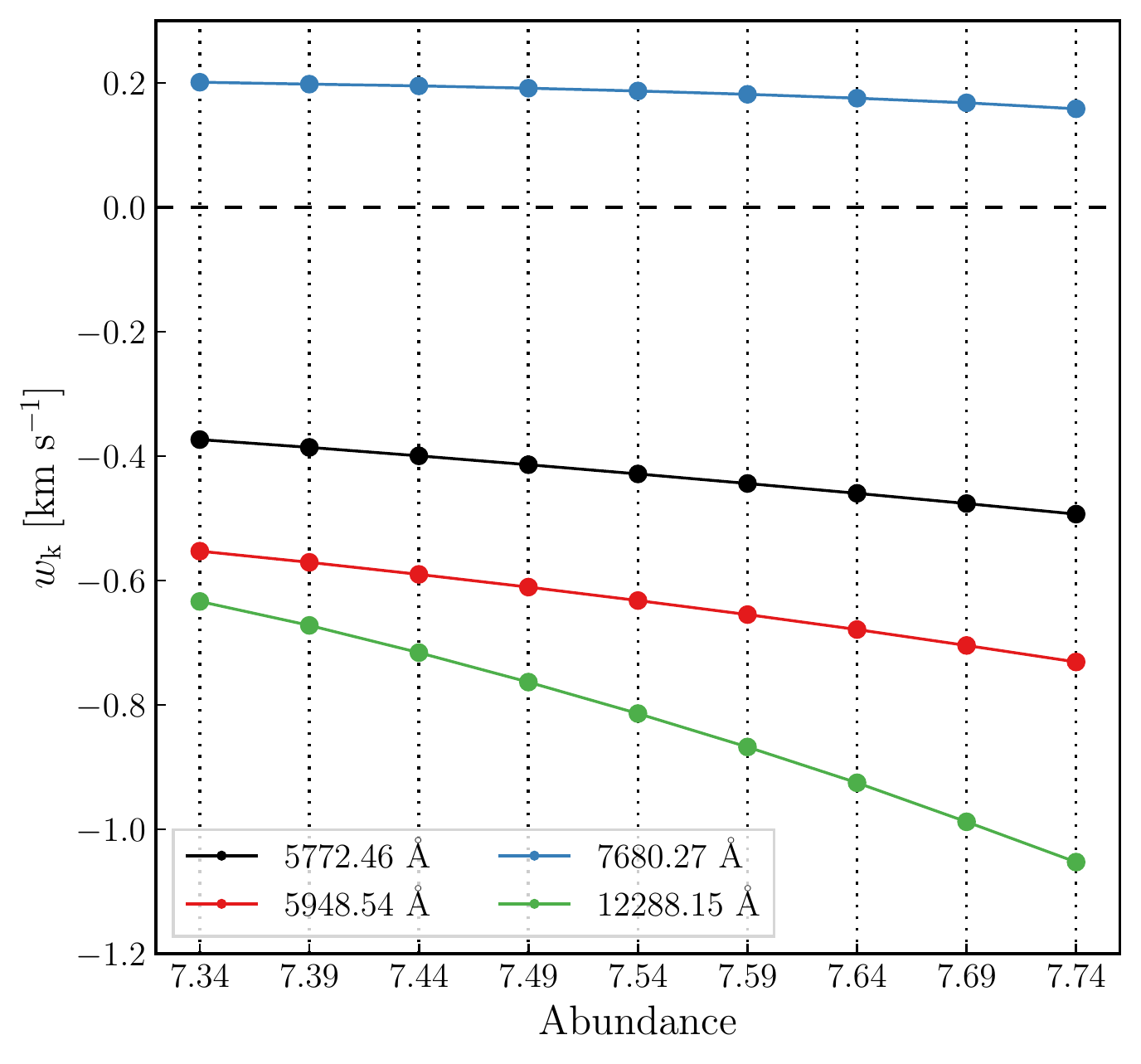}
        \caption{Fitted broadening values for four lines,
            fitting 1D LTE to 1D NLTE profiles. Larger NLTE effects are present in
            \lineAng{5948.65} and \lineAng{12228.15} lines than the
            \lineAng{5772.46} and \lineAng{7680.27} lines.}
        \label{fig:matthias_comparison}
    \end{figure}

\begin{figure}[h]
        \centering
        \includegraphics[width=90mm]{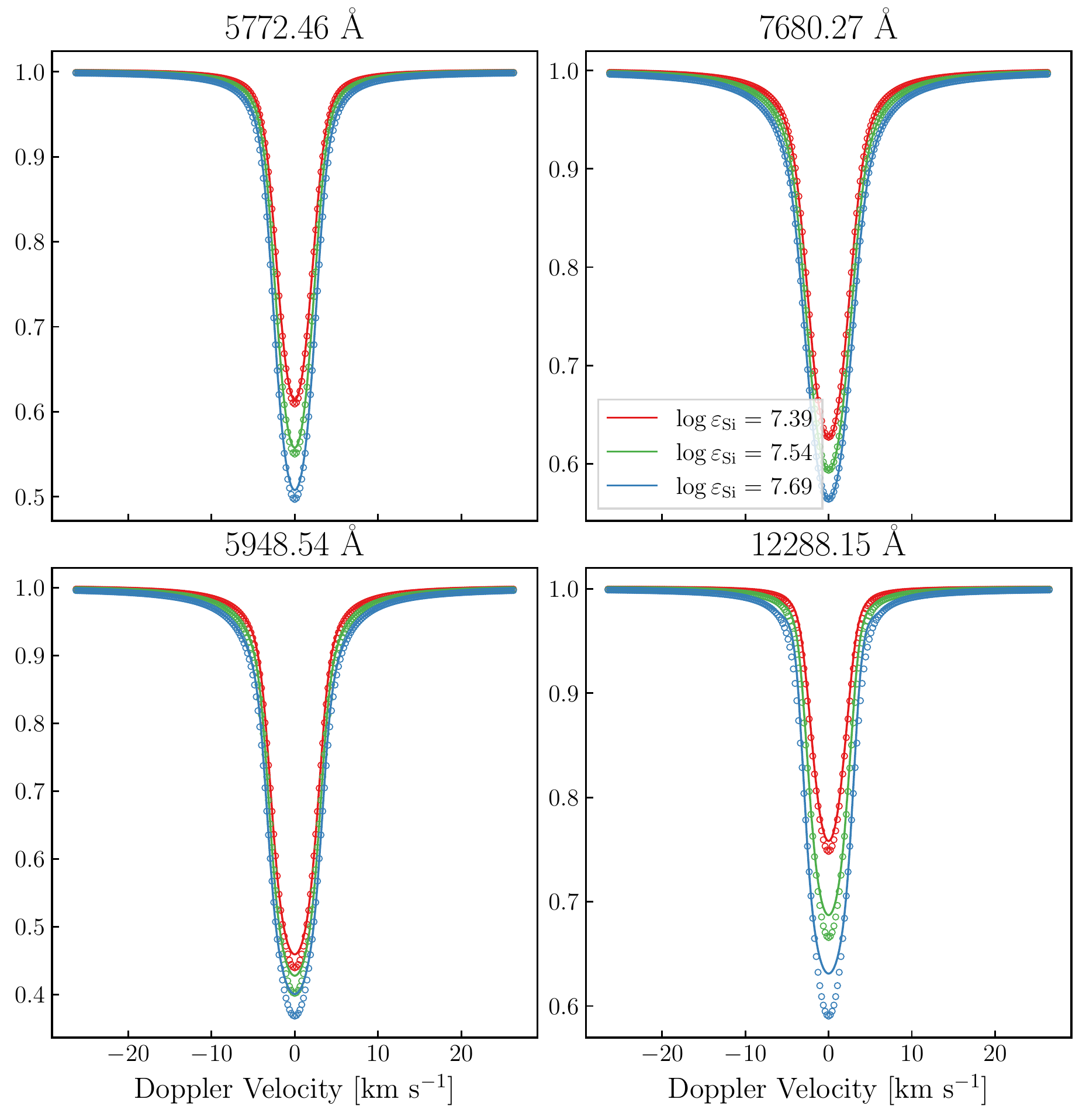}
        \caption{
            LTE (points) versus NLTE (solid lines) profiles for three representative
            abundance values. The \lineAng{5948.54} and \lineAng{12288.15} show
            visible NLTE effects in the core that become stronger with increasing
            abundance.
        }
        \label{fig:lte_nlte_syn}
    \end{figure}

    \noindent
    Following the comparison between model atmospheres and spectral synthesis
    routines, four lines were synthesised in
    both 1D LTE and 1D NLTE as an additional investigation
    into the potential extra broadening seen in \linfor. These lines were
    $5772.46$, $5948.54$, $7680.27$ and \lineAng{12288.15}. We fit the 1D NLTE line
    profiles with the 1D LTE syntheses for nine different abundance values to
    investigate whether the consequent fitted broadening is strongly affected. The
    abundance corrections generally remain within the prescribed $-0.01$ dex; the
    \lineAng{12288.28} requires slightly higher corrections at higher abundances.
    The employed model atom is the same as that in
    \citet{wedemeyerStatisticalEquilibrium2001} with $115$ energy levels for
    \ion{Si}{I} and \ion{Si}{II} with $84$ transitions. Moreover, the collisional
    cross-sections for neutral particles follow \citet{drawinCOLLISIONTRANSPORT1967}
    and \citet{steenbockStatisticalEquilibrium1984} using a correction factor of
    $S_\mathrm{H}=0.1$.
    Fig.~\ref{fig:matthias_comparison} shows the fitted broadening at each abundance
    point for each line, and Fig.~\ref{fig:lte_nlte_syn} shows the LTE versus NLTE line
    profiles. In this small sample, NLTE effects become stronger with increasing
    abundance resulting in stronger negative broadening for
    the \lineAng{5772.46}, \lineAng{5948.54} and \lineAng{12288.15} lines, and less
    positive broadening for the \lineAng{7680.27} line. It should be noted that the
    trends are reversed if we fit NLTE to LTE profiles, i.e. lines that showed
    negative broadening would instead show positive broadening.
    The three lines with a negative NLTE correction also require negative broadening,
    suggesting that part of the overly broadened line profiles could be attributed
    to NLTE effects. We also found that removing the core of the line (within
    $\pm 3$ \kms) removes the necessity for negative broadening, illustrating that
    the NLTE effects are concentrated in the cores of these lines.

    \clearpage
    \newpage
    %
    \section{De-broadening of spectral lines}
    \label{appendix:ddsmooth}
    \noindent
    Mathematically, the effect of broadening is described as convolution and
    de-broadening as deconvolution. It is well-known that deconvolution is an
    ill-posed problem. A robust solution to a deconvolution problem can only be
    obtained by suitable regularisation. The problem becomes immediately apparent
    when considering a Gaussian (the ``kernel'') with which a spectrum is to be
    de-convolved. In Fourier space, deconvolution corresponds to a division with
    the Fourier transform of the kernel, which is again a Gaussian in this case. The rapid decline
    of the wings of a Gaussian leads to a division by almost zero at distances of a
    few widths of the Gaussian away from its centre. Any numerical or physical
    imprecision leads to large disturbance under such circumstances. This means even
    noise-free synthetic line profiles cannot be de-convolved by a naive division
    of the raw spectrum by the Fourier transform of a Gaussian kernel.

    The key question is now whether one can suitably regularise the problem or
    reduce the level of noise amplification. We followed the latter approach by
    seeking a different kernel with less steeply falling wings than a Gaussian. As
    starting point we chose the kernel function
    \begin{equation}
        G_1(x) = \frac{\alpha}{2} e^{-\ddx},
        \label{kernelG1}
    \end{equation}
    normalised to one according to
    \begin{equation}
        \int_{-\infty}^{\infty} dx \,\, G_1(x) = 1\,.
        \label{G1norm}
    \end{equation}
    The parameter~$\alpha$ controls the width of the kernel. The kernel is
    symmetric, and has a peaked shape at the origin. Later, we shall discuss
    higher powers in terms of convolutions of the kernel with itself. We index the
    resulting kernels by the number of convolutions, which in the present case is
    one. Convolving a function~$f$ with kernel $G_1$ results in a function~$g$
    according to
    \begin{equation}
        g(\x) = \int_{-\infty}^{\infty}  d\xx\,G_1(\xx-\x)\,f(\xx)\,.
        \label{basicconvol}
    \end{equation}
    Our choice of the kernel function was inspired by work of \citet{dorfiSimpleAdaptive1987}.
    The authors pointed out that the above kernel is the Green's
    function associated with the differential operator
    \begin{equation}
        \mathbf{1} - \frac{1}{\alpha^2} \frac{d^{\,2}}{dx^2}\,,
    \end{equation}
    where $\mathbf{1}$ indicates the identity operator. This allows one to
    formulate the convolution expressed by Eq. \ref{basicconvol} as
    solution of a ordinary differential equation of second order. Discretising
    the differential operator, as well as the functions $f$ and $g$ (simplest
    on an equidistant \x-grid), results in a set of linear equations of the form
    \begin{equation}
        \mathbf{A}\,g = f
        \label{matrixeq}
    \end{equation}
    where $\mathbf{A}$ is a tri-diagonal matrix which can be inverted
    efficiently. Remarkably, in this formulation a deconvolution appears even
    simpler than convolution: if $g$ is given, a simple matrix multiplication
    with matrix~$\mathbf{A}$ yields the de-convolved function $f$. It was this
    feature that made us choose $G_1$ as kernel for deconvolution, and for consistency, also for
    convolution. Using the same kernel when convolving a line profile has the
    advantage that during fitting there is a continuous transition from
    convolution to deconvolution and vice versa. This improves the
    stability of the fitting operation when having to deal with a situation where
    the broadening or de-broadening is around zero. It should be noted that the
    $G$ kernels are in fact the Mat\'ern functions with half-numbered indices
    \citep{gentonClassesKernels2002}.

    One may question the suitability of the function $G_1$ for describing
    broadening or de-broadening effects, and might prefer a more Gaussian-shaped
    kernel. The Central Limit Theorem states that repeated convolution of a
    function with itself (subject to certain regularity conditions) approaches a
    Gaussian. We used this property to construct further kernel functions by
    convolving $G_1$ with itself. We obtained the following sequence of functions
    \begin{equation}
        G_2(x) \equiv G_1^2(x) =\frac{\alpha}{4} e^{-\ddx} (1+\ddx),
    \end{equation}
    \begin{equation}
        G_3(x) \equiv G_1^3(x) =  \frac{\alpha}{16} e^{-\ddx} \left[3(1+\ddx) + \ddxsq\right],
    \end{equation}
    \begin{equation}
        G_4(x) \equiv G_1^4(x) =  \frac{\alpha}{96} e^{-\ddx} \left[15(1+\ddx) + \ddxsq(6+\ddx)\right].
    \end{equation}
    All functions are normalised to one according to Eq. \ref{G1norm}.
    The possibility of formulating the convolution operation in Fourier space
    let it appear desirable to have the Fourier transforms of the $G_n$ function
    at hand. Defining the Fourier transform via
    \begin{equation}
        \widehat{G}(k) \equiv \frac{1}{\sqrt{2\pi}} \int_{-\infty}^{\infty} G(x)\,\exp(ikx)\,dx
    \end{equation}
    we obtained for the (purely real) transforms of the kernel functions
    \begin{equation}
        \widehat{G}_n(k) = \frac{\alpha^{2n}}{\sqrt{2\pi}\,(\alpha^2+k^2)^{n}}
        \hspace{2em} n \in \{1,2,3,4\}\,.
    \end{equation}

\begin{figure}[h]
        \resizebox{\hsize}{!}{\includegraphics{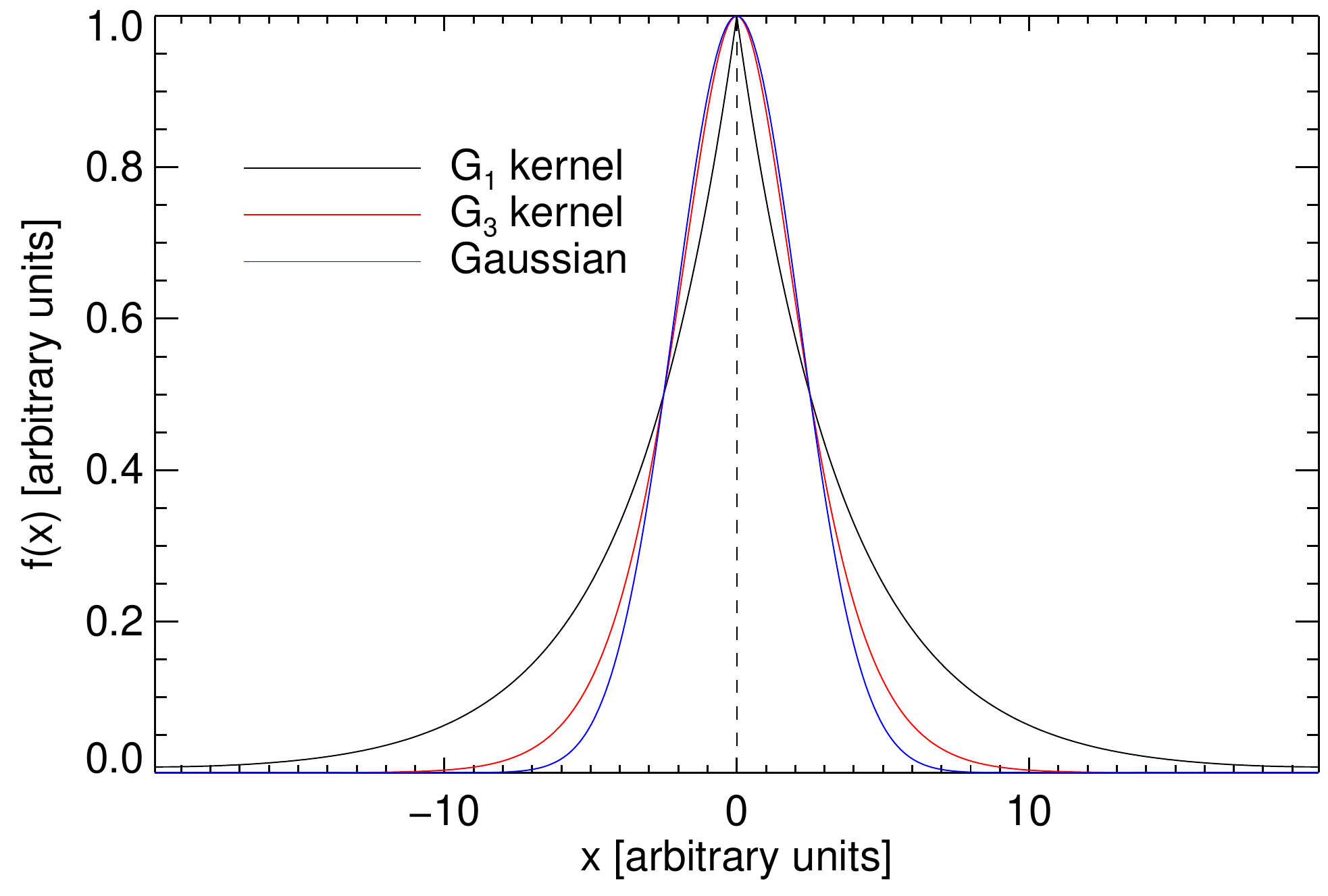}}
        \caption{Shapes of the kernels $G_1$ and $G_3$ in comparison with a
            Gaussian. For display purposes all functions
            were normalised to $G_i(0)=1$. All share the same FWHM.}
        \label{f:ddkernelshapes}
    \end{figure}

    Figure~\ref{f:ddkernelshapes} illustrates the shapes of kernels $G_1$ and
    $G_3$. $G_3$ already resembles a Gaussian quite well, having a smooth maximum
    at the origin and wings that are moderately wider than in the case of a
    Gaussian. One should keep in mind
    that approaching a Gaussian brings back in the problems that we intended to
    mitigate, namely significant amplification of noise. Hence, one should limit
    the number of repeated convolutions.
    A last point concerns the choice of the broadening parameter $\alpha$. While
    the formulae take a simple form by using $\alpha$, physically one likes to
    specify the width of the kernel more directly. Fortunately, except for the
    pre-factor, all kernels are functions of the product \ddx\ alone, resulting
    in a simple (inverse) relation between the full width at half maximum (FWHM)
    of a kernel and $\alpha$. For kernel $G_1$, the relation reads $\alpha =
        2\sqrt{\ln 2}/\mathrm{FWHM}$. The other kernels can be obtained from $G_1$ by
    two-, three-, and four-fold application of $G_1$ with a FWHM of 0.42243,
    0.29746, and 0.24325 times the targeted FWHM, respectively. The numbers were
    obtained by solving numerically the transcendent equations for the widths of
    the three kernels in question, and the approach was applied when creating
    Figure \ref{f:ddkernelshapes}.
    \begin{figure}[h]
        \resizebox{\hsize}{!}{\includegraphics{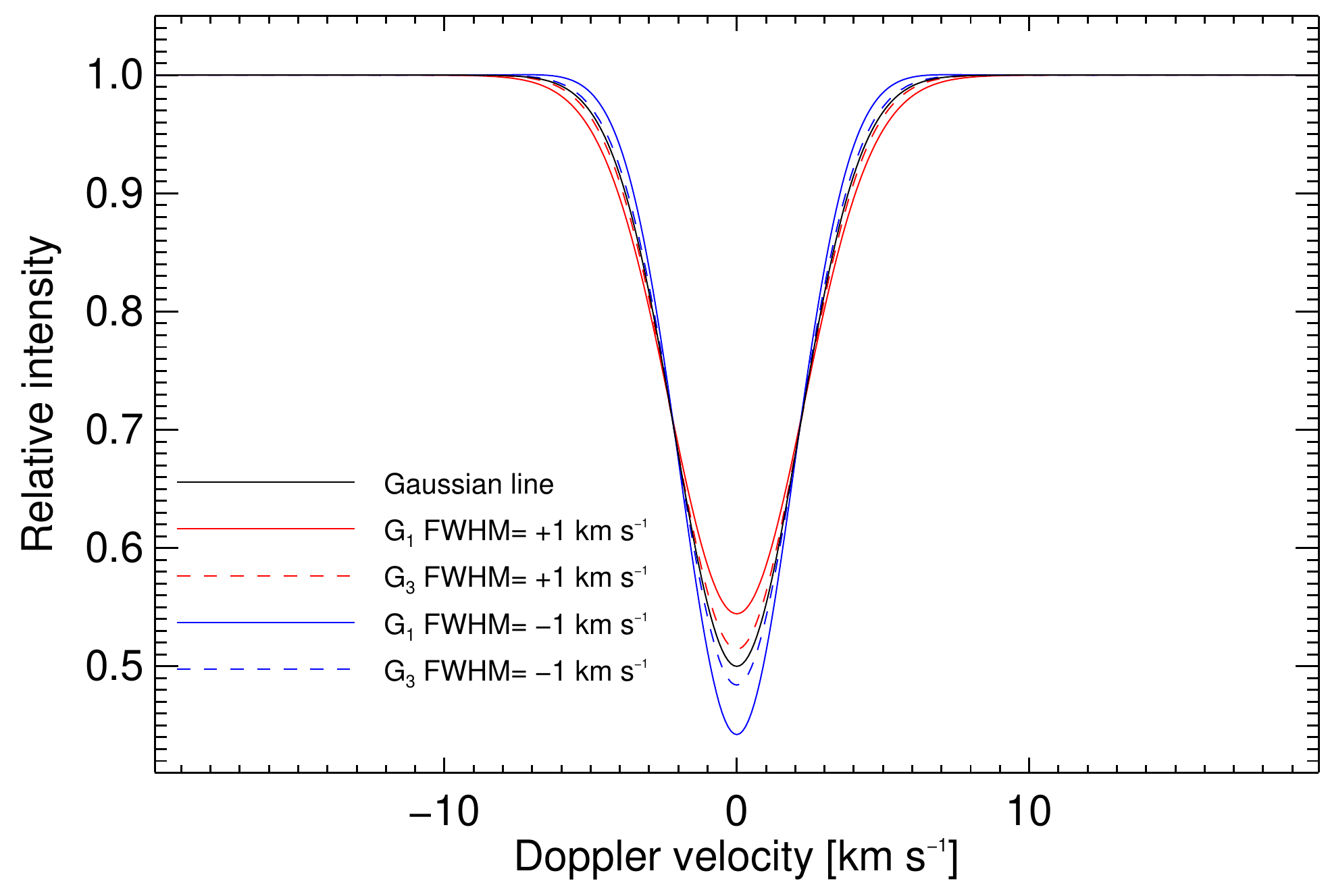}}
        \caption{Broadening and de-broadening effects of kernels G$_1$ and G$_3$:
            a Gaussian-shaped line with FWHM of \mbox{5\,km\,s$^{-1}$} was
            convolved and de-convolved with kernels having an absolute width of
            \mbox{1\,km\,s$^{-1}$}. Negative broadening values indicate a
            deconvolution.}
        \label{f:ddkerneleffects}
    \end{figure}

    Figure \ref{f:ddkerneleffects} illustrates the impact of convolving and
    de-convolving a Gaussian-shaped line profile with kernels $G_1$ and $G_3$. The
    FWHM of the line and the levels of broadening or de-broadening were roughly tailored
    after the silicon lines we observe in solar disk-centre spectra.
    $G_3$ has a milder effect on the line shape than $G_1$ at the same
    FWHM. While not clear from the plot, it turns out that one can largely mimic
    the effect of $G_1$ by using $G_3$ with a greater width. For the given
    parameters the ``peaky'' shape of $G_1$ leaves no imprint in the line
    shape. So far, the kernels described above worked in practice, but deconvolution can be handled only for
    kernels significantly narrower than the spectral line to be de-broadened.
    \clearpage
    \newpage
    %
    \section{Centre-to-limb variation of the continuum and line shapes predicted by
      the 3D models}
    \label{appendix:clv}
    On request by the referee here we report on the performance of our 3D models
    in combination with our spectral synthesis codes ({\tt Linfor3D} for line
    syntheses, {\tt NLTE3D} for spectral energy distributions) when representing
    the solar centre-to-limb variation (CLV) of its continuum radiation and the
    shape of lines. We start with the CLV, and later point to investigations in
    the literature which use features of line shapes that were calculated with the
    help of \cobold\ models.

    Figure~\ref{f:clv} shows a comparison between prediction of our 3D models with
    observations by \citet[][hereafter NL]{neckelSolarLimb1994}. NL provide 30
    wavelengths from the near-UV to the near-IR for which they measured relative
    intensities over narrow bandwidths (in equivalent Doppler velocity between 1.4
    to 1.9\,\kms). The wavelength points were chosen to be largely free of
    absorption lines. We ran pure continuum syntheses for the NL wavelengths for
    our four 3D models, and for further comparison for three 1D model
    atmospheres. The lower panels (left panels in rotated view) depict the
    result. For the non-magnetic models msc600 and n59 the correspondence between
    observation and model prediction is very good for
    $\lambda>0.6\,\mu\mathrm{m}$. We note that the results for models msc600 and n59
    are almost identical so that they are difficult to distinguish on the scale of
    the plot. 1D models are added for setting the scale for a ``very good''
    correspondence: a standard {\tt ATLAS9}
    \citep{kuruczModelAtmospheres1979, sbordoneATLASSYNTHE2004, kuruczATLAS12SYNTHE2005}
    model
    shows a significantly steeper CLV than found in the observations. Similarly,
    an {\tt LHD} model ({\tt LHD} is a home-grown 1D stellar atmosphere code that
    uses the same microphysics as applied in our 3D models) shows a similar
    behaviour. Remarkably, our msc600 model performs even slightly better than the
    well-known Holweger-M\"uller solar model
    \citep{holwegerEmpirischesModell1967,holwegerPhotosphericBarium1974}
    despite the fact that this
    semi-empirical model was constructed to match the CLV.

    At $\lambda < 0.6\,\mu\mathrm{m}$ the good correspondence
    deteriorates. However, as NL point out themselves the notion of observing pure
    continuum over the bandwidth used in the measurements becomes
    questionable. At wavelengths shorter than $0.6\,\mu\mathrm{m}$
    the previously negligible contribution of line absorption to the observed
    intensity becomes non-zero but otherwise undefined. To address this issue we
    took an extreme stand and calculated the intensity over $20\,\mbox{\AA}$ wide
    intervals including \textit{all} line absorption with the help of opacity
    distribution functions (ODFs). We combined line ODFs from Kurucz' {\tt ATLAS}
    suite \citep{kuruczATLAS9Model2017} with continuous opacities from our own
    opacity package. In each ODF interval the distribution of the line opacity was
    represented by a step-function of 12 steps. For each step, the emergent
    intensity was calculated and finally integrated over the whole ODF interval so
    that we obtained the intensity emerging in the $20\,\mbox{\AA}$ wide ODF
    interval. From the construction is is clear that one obtains the average
    effect on the intensity of the absorbers present in the ODF interval.
    The upper panels (right panels in rotated view) of Fig.~\ref{f:clv} illustrate
    that the correspondence between model predictions and observations clearly
    improves for wavelengths $\lambda<0.6\,\mu\mathrm{m}$, strongly suggesting that
    indeed missing line absorption is the reason for the mismatch at these shorter
    wavelength in the pure continuum calculations. In fact, the good
    correspondence in the near-UV is striking. One has to admit that this is
    certainly in part fortuitous since the true contribution of lines to the
    observations is unclear, and also whether the line lists going into the
    construction of the ODFs are sufficiently complete in the near-UV.  To
    illuminate the influence of line absorption a bit further we write the
    intensity ratio between a location~$\mu$ and and disk centre $\mu=1$ as
    \beq
    \frac{I_\mu}{I_1}
    = \frac{c_\mu-l_\mu}{c_1-l_1}\approx \frac{c_\mu}{c_1}\left(1-\frac{l_\mu}{c_\mu}+\frac{l_1}{c_1}\right)
    \label{e:iratio}
    \eeq
    where $c_\mu$ is the continuum intensity at location~$\mu$ and $l_\mu$ the
    intensity reduction by line absorption at this point. The approximate equality
    holds for weak line absorption. Equation~\eref{e:iratio} shows that there is a
    compensatory effect by considering intensity ratios. Moreover, since there is
    typically a reduction of the intensity ratio with respect to pure continuum
    calculations the contribution of line absorption must become larger towards
    the stellar limb. It is not straight forward to see why this is: the silicon
    lines synthesised for this investigation generally show a mild decrease in
    strength towards the limb -- particularly the medium to strong lines. However,
    we conjecture that in fact molecular lines are a major player here which
    significantly increase in strength due to the lower temperatures at which line
    formation takes place towards the limb.

    Since the contribution of lines is observationally not well-defined we have
    also plotted intensity ratios considering ODF sub-intervals
    only. We expected that especially the sub-interval with the smallest line
    contribution would result in a closer match to the observations which was,
    however, not immediately apparent (not shown). We finally remark that the
    findings discussed here coincide with results on the CLV based on
    a \cobold\ model of an earlier generation \citep{ludwigSolarAbundances2010}.

    For completeness we also investigated the CLV of the magnetic models b000 and
    b200. Figure~\ref{f:clv} illustrates that the field-free model provides a
    reasonable match to the observations while the 200\,G model is clearly
    off. For the given (somewhat artificial) field configuration one may conclude
    that the observed CLV does not permit a mean field strength
    $\gtrsim 50\,\mathrm{G}$ on the Sun. \citet{pereiraHowRealistic2013} already
    arrived at a similar conclusion by comparing a 100\,G model with a field-free
    case.

\begin{figure}
        \resizebox{\hsize}{!}{\includegraphics{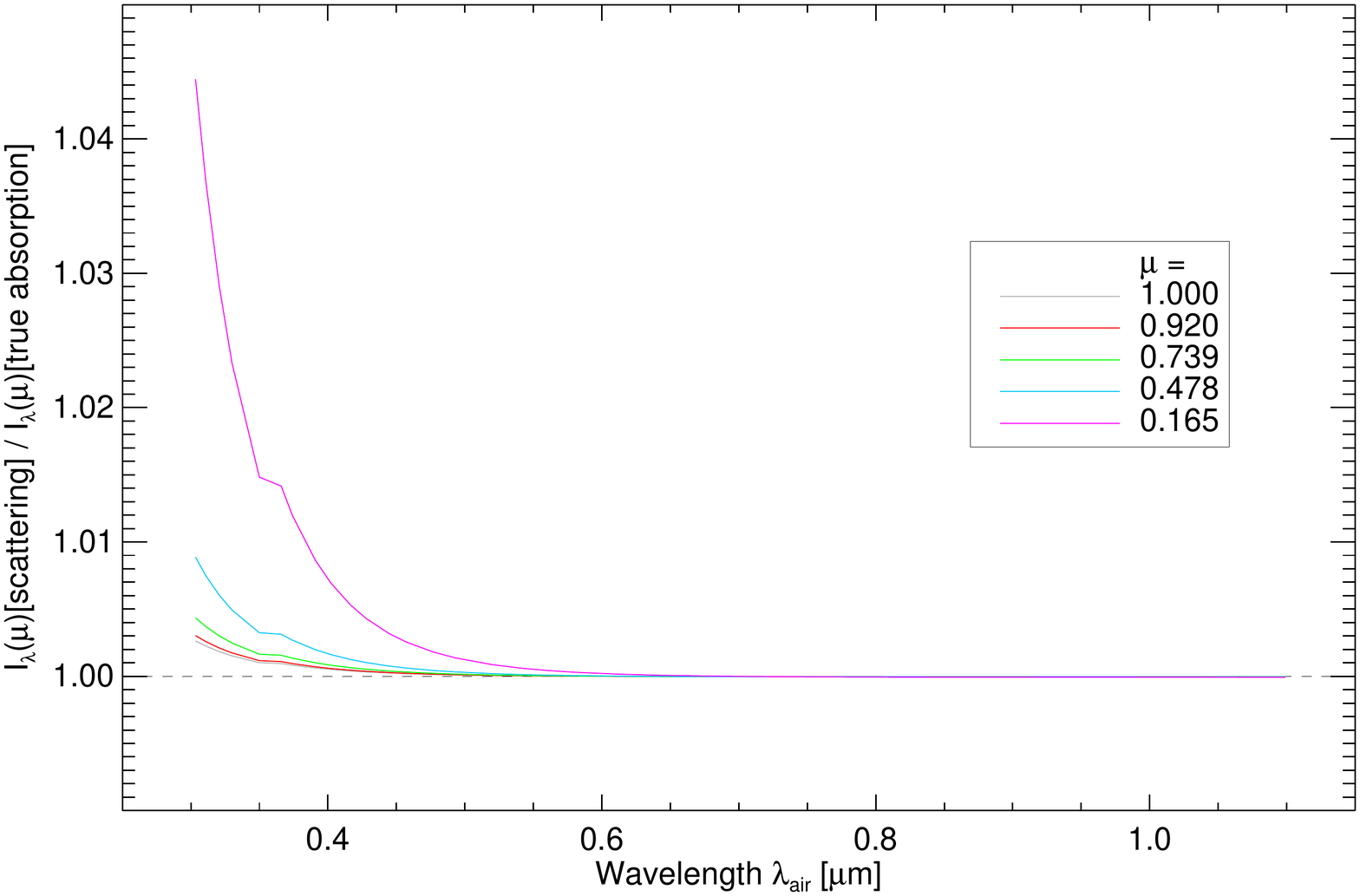}}
        \caption{Ratio of the emergent intensities as function of wavelength and
            limb angle cosine $\mu$ between cases when continuum scattering is treated
            exactly or as true absorption. The average vertical
            profile of model~msc600 was
            used as structure in the spectral synthesis calculations.}
        \label{f:scatotabs}
    \end{figure}

\begin{figure}
        \resizebox{\hsize}{!}{\includegraphics{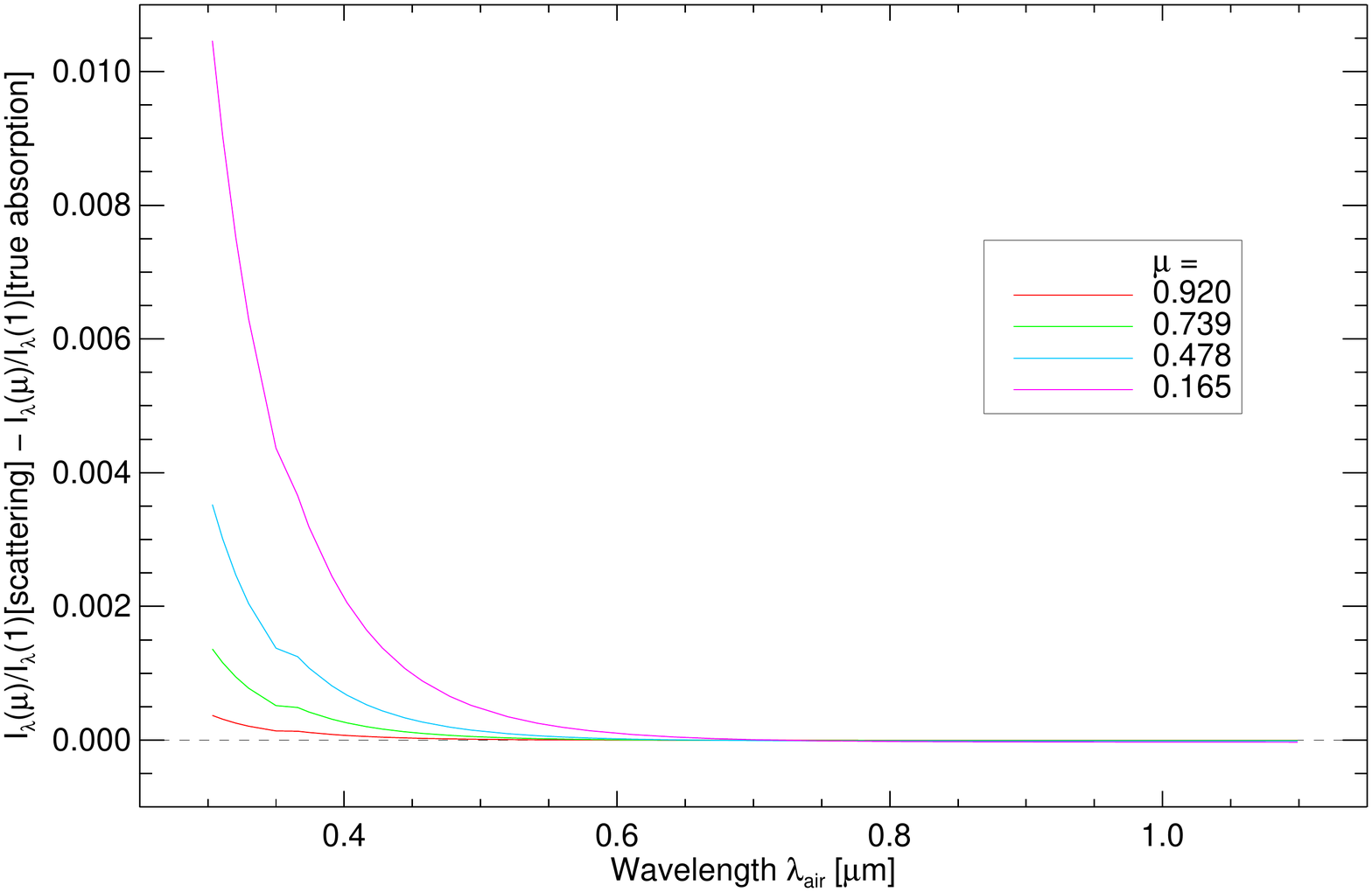}}
        \caption{Difference of the centre-to-limb variation when continuum
            scattering is treated exactly or as true absorption. Note the reduction of
            the impact of scattering in comparison to Fig.~\ref{f:scatotabs}.
        }
        \label{f:clvscattering}
    \end{figure}

\begin{figure*}
        \mbox{}\hfill%
        \resizebox{0.43\hsize}{!}{\includegraphics[angle=90]{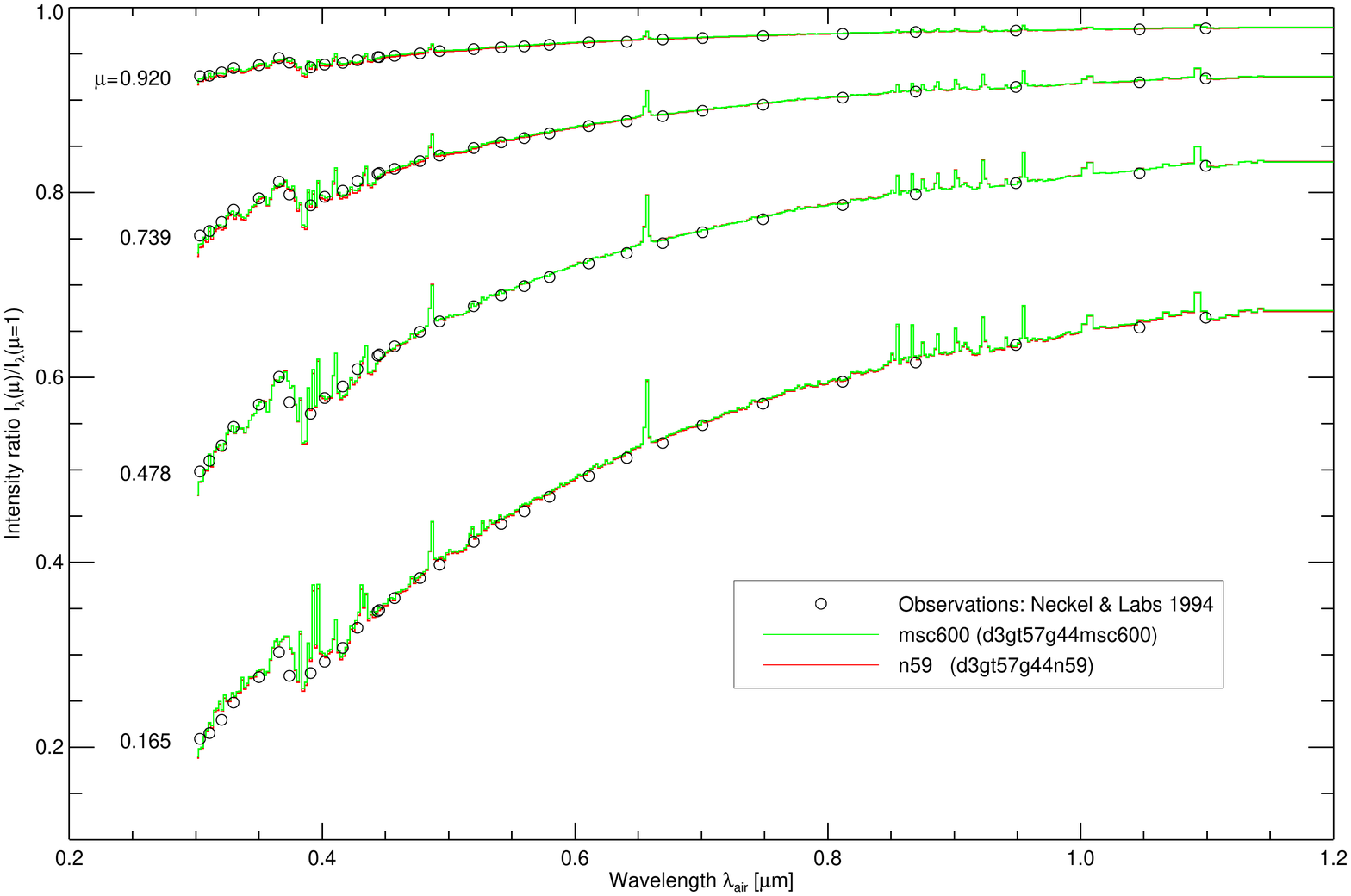}}\hspace{2ex}%
        \resizebox{0.43\hsize}{!}{\includegraphics[angle=90]{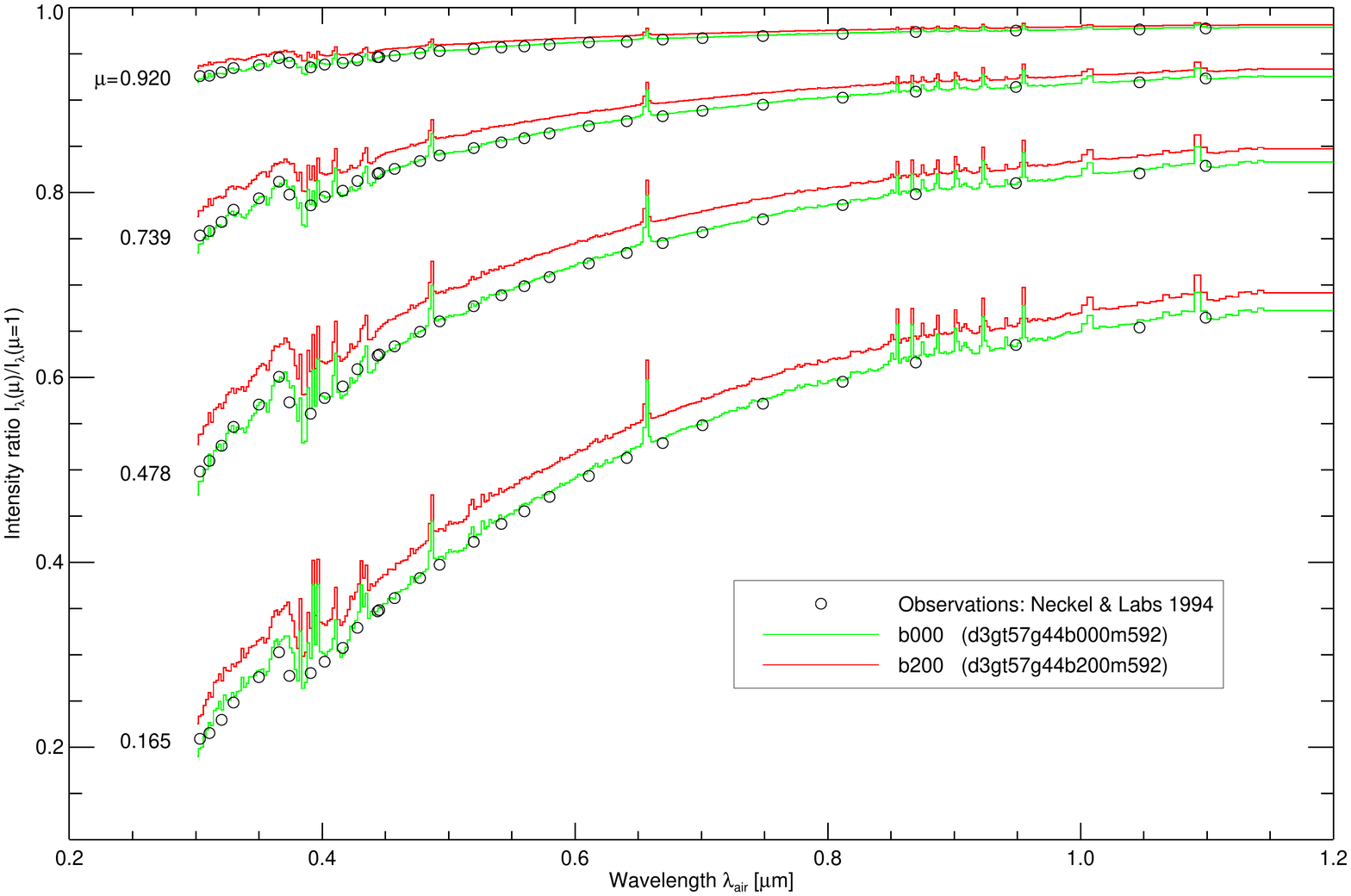}}\hfill\mbox{}\\[-1ex]
        \mbox{}\hfill%
        \resizebox{0.43\hsize}{!}{\includegraphics[angle=90]{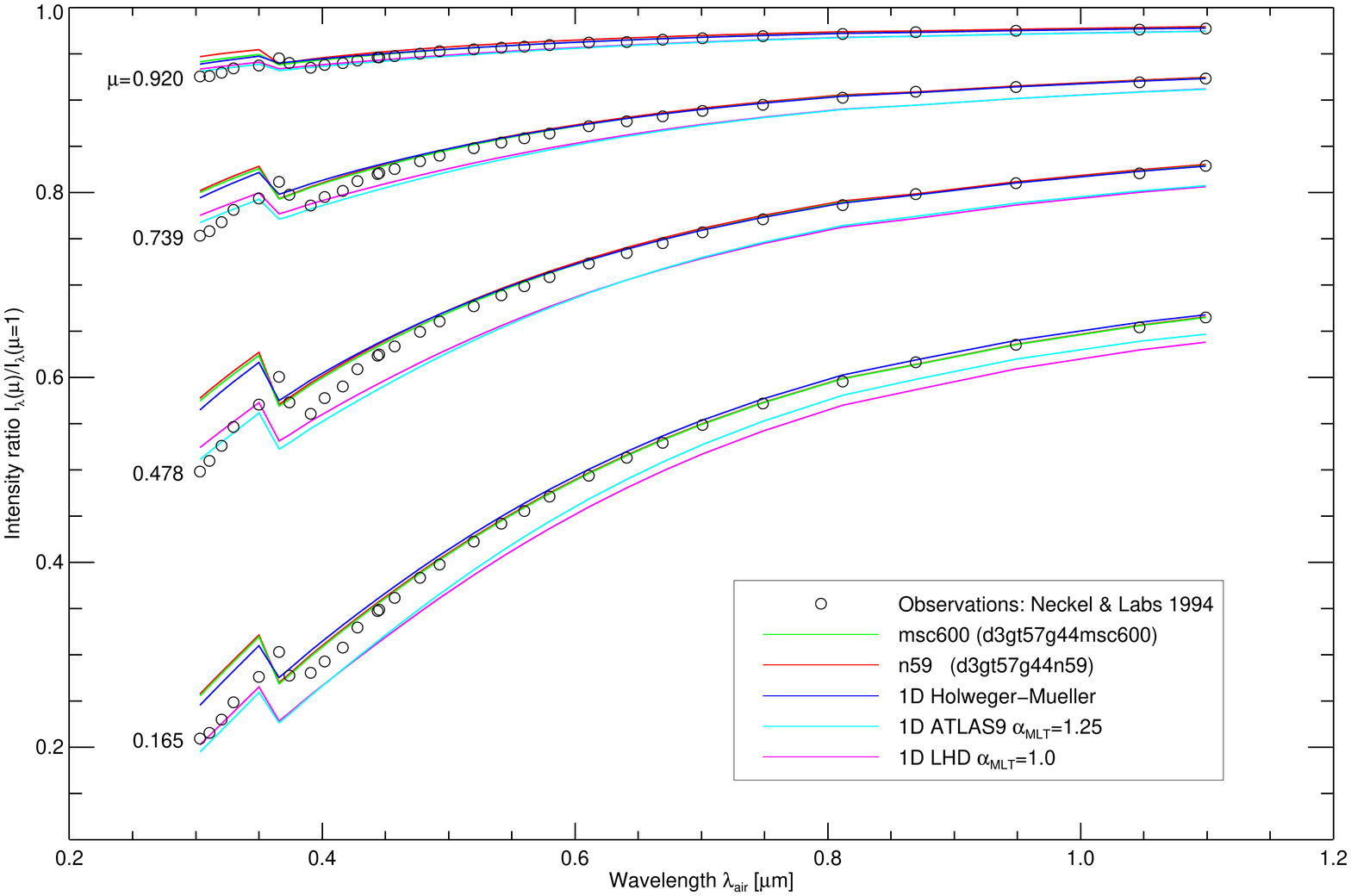}}\hspace{2ex}%
        \resizebox{0.43\hsize}{!}{\includegraphics[angle=90]{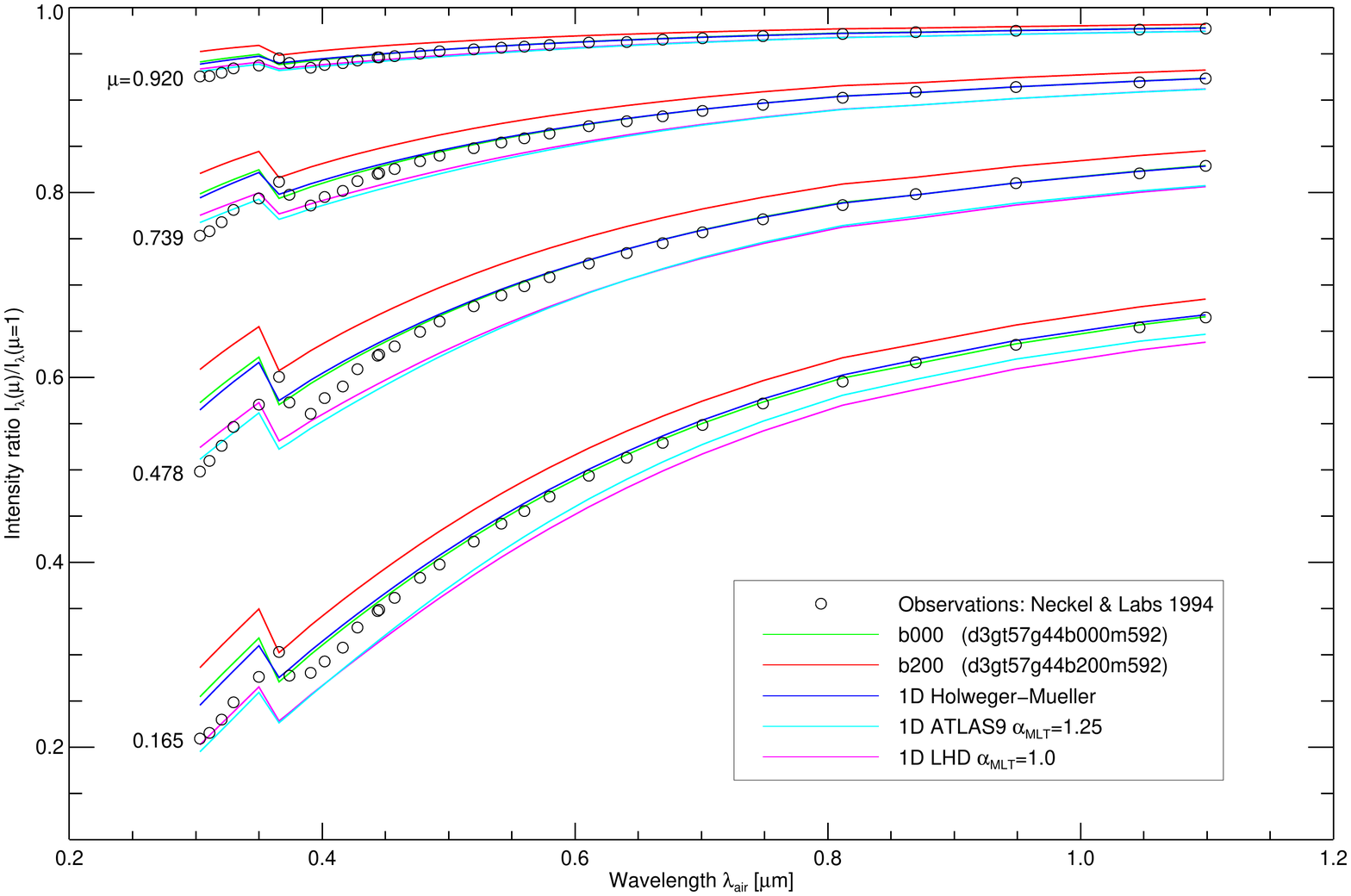}}\hfill\mbox{}
        \caption{Centre-to-limb variation as predicted by our 3D models in
            comparison to observations and 1D results. Syntheses excluding (bottom)
            and including (top) line
            absorption are shown. Left
            panels: models msc600 and n59. Right panels: models b000 and b200.}
        \label{f:clv}
    \end{figure*}

    All spectral synthesis calculations for the 3D models underlying
    Fig.~\ref{f:clv} approximate scattering in the continuum as well as lines as
    true absorption. We investigated the effect of this approximation on the CLV
    in the continuum by comparing the cases of isotropic coherent scattering and
    true absorption in a 1D stratification obtained when averaging (over optical
    depth surfaces and time) the msc600 model. As a first step,
    Fig.~\ref{f:scatotabs} illustrates that scattering generally leads to an
    increase of the emergent intensity for wavelength $\lesssim 0.65\,\mu\mbox{m}$,
    and that effects becomes more pronounced towards the solar limb. However,
    quantitatively the changes of the intensity are modest ($\lesssim
        4.5\,\%$). Figure~\ref{f:clvscattering} shows that this translates into small
    differences ($\lesssim 0.01$) in the CLV, again, mostly at short wavelengths
    and close to the limb. The overall limb darkening reduces the effect apparent in
    the intensity ratios directly. To see this, we denote by $S$ the intensity when scattering
    is treated exactly, by $A$ the intensity when scattering is treated as true
    absorption. The difference in the CLV can then be written as
    \beq
    \frac{S(\mu)}{S(1)} - \frac{A(\mu)}{A(1)} \approx \frac{S(\mu)-A(\mu)}{A(1)}
    = \frac{A(\mu)}{A(1)}\,\left(\frac{S(\mu)}{A(\mu)} - 1\right)\, .
    \eeq
    The approximate equality comes from the observation that $S(1)\approx A(1)$.
    All this means that the very good correspondence between models and
    observations shown in Fig.~\ref{f:clv} gets only very slightly worse when
    scattering is treated correctly but remains very satisfactory.

    We now briefly turn to line shapes, particularly line shifts, as predicted by
    model n59. As a first example we point to
    \citet[][Fig.~6]{gonzalezhernandezSolarGravitational2020} where absolute core
    shifts of 144
    \ion{Fe}{I} lines are compared to observations. In their work a very accurate
    wavelength calibration of the observed spectra was achieved by utilising a
    laser frequency comb.  Lines with an equivalent width less than
    $60\,\mbox{m\AA}$ show a good correspondence with their observed shifts. One
    has to keep in mind here that a comparison on an absolute scale also relies on
    accurately known laboratory wavelengths. The reason for the mismatch of lines
    with equivalent widths greater than $60\,\mbox{m\AA}$ is not clear but not
    necessarily related to shortcomings in the model structure. Moreover, it was
    also seen in LTE line syntheses based on {\tt STAGGER} models \citep[see][for
        further discussion]{gonzalezhernandezSolarGravitational2020}.
    As second example, in \citet[][Fig.~19]{lohner-bottcherConvectiveBlueshifts2019}
    the observed
    shape of the \ion{Fe}{I} 6173\,\AA\ line is compared to synthetic line
    profiles computed with a \cobold\ model as a function of limb angle. While not
    all details are matched the overall correspondence is satisfactory.
    \clearpage
    \newpage

    \section{The partition functions of silicon as implemented  \linfor}
    \label{appendix:partition}
\begin{figure}[h]
        \resizebox{\hsize}{!}
        {\includegraphics{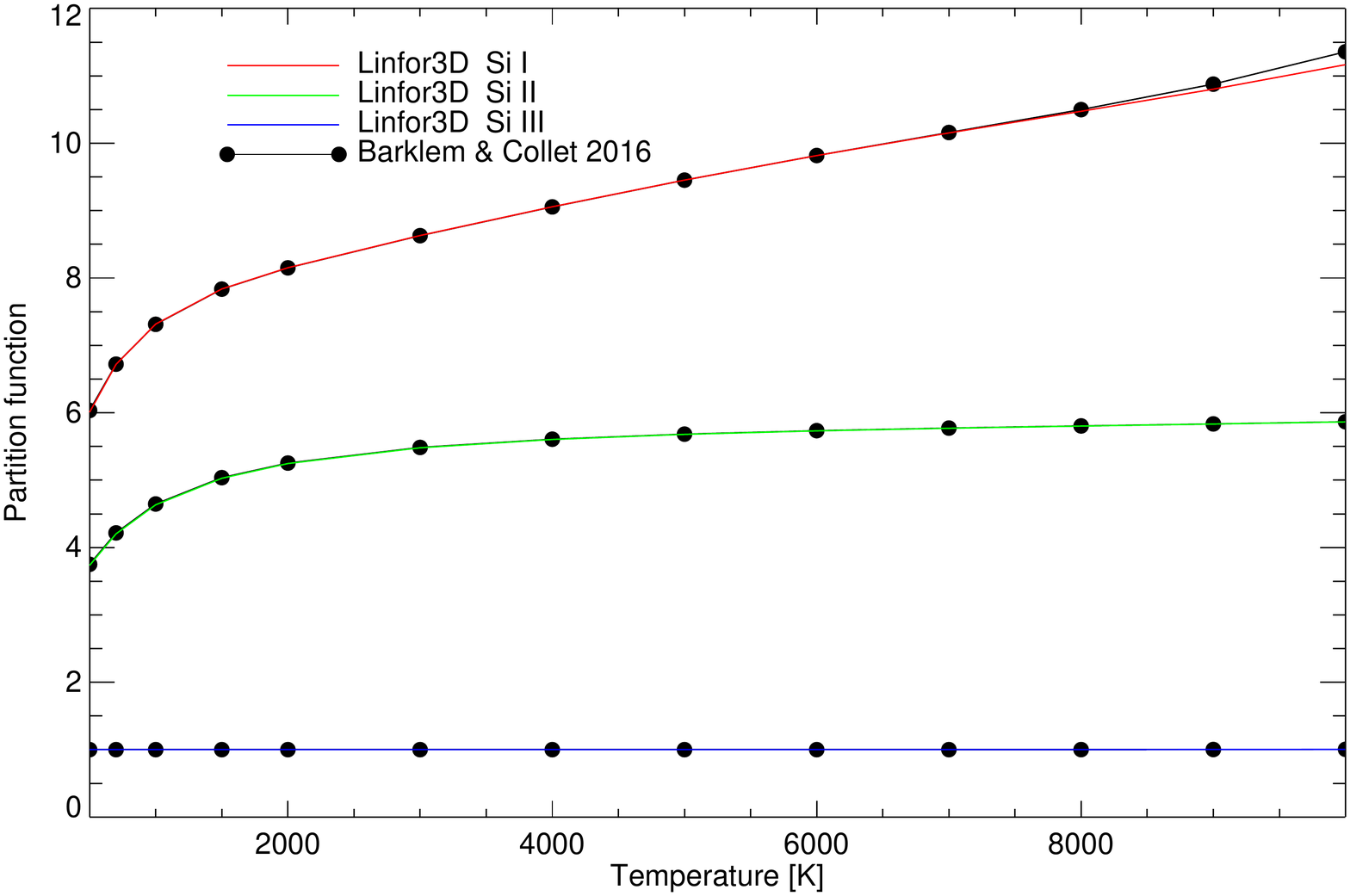}}
        \caption{Partition functions implemented in \linfor\ compared to data of
            \citet{barklemPartitionFunctions2016}.
        }
        \label{fig:partition}
    \end{figure}

    \noindent
    We compared the partition functions of the first three ionisation stages of
    silicon as implemented in our spectral synthesis code to data given in the
    recent compilation of \citet{barklemPartitionFunctions2016}. We find a close to
    perfect agreement in the temperature range relevant for the formation of
    silicon lines in the solar photosphere, as seen in Fig.~\ref{fig:partition}.

\end{appendix}

\end{document}